\documentclass[acmtog,timestamp]{acmart}

\usepackage{epsfig}
\usepackage{graphicx}
\usepackage{amsmath}
\usepackage{amssymb}
\usepackage{float}
\usepackage{amsfonts}
\usepackage{bm}
\usepackage{url}
\usepackage{verbatim}
\usepackage{color}
\usepackage{subcaption}
\usepackage{enumitem}
\usepackage{tabularx}
\usepackage{amsthm}
\usepackage{icomma}
\usepackage{booktabs}
\usepackage{multirow}
\usepackage{array}
\usepackage{enumitem}
\usepackage{soul}
\usepackage{placeins}
\usepackage{appendix}
\usepackage[framemethod=tikz]{mdframed}
\newcolumntype{P}[1]{>{\raggedright\arraybackslash}p{#1}}

\usepackage{algorithm}
\usepackage{algpseudocode}

\newcommand{\bfx}{{\bf x}}
\newcommand{\bfy}{{\bf y}}
\newcommand{\bfs}{{\bf s}}

\newcommand{\bfv}{{\bf v}}

\newcommand{\bfa}{{\bm \alpha}}

\newcommand{\rdim}[1]{\mathbb{R}^{#1}}

\setcitestyle{square}

\acmJournal{TOG}
\acmVolume{38}
\acmNumber{5}
\acmArticle{148}
\acmYear{2019}
\acmMonth{10}

\title[KRISM]{KRISM --- Krylov Subspace-based Optical Computing of Hyperspectral Images}

\author{Vishwanath Saragadam}
\author{Aswin C.\ Sankaranarayanan}
\affiliation{%
  \institution{Carnegie Mellon University}
  \department{Electrical and Computer Engineering}
  \city{Pittsburgh}
  \state{PA}
  \postcode{15213}
  \country{USA}
}
\email{saswin@andrew.cmu.edu}


\setcopyright{rightsretained}

\acmDOI{https://doi.org/10.1145/3345553}
\hypersetup{draft}
\begin{document}


\begin{teaserfigure}
   \includegraphics[width=\textwidth]{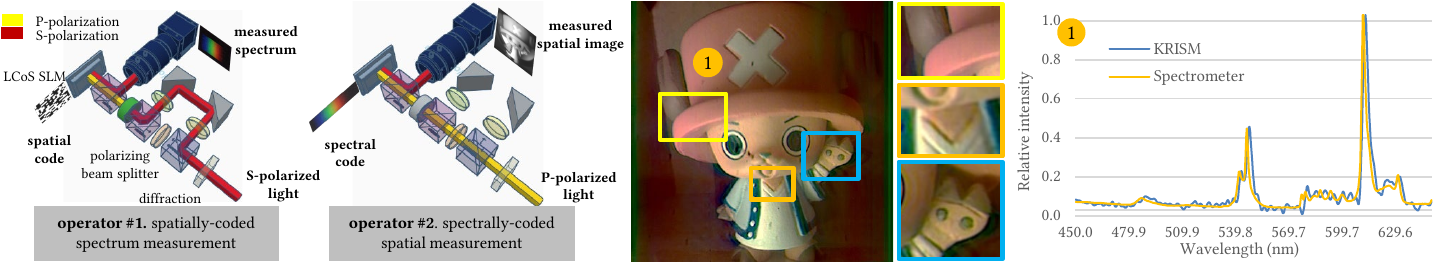}
   \caption{Hyperspectral imagers resolve scenes at high spatial and spectral resolutions.  We propose a novel architecture called KRISM that provides the ability to implement two operators: a spatially-coded spectrometer and a spectrally-coded spatial imager. By iterating between the two, we can acquire a low rank approximation of the hyperspectral image in a light efficient manner with very few measurements. The left image shows optical schematics for implementing the two operators. On the right, we show a hyperspectral image of a scene illuminated with a compact fluorescent lamp (CFL) acquired  using our lab prototype. The proposed method enables high spatial and spectral resolution  as observed  in  the zoomed-in image patches and CFL peaks, respectively.}
   \label{fig:chopper}
\end{teaserfigure}

\begin{abstract}
We present an adaptive imaging technique that optically computes a low-rank approximation of a scene's hyperspectral image, conceptualized as a matrix.
Central to the proposed technique  is the optical implementation of two measurement operators: a spectrally-coded imager and a spatially-coded spectrometer. 
By iterating between the two operators, we show that the top  singular vectors and singular values of a hyperspectral image can be adaptively and optically computed with only a few iterations.
We  present an optical design that uses pupil plane coding for implementing the two operations and show several compelling results using a lab prototype to demonstrate the effectiveness of the proposed hyperspectral imager.
\end{abstract}
%


\maketitle
%
%





\newtheorem{theorem1}{Theorem}
\newtheorem{theorem2}{Theorem}
\newcommand{\mt}{KRISM}

\section{Introduction} \label{section:intro}
Hyperspectral images (HSIs) capture light intensity of a scene as a function of space and wavelength and have been used in numerous  vision \cite{pan2003face,tarabalka2010segmentation,kim20123d}, geo-science and remote sensing applications \cite{cloutis1996review,harsanyi1994hyperspectral}.
Traditional approaches for hyperspectral imaging, including tunable spectral filters and pushbroom cameras, rely on sampling the HSI, i.e., measuring the photon counts in each spatio-spectral voxel.
When imaging at high-spatial and spectral resolutions, the amount of light in a voxel can be quite small, thus requiring long exposures to mitigate the effect of noise.

HSIs are often endowed with rich structures that can be used to alleviate the challenges faced by traditional imagers.
For example, natural scenes are often comprised of a few materials of distinct spectra and further, illumination of limited spectral complexity \cite{parkkinen1989characteristic,lee2000spectral}.
This implies that the collection of spectral signatures observed at various locations in a scene lies close to a low-dimensional subspace.
Instead of sampling the HSI of the scene one spatio-spectral voxel at a time, we can dramatically speed-up acquisition and increase light throughput by measuring only projections on this low-dimensional subspace.
However, such  a measurement scheme requires a priori knowledge of the scene since this subspace is entirely scene dependent.
This paper introduces an optical computing technique that identifies this subspace using an iterative and adaptive sensing strategy and constructs a low-rank approximation to the scene's HSI.

The proposed imager senses a low-rank approximation of a HSI by optically implementing the so-called Krylov subspace method~\cite{golub1965calculating}.
We show that this requires two operators: a spatially-coded spectrometer and a spectrally-coded spatial imager; when we interpret the HSI as a 2D matrix, these two operators correspond to left and right multiplication of the matrix with a vector.
The two operators are subsequently used in an iterative and adaptive imaging procedure whose eventual output is a low-rank approximation to the HSI.
The proposed imager is adaptive, i.e., the measurement operator used to probe the scene's HSI at a given iteration depends on previously made measurements.
This is a marked departure from current hyperspectral imaging strategies where the signal model is merely used as a prior for recovery from non-adaptive  measurements \cite{arce2014compressive}.

\paragraph{Contributions.} We propose an optical architecture that we refer to as KRylov subspace-based Imaging and SpectroMetry (KRISM) and make the following three contributions:
\begin{itemize}[leftmargin=*]
	\item \textit{Optical computation of HSIs.} We show that optical computing of HSIs to estimate its dominant singular vectors provides significant advantages in terms of  increased light throughput and reduced measurement time.
	\item \textit{Coded apertures for resolving space and spectrum.} Sensing architectures typically  used in spectrometry and imaging are mutually incompatible due to use the of slits in spectral imaging and  open apertures in conventional imaging.  To mitigate this, we study the effect of pupil plane coding on the HSI and propose a coded aperture design that is  capable of simultaneously achieving high spatial and spectral resolutions.
	\item \textit{Optical setup.} We design and validate a novel and versatile optical implementation for KRISM that uses a single camera and a single spatial light modulator (SLM) to efficiently implement spatially-coded spectral and spectrally-coded spatial measurements.
\end{itemize}
The contributions above are supported via an extensive  set of simulations as well as  real experiments performed using the lab prototype.

\paragraph{Limitation.} The benefits and contributions described above come with a key limitation. Our method is only advantageous if there are a sufficient number of spectral bands and the hyperspectral image is sufficiently low rank. If we only seek to image with very few spectral bands or if the scene is not well approximated by a low-rank model, then the proposed method performs poorly against traditional sensing methods.

\section{Prior work} \label{section:prior}

\paragraph{Nyquist sampling of HSIs.} Classical designs for hyperspectral imaging based on Nyquist sampling include the tunable filter --- which scans one narrow spectral band at a time, measuring the image associated with spectral bands at each instant --- or using a pushbroom camera --- which scans  one spatial  row at a time, measuring the entire spectrum associated with each pixel on the row.
Both approaches are time-consuming as well as  light inefficient since each captured image wastes a large percentage of light incident on the camera.

\paragraph{Multiplexed sensing.} 
The problem of reduced light throughput can be mitigated by the use of multiplexing.
One of the seminal results in computational imaging is that  the use of multiplexing codes including the Hadamard transform can often lead to significant efficiencies either in terms of increased SNR or faster acquisition \cite{harwit1979hadamard}.
This can either be spectral multiplexing \cite{mohan2008agile} or spatial multiplexing \cite{sun2009compressive}.
While multiplexing mitigates light throughput issues, it does not reduce the number of measurements required. 
Sensing at high spatial and/or spectral resolution still requires long acquisition times to maintain a high SNR.
Fortunately, HSIs have  concise signal models that can be exploited to reduce the number of measurements.

\begin{figure}[!tt]
	\centering
	\includegraphics[width=\columnwidth]{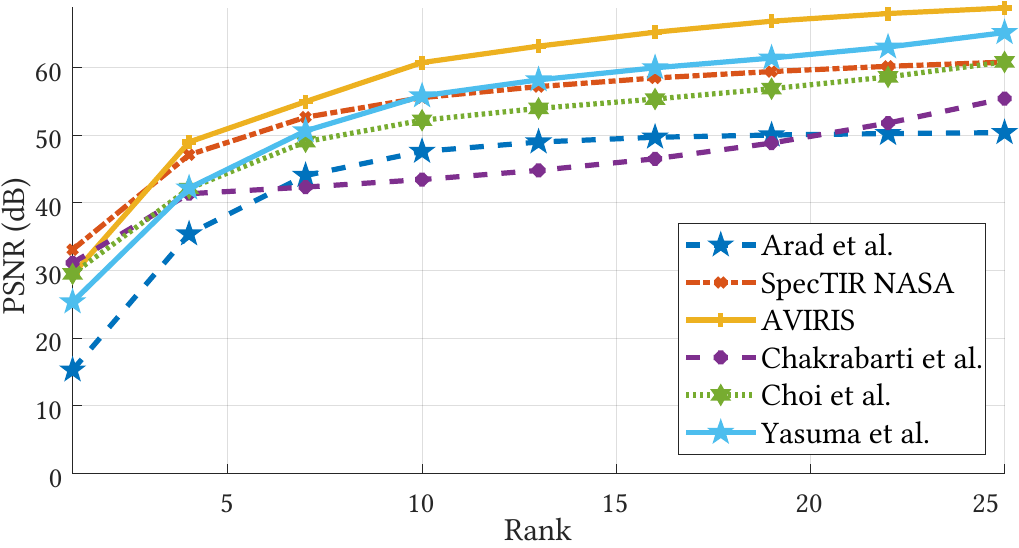}
	\caption{HSIs, interpreted as a matrix, are often low rank. We validate this observation by plotting accuracy in terms of peak SNR (PSNR) as a function of the rank of the approximation. We do this for many commonly used HSI datasets and  observe that the PSNR is higher than 40dB for a rank 10 approximation across all datasets.}
	\label{fig:snr_vs_rank}
	\vspace{-1.5em}
\end{figure}

\paragraph{Low-rank models for HSIs.}
There are many approaches to approximate HSIs using low-dimensional models; this includes group sparsity in transform domain \cite{rasti2013hyperspectral}, low rank model \cite{li2012compressive,golbabaee2012hyperspectral}, as well as low-rank and sparse model \cite{waters2011sparcs,saragadam2017compressive}.
Of particular interest to this paper is the  low-rank modeling of HSIs when they are represented as a 2D matrix (See Figure \ref{fig:snr_vs_rank}).
These models have  found numerous uses in vision and graphics including color constancy  \cite{finlayson1994color}, color displays \cite{kauvar2015adaptive}, endmember detection  \cite{winter1999nfind}, source separation \cite{hui2018illuminant}, anomaly detection \cite{saragadam2017compressive}, compressive imaging \cite{golbabaee2012hyperspectral} and denoising \cite{zhao2015hyperspectral}.
Chakrabarti and Zickler \citeNN{chakrabarti2011statistics} also provide empirical justification that HSIs of natural scenes are well represented by low dimensional models.
\begin{table*}[!ttt]
	\includegraphics[width=\textwidth]{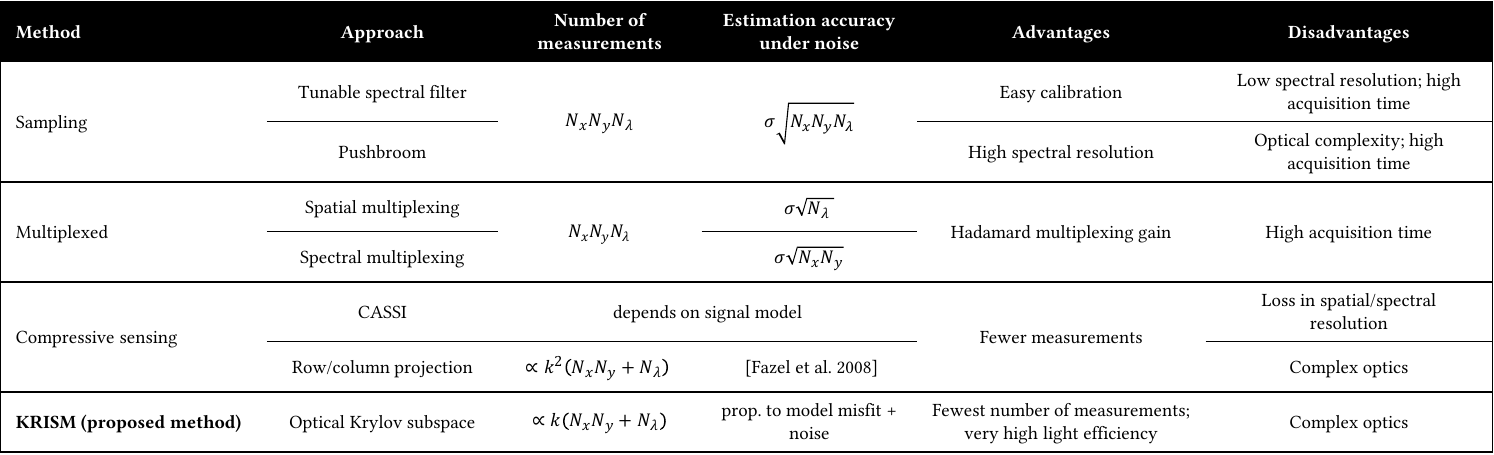}
	\caption{Various sensing strategies for hyperspectral imaging of $N_x \times N_y$ spatial dimension and $N_\lambda$ spectral bands. Noise in measurement is assumed to be AWGN with $\sigma^2$ variance. The expressions in third column represent the number of measurements required, while those in fourth column represent the error in reconstruction.} 
	\label{tab:meas_comp}
	\vspace{-2em}
\end{table*}
\paragraph{Compressive hyperspectral imaging.} The low-rank model has also been used for compressive sensing (CS) of HSIs.
CS aims to recover a signal from a set of linear measurements that are fewer than its dimensionality  \cite{baraniuk2007compressive}.
This is achieved  by modeling the sensed signal using lower dimensional representations --- low-rank matrices being one such example.
The technique most relevant to this paper is that of row/column projection \cite{fazel2008compressed} where the measurement model is restricted to obtaining row and column projections of a matrix.
Given a  matrix $X\in \rdim{m\times n}$, and measurement operators $S_\text{row} \in \rdim{p \times m}, S_\text{column} \in \rdim{n \times p}$, the measurements acquired are of the following form,
\begin{equation*}
	Y_\text{row} = S_\text{row}X,  \quad
	Y_\text{column} = XS_\text{column}.
\end{equation*}
When the matrix $X$ has a rank $k$, it can be shown that it is sufficient to acquire $p$ images and $p$ spectral profiles with  $p\propto k^2$.
In contrast, the method proposed in this paper requires only a number of measurements proportional to the rank of the matrix; however, these measurements are adaptive to the scene.
At an increased cost of optical complexity, adaptive sensing promises  accurate results with  fewer  measurements than non-adaptive measurement strategies. 

\paragraph{Hyperspectral imaging architectures.} 
Several architectures have been proposed for CS acquisition of HSIs.
The Dual-Disperser Coded Aperture Snapshot Spectral Imager (DD-CASSI) \cite{gehm2007single} obtains a single image multiplexed in both spatial and spectral domains by dispersing the image with a prism, passing it through a coded aperture, and then recombining with a second prism.
In contrast, the Single Disperser CASSI (SD-CASSI) \cite{wagadarikar2008single}  relies on a single prism that performs spatial coding using a binary mask followed by spectral dispersion with a prism.
Baek et al.\ \citeNN{baek2017compact} disperse the image by placing a  prism right before an SLR camera.
The HSI is then reconstructed by studying the dispersion of color at the edges in the obtained RGB image.
Takatani et al.\ \citeNN{takatani2017one} instead propose a snapshot imager that uses a  faced reflectors  overlaid with color filters.
Various other snapshot techniques have been proposed which rely on space-spectrum multiplexing \cite{cao2016computational,lin2014spatial,jeon2016multisampling}.
While snapshot imagers require only a single image, they often produce HSIs with reduced spatial or spectral resolutions.
Data-driven approaches such as overcomplete dictionaries \cite{lin2014spatial} and convolutional neural networks \cite{deepcassi2017} partially alleviate the loss in resolution by building priors for the HSI.
However, they require complex optimization that can often be time consuming. 

Resolution and accuracy of  the  HSI can be improved by acquiring multiple measurements instead of a single snapshot image.
Examples include multiple spatio-spectrally encoded images \cite{kittle2010multiframe}, spatially-multiplexed spectral measurements \cite{li2012compressive,sun2009compressive} or separate spatial and spectral coding \cite{lin2014dual}.
While multi-measurement techniques overcome spatial and spectral resolution limits, the price is paid in the form of increased number of measurements and hence, reduced time resolution.

Performance of snapshot techniques can be improved by tailoring the spatial masks to a given HSI dataset \cite{rueda2016compressive,rueda2017high} or by optimizing spatial masks for sensing a selected subset of spectral bands \cite{arguello2013rank}.
Optimizing the spatial masks results in increased accuracy, but still requires long reconstruction times.
A key insight into the existing methods is that the measurements are either non-adaptive and random, or adapted to a fixed signal class.
In contrast, the proposed method is \textit{adapted to the specific instance of the signal}, requires fewer measurements,  and has practically no post-processing for reconstruction.
Table \ref{tab:meas_comp} compares and contrasts various HS imaging strategies and their relative merits in terms of number of measurements and error in reconstruction.
We next discuss the concept of Krylov subspaces for low-rank approximation of matrices, which motivates iterative and adaptive techniques and paves the way to the proposed method.

\paragraph{Krylov subspaces.}
Central to the proposed method is a class of techniques, collectively referred to as Krylov subpaces, for estimating singular vectors of matrices.
Recall that the singular value decomposition (SVD) of a matrix $X\in \rdim{m\times n}, m\le n$ is given as $X = U \Sigma V^\top$, where $U\in \rdim{m\times m}$ and $V\in\rdim{n\times n}$ are orthonormal matrices, referred to as the singular vectors, and $\Sigma\in\rdim{m \times n}$ is a diagonal matrix of singular values. 
Krylov subspace methods allow for efficient estimation of the singular values and vectors of a matrix and enjoy two key properties.
First, we only need access to the matrix $X$ via left and right multiplications with vectors, i.e., we do not need explicit access to the elements of the matrix $X$.
Second,  the top singular values and vectors of a low-rank matrix can be estimated using a small set of matrix-vector multiplications. 
These two properties are invaluable when the matrix is very large or when it is implicitly represented using operators or, as is the case in this paper, the matrix is the scene's HSI and we only have access to optical implementations of the underlying matrix-vector multiplications.

There are  many variants of Krylov subspace techniques which differ mainly  on their robustness to noise and model mismatch.
The techniques in this paper are based on the so-called Lanczos bidiagonalization with full orthogonalization \cite{golub1965calculating,hernandez2007restarted}.
Such iterative operations to reduce the complexity of matrix-vector multiplications  have found use in communication theory in the form of reduced-rank filtering \cite{tian2005low,ge2004reduced} and adaptive beam forming \cite{ge2006data}.
Our goal is to leverage the benefits of iterative operations for low-rank approximation of high dimensional optical signals, in particular HSIs.

\paragraph{Optical computing of low-rank signals.}
Matrix-vector and matrix-matrix multiplications can often be implemented as optical systems.
Such systems have been used for matrix-matrix multiplication \cite{athale1982optical}, matrix inversion \cite{rajbenbach1987optical}, as well as computing eigenvectors \cite{kumar1981eigenvector}.
Of particular interest to our paper is the optical computing of the light transport operator using Krylov subspace methods \cite{o2010optical}.
The light transport matrix $T$ represents the linear mapping between scene illumination and a camera observing the scene.
Each column of the matrix $T$ is the image of the scene when only a single illuminant is turned on.
Hence, given a  vector ${\bm \ell}$ that encodes the scene illumination, the image captured by the camera is given as ${\bf r} = T {\bm \ell}$.
By Helmholtz reciprocity, if we replaced every pixel of the camera by a light source and every illuminant with a camera pixel, then the light transport associated with the reversed illumination/sensing setup is given  as $T^\top$.
Hence, by co-locating a projector with the camera and a camera with the scene's illuminants, we have access to both left- and right-multiplication of the light transport matrix with vectors; we can now apply Krylov subspace techniques for \textit{optically} estimating a low-rank approximation to the light transport matrix.
This delightful insight is one of the key results in \cite{o2010optical}. 

This paper proposes a translation of the ideas in \cite{o2010optical} to hyperspectral imaging.
However, as we will see next, this translation is not straightforward and requires  the construction of novel imaging architectures. 

\section{Optical Krylov Subspaces for Hyperspectral Imaging} \label{section:overview}
In this section, we provide a high-level description of optical computing of HSIs using Krylov subspace methods.

\paragraph{Notation.} We represent HSIs in two different ways:
\begin{itemize}[leftmargin=*]
\item $H(x, y, \lambda)$ ---  a real-valued function over 2D space $(x, y)$ and 1D spectrum $\lambda$,
\item $X \in \rdim{N_x N_y \times N_\lambda}$ --- a matrix with $N_xN_y$ rows and $N_\lambda$ columns, such that each column corresponds to the vectorized image at a specific spectrum.
\end{itemize}
The goal is to optically build the following two operators:
\begin{itemize}[leftmargin=*]
\item \textit{Spectrally-coded imager $\mathcal{I}$ } --- Given a spectral code $\bfx \in \rdim{N_\lambda}$, we seek to measure the image $\bfy \in \rdim{N_x N_y}$ given as
\begin{align}
	\bfy = \mathcal{I}(\bfx) = X\bfx.
\end{align}
The image $\bfy$  corresponds to a grayscale image of the scene with a camera whose spectral response is $\bfx$.
\item \textit{Spatially-coded spectrometer $\mathcal{S}$ } --- Given a spatial code $\widetilde{\bfx} \in \rdim{N_xN_y}$, we seek to measure a spectral measurement $\widetilde{\bfy} \in \rdim{N_\lambda}$ given as
\begin{align}
	\widetilde{\bfy} = \mathcal{S}(\widetilde{\bfx}) = X^\top \widetilde{\bfx}.
\end{align}
The  measurement $\widetilde{\bfy}$ corresponds to the spectral measurement of the scene, where-in the spectral profile of each pixel is weighted by the corresponding entry in the spatial code $\widetilde{\bfx}$.
\end{itemize}
Since the two operators correspond to left and right multiplication of a vector to the HSI matrix $X$, we can implement any Krylov subspace technique to estimate the top singular vectors and values.
\paragraph{Number of measurements required.} To obtain a rank-$k$ approximation of the matrix $X$, we would require at least $k$ spatially-coded spectral measurements --- each of dimensionality $N_\lambda$, and $k$ spectrally-coded images --- each of dimensionality $N_x N_y$.
Hence, the number of measurements required by the approach is proportional to $k (N_x N_y + N_\lambda)$ and, over traditional Nyquist sampling, it represents a reduction in measurements by a factor of
\begin{align}
	\frac{k(N_x N_y + N_\lambda)}{N_x N_y N_\lambda} = k \left( \frac{1}{N_\lambda} + \frac{1}{N_x N_y} \right). 
\end{align}
For low-rank HSIs, we can envision dramatic reductions in measurements required over Nyquist sampling  especially when sensing at high spatial and spectral resolutions (see Table \ref{tab:meas_comp}).
\begin{figure*}
\includegraphics[width=\textwidth]{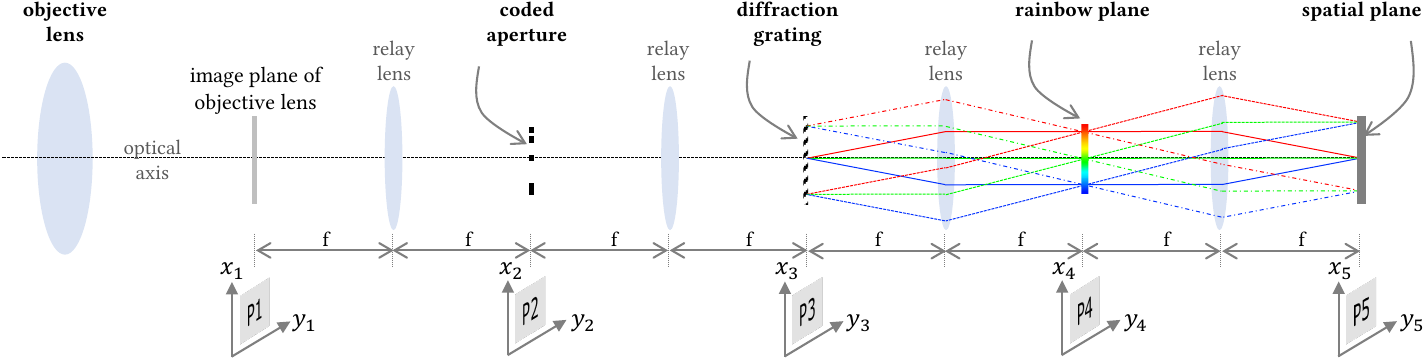}
\caption{Schematic diagram of simultaneous spatio-spectral measurements with a coded aperture. The diffraction grating disperses light along x-axis. The image of the scene is formed on plane P1. The coded aperture is placed in P2, which introduces a diffraction blur in spatial plane P3, and dictates the spectral profile formed on the plane P4. A slit or an open aperture on P2 is not a good choice for simultaneously high spatial and spectral resolution. Instead, we rely on design of a novel pupil aperture that enables simultaneous high spatial and spectral resolution.}
\label{fig:schematic}
\end{figure*}

\paragraph{Challenges in implementing operators $\mathcal{I}$ and $\mathcal{S}$.} 
Spatially-coded spectral measurements have been implemented in the context of compressive hyperspectral imaging \cite{sun2009compressive}.
Here, light from a scene is first focused onto an SLM that performs spatial coding, and then directed into a spectrometer.
For spectral coding at a high-resolution, we could replace the sensor in a spectrometer with an SLM; subsequently, we can form and measure an image of the coded light  using a lens.
However, high-resolution spectrometers invariably use a slit aperture  that produces a large one-dimensional blur in the spatial image due to diffraction.
We  show in Section \ref{section:spacespectrum} that  simultaneous spatio-spectral localization is not possible with either a slit or an open aperture.
This leads to the design of optimal binary coded apertures which enable  high spectral and spatial resolutions.
Subsequently, in Section \ref{section:optical}, we present the design of KRISM and validate its performance in Section \ref{section:real}.

\section{Coded apertures for simultaneous sensing of space and spectrum} \label{section:spacespectrum}
In this section, we introduce an optical system capable of simultaneously resolving space and spectrum at high resolutions. 
%
%
\subsection{Optical setup}
The ideas proposed in this paper rely on the optical setup shown in Figure \ref{fig:schematic} which is a slight modification of a traditional spectrometer.
An objective lens focuses a scene onto its image plane, that we denote as P1.
This is followed by two 4$f$ relays with a coded aperture placed on the first pupil plane, P2, and a diffraction grating placed at the plane marked as P3.
We are interested in the intensity images formed at the planes marked at the ``rainbow plane'' P4 and the  ``spatial plane'' P5, and their relationship to the image formed on P1, the coded aperture, and the grating parameters.

We assume that the field formed on the plane P1 is  \textit{incoherent} and, hence, we only need to consider its intensity and how it propagates, and largely ignore its phase.
Let $H(x, y, \lambda)$ be the intensity of the field as a function of spatial coordinates $(x, y)$ and wavelength $\lambda$.
Let $a(x, y)$ be the aperture code placed at the plane P2, $v_0$ be the density (measured in grooves per unit length) of the diffraction grating in P3, and $f$ be the focal length of the lenses that form the 4$f$ relays.
The hyperspectral field intensity at the plane P4 is given as
\begin{align}
F_4(x, y, \lambda) =  \frac{1}{\lambda^2 f^2}a^2(-x+ f \lambda v_0, -y)\,\, S(\lambda),
\end{align}
where $S(\lambda)$ is  the scene's overall spectral content defined as
\begin{equation}
S(\lambda) = \int_x \int_y H(x, y, \lambda) dx dy. \nonumber
\end{equation}
The intensity field at the spatial plane P5 is given as
\begin{align}
F_5(x, y, \lambda)  = H(x, y, \lambda) \ast \left|\frac{1}{\lambda^2 f^2} A\left(-\frac{x}{\lambda f}, -\frac{y}{\lambda f} \right) \right|^2,
\end{align}
where $A(u, v)$ is the 2D spatial Fourier transform of the aperture code $a(x, y)$, and $\ast$ denotes two-dimensional spatial convolution along $x$ and $y$ axes.
These expressions arise from Fourier optics  \cite{goodman2005introduction} and their derivation is provided in the supplemental material.

\begin{figure}[!ttt]
\centering
\includegraphics[width=\columnwidth]{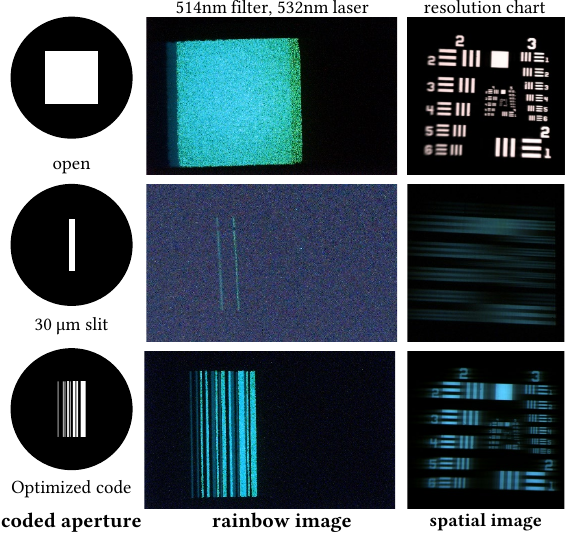}
\caption{We implemented the setup shown in Figure \ref{fig:schematic} to verify the effect of different pupil codes. The scene consists of a resolution chart illuminated by two distinct narrowband light sources. An open aperture leads to sharp spatial images, but the spectrum is blurred. On the other hand, a slit offers high spectral resolution, but the spatial image is blurred. Optimized codes offer invertible spectral blur, and at the same time, invertible spatial blur.}
\label{fig:apertures}
\end{figure}

\paragraph{Image formed at the rainbow plane P4.} A camera with spectral response $c(\lambda)$ placed at the rainbow plane would measure 
\begin{align} 
I_R(x, y) &=  \int_\lambda  a^2(-x+ f \lambda v_0, -y) \frac{1}{\lambda^2 f^2} S(\lambda) c(\lambda) d\lambda \nonumber\\
&\propto a^2(-x, -y) \ast \left( S\left(\frac{x}{f v_0}\right) \widetilde{c} \left(\frac{x}{f v_0}\right) \right),
\label{eq:spectralblur}
\end{align}
where $\widetilde{c}(\lambda) = c(\lambda)/\lambda^2f^2$.
Here, the dimensionless term $fv_0$, that scales of the spectrum $S(\cdot)$, indicates the resolving power of the diffraction grating.
For example, we used a focal length $f = 100$ mm and a grating with groove density $v_0 = 300$ grooves/mm for the prototype discussed in Section \ref{section:optical}; here,  $f v_0 = 30,000$. This implies that the spectrum is stretched by a factor of $30,000$. 
Therefore, a $1$ nm of the spectrum maps to 30 $\mu$m, which is about 6-7 pixel-widths on the camera that we used.
The key insight  this expression provides is that the  image $I_R$ is the convolution of the scene's spectrum --- denoted as a 1D image --- with the aperture code $a(\cdot, \cdot)$ (see Figure \ref{fig:apertures}).
This implies that we can measure the spectrum of the scene, albeit convolved with the aperture code on this plane; this motivates our naming of this plane as the rainbow plane.

\paragraph{Image at the spatial plane P5.} A camera with the spectral response $c(\lambda)$ placed at the spatial plane P5 would measure
\begin{align} 
I_S(x, y) = \int_\lambda \left( H(x, y, \lambda) \ast  \left|\frac{1}{\lambda^2 f^2} A\left(-\frac{x}{\lambda f}, -\frac{y}{\lambda f} \right) \right|^2 \right) c(\lambda) d\lambda
\label{eq:spatialblur}
\end{align}
$I_S$ is a ``spatial image'' in that spectral components of the HSI have been integrated out. Hence, we refer to P5 as the spatial plane.
Figure  \ref{fig:apertures} shows the image formed at P5 for different choices of the coded apertures, including slits and open apertures.

\paragraph{Implementing KRISM operations.} The derivation above suggests that we get a spatial image of the scene formed at the spatial plane P5 and a spectral profile at the rainbow plane P4. We can therefore build the two operators central to KRISM by coding light on one of the planes while measuring it at the other.
For the spectrally-coded imager $\mathcal{I}$, we will place an SLM on the rainbow plane P4 while measuring the image, with a camera, at P5.
For the spatially-coded spectrometer $\mathcal{S}$, we place an SLM on P3 --- which is optically identical to P5 --- while measuring the image formed at P4.

\paragraph{Effect of the aperture code on the scene's HSI} 
Introducing an aperture code $a(x, y)$ on the plane P2 can be interpreted as distorting the scene's HSI in two distinct ways.
First, a spectral blur is introduced whose point spread function (PSF) is a scaled copy of the aperture code $a(x, y)$.
Second, a spatial blur is introduced for each spectral band whose  PSF is the power spectral density (PSD) of the aperture code, suitably scaled.
With this interpretation,  the images formed on planes P4 and P5 are a spectral and spatial projection, respectively, of this new blurred HSI. 
Our proposed technique measures a low-rank approximation to this blurred HSI and we can, in principle, deblur it to obtain the true HSI of the scene.
However, the spatial and spectral blur kernels may not always be invertible. 
As we show next, the choice of the aperture is critical and that traditional apertures such as a slit in spectrometry and an open aperture in imaging will not lead to invertible blur kernels. 
\subsection{Failure of slits and open apertures}
We now consider the effect of the traditional apertures used in imaging and spectrometry ---  namely, an open aperture and a slit, respectively --- on the images formed at the rainbow and the spatial planes.
Suppose that the aperture code $a(x, y)$ is a box function of width $W$ mm and height $H$ mm, i.e.,
\[ a(x, y) = \textrm{rect}_W(x)\, \textrm{rect}_H(y).\]
Its Fourier transform $A(u, v)$ is the product of two sincs
\[ A(u, v) = \textrm{sinc}(Wu)\, \textrm{sinc}(Hv). \]
The spatial image $I_S$ is convolved with the PSD $|A(u, v)|^2$ scaled by $f\lambda$, so the blur observed on it has a spatial extent of $f\lambda/W \times f\lambda/H$ units.
Suppose that $f = 100 \textrm{ mm}$ and $\lambda = 0.5 \mu$m, the observed blur is $50/W \times 50/H\ (\mu \textrm{m})^2$. 
The rainbow plane image $I_R$, on the other hand, simply observes a box blur whose spatial extent is $W  \times H$ $\textrm{mm}^2$.
Armed with these expressions, we can study the effect of an open and a slit apertures on the spatial and rainbow images.

\begin{figure*}[!ttt]
	\centering
	\begin{subfigure}[b]{0.32\textwidth}
		\centering
		\begin{subfigure}[c]{0.5\textwidth}
			\centering
			\includegraphics[width=\textwidth]{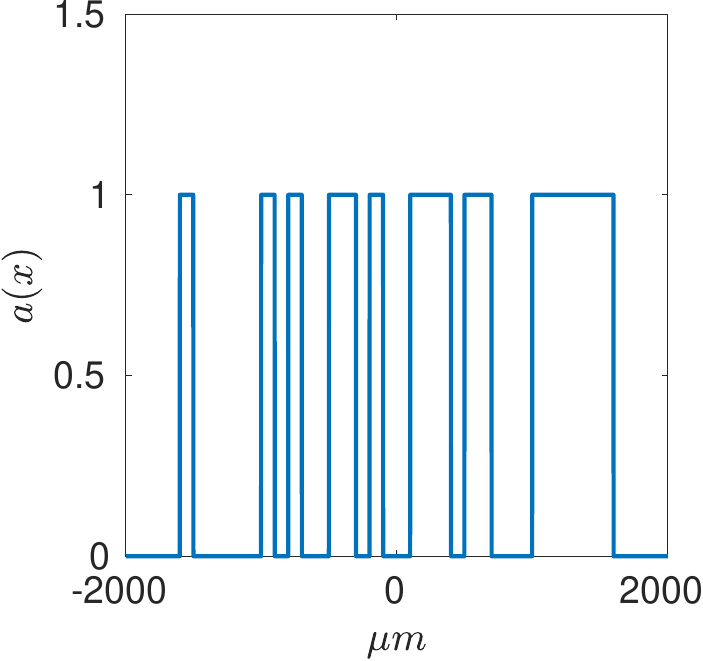}
		\end{subfigure}
		\begin{subfigure}[c]{0.48\textwidth}
			\centering
			\includegraphics[width=\textwidth]{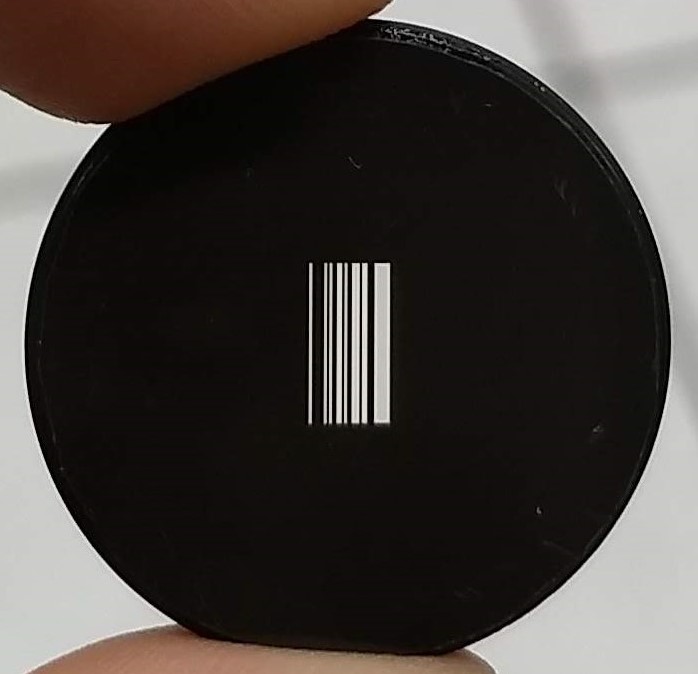}
		\end{subfigure}
		\caption{Optimized code and aperture for $N=32$.}
	\end{subfigure}
		\begin{subfigure}[b]{0.32\textwidth}
		\centering
		\includegraphics[width=0.50\textwidth]{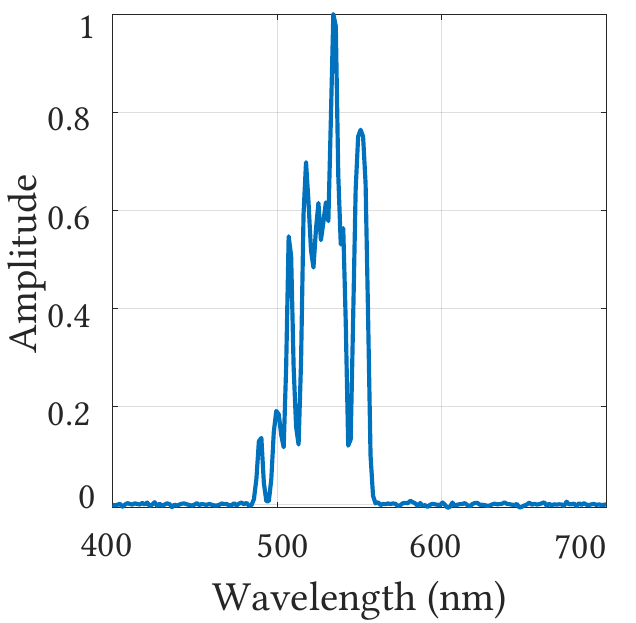}
		\includegraphics[width=0.45\textwidth]{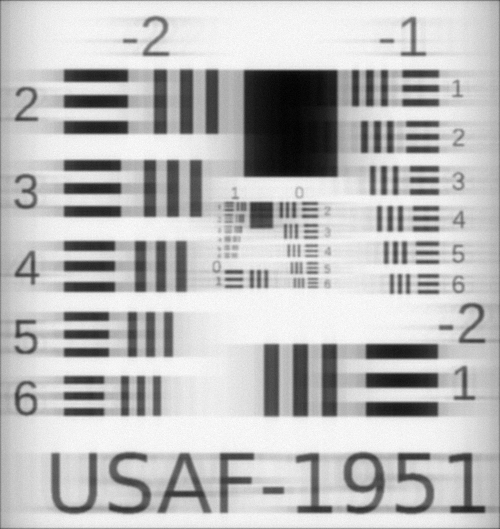}
		\caption{Raw measurements.}
	\end{subfigure}
	\begin{subfigure}[b]{0.32\textwidth}
		\centering
		\includegraphics[width=0.5\textwidth]{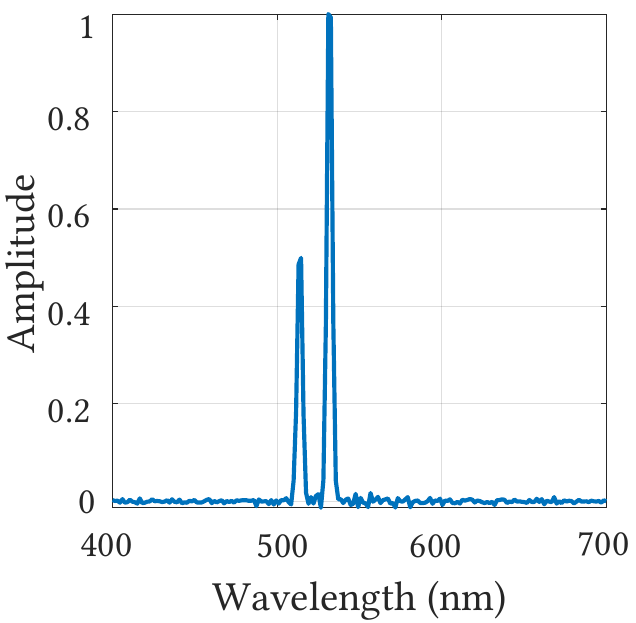}
		\includegraphics[width=0.45\textwidth]{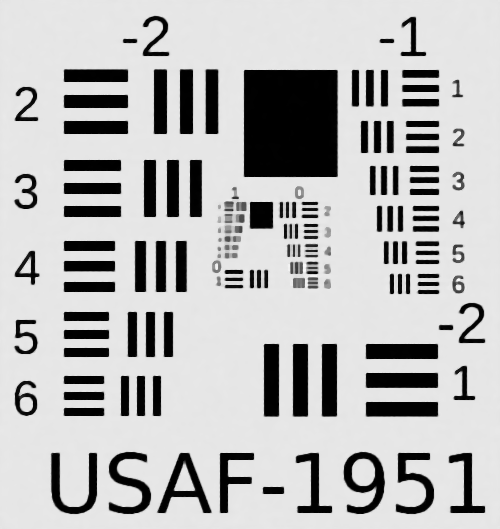}
		\caption{Deconvolved spectrum and image respectively.}
	\end{subfigure}
	\caption{Optimized codes ensure that the spectral as well as spatial blur can be deconvolved stably. We simulate the performance of optimal code on a spatial and spectral target similar to Figure \ref{fig:apertures}. Spectrum was deconvolved using Wiener deconvolution, and spatial images were deconvolved using TV prior. Optimized codes offer high spatial as well as spectral resolution. }
	\label{fig:code_compare}
\end{figure*}

\paragraph{Scenario \#1 --- An open aperture.} Suppose that $W = H = 10$ mm, then we can calculate the spatial blur to be $5 \mu$m in both height and width and hence, we can expect a very sharp spatial image of the scene.
The blur on the rainbow image has a spread of $10$ mm; for  relay lenses with focal length $f = 100$mm and grating with groove density $v_0 = 300$ grooves/mm, this would be equivalent of a spectral blur of $10,000/30 \approx 333$ nm.
Hence, we cannot hope to achieve high spectral resolution with an open aperture.

\paragraph{Scenario \#2 --- A slit.} A slit is commonly used in spectrometers; suppose that we use a slit of width $W = 100 \mu$m and height $H = 10$mm.
Then, we expect to see a spectral blur of $100/30 \approx 3.3$ nm.
The spatial image is blurred along the y-axis by a $5 \mu$m blur and along the x-axis by a 
$50/0.1 = 500 \mu$m blur; effectively, with a $5 \mu$m pixel pitch, this would correspond to a 1D blur of 100 pixels.
In essence, the use of a slit leads to severe loss in spatial resolution.

Figure \ref{fig:apertures} shows images formed at the rainbow and spatial planes for various aperture codes. 
This validates our claim that  conventional imagers are unable to simultaneously achieve high spatial and spectral resolutions due to the nature of the apertures used.
We next design  apertures with carefully engineered spectral and spatial blurs, which can be  deblurred in post-processing.

\subsection{Design of aperture codes}
We now design an aperture code that is capable of resolving both space and spectrum at high-resolutions.
Our use of coded apertures is inspired by seminal works in coded photography for motion and defocus deblurring \cite{raskar2006coded,veeraraghavan2007dappled,levin2007image}.
\paragraph{Observation.} Recall that the rainbow plane image $I_R$ is a convolution between a 1D spectral profile $s(\cdot)$ and a 2D aperture code $a(x,y)$.
This convolution is one dimensional, i.e., along the $x$-axis; hence, we can significantly simplify the code design problem by choosing an aperture of the form
{
\setlength{\abovedisplayskip}{3pt}
\setlength{\belowdisplayskip}{3pt}
\begin{equation}
	a(x, y) = a(x)\,\, \textrm{rect}_H(y),
\end{equation}
}
with $H$ being as large as possible.
The choice of the $\textrm{rect}$ function along the $y$-axis leads to a high light throughput as well as a compact spatial blur along the $y$-axis.
\sloppy
For ease of fabrication, we further restrict the  aperture code to be binary and of the form
{
\setlength{\abovedisplayskip}{3pt}
\setlength{\belowdisplayskip}{3pt}
\begin{align}
	a(x) &= \sum\limits_{k=0}^{N-1} a_k\, \mathbb{I}_{[k \Delta, (k+1) \Delta]}(x),
	\label{eq:code}
\end{align}
}
where $\mathbb{I}_{[p, q]}(x) = 1$ when  $x \in [p, q]$ and zero otherwise.
Hence, the mask design reduces to finding an $N$-bit codeword ${\bf a} = \{a_0, \ldots, a_{N-1}\}$.
The term $\Delta$, with units in length, specifies the physical dimension of each bit in the code.
We fix its value based on the desired spectral resolution.
For example, for $f = 100$mm and $v_0 = 300$ grooves/mm, a desired spectral resolution of $1$nm would require $\Delta \le 30 \mu$m.

Our goal is to design masks that enable the following:
\begin{itemize}[leftmargin=*]
\item \textit{High light throughput.} For a given code length $N$, we seek codes with large light throughput which is equal to the number of ones in the code word ${\bf a}.$
\item \textit{Invertibility of the spatial and spectral blur.} The code is designed such that the resulting spatial and spectral blur are both invertible.
\end{itemize}

An invertible blur can be achieved by engineering its PSD to be flat.
%
Given that the spectrum is linearly convolved with $a(x)$, a $(N + N_\lambda -1 )$-point DFT of the code word ${\bf a}$ captures all the relevant components of the PSD of $a(x)$.
Denoting this $(N + N_\lambda -1)$-point DFT of ${\bf a}$ as ${\bf A}[k]$, we aim to maximize its minimum value in magnitude.
Recall from (\ref{eq:spatialblur}) that the spatial PSF is the power spectral density (PSD) of $a(x)$, with suitable scaling.
Specifically, the Fourier transform of spatial blur is given by $c(\lambda f u)$, where $c(x) = a(x)\ast a(-x)$ is the linear autocorrelation of $a(x)$ and $u$ represents spatial frequencies.
From (\ref{eq:code}), we get,
{ \begin{align}
\small
	c(x) &= a(x)*a(-x) \nonumber \\
		 & = \sum\limits_{k=-N}^{N-1}c_k \left(\mathbb{I}_{[k\Delta, (k+1)\Delta]}(x) \ast \mathbb{I}_{[k\Delta, (k+1)\Delta]}(x)\right),
\end{align}
}
where $c_k$ is the discrete linear autocorrelation of $a_k$.
Thus, it is sufficient to maximize $c_k$ to obtain an invertible spatial blur.

\begin{figure}[!tt]
	\centering
	\begin{subfigure}[c]{0.475\columnwidth}
		\centering
		\includegraphics[width=\textwidth]{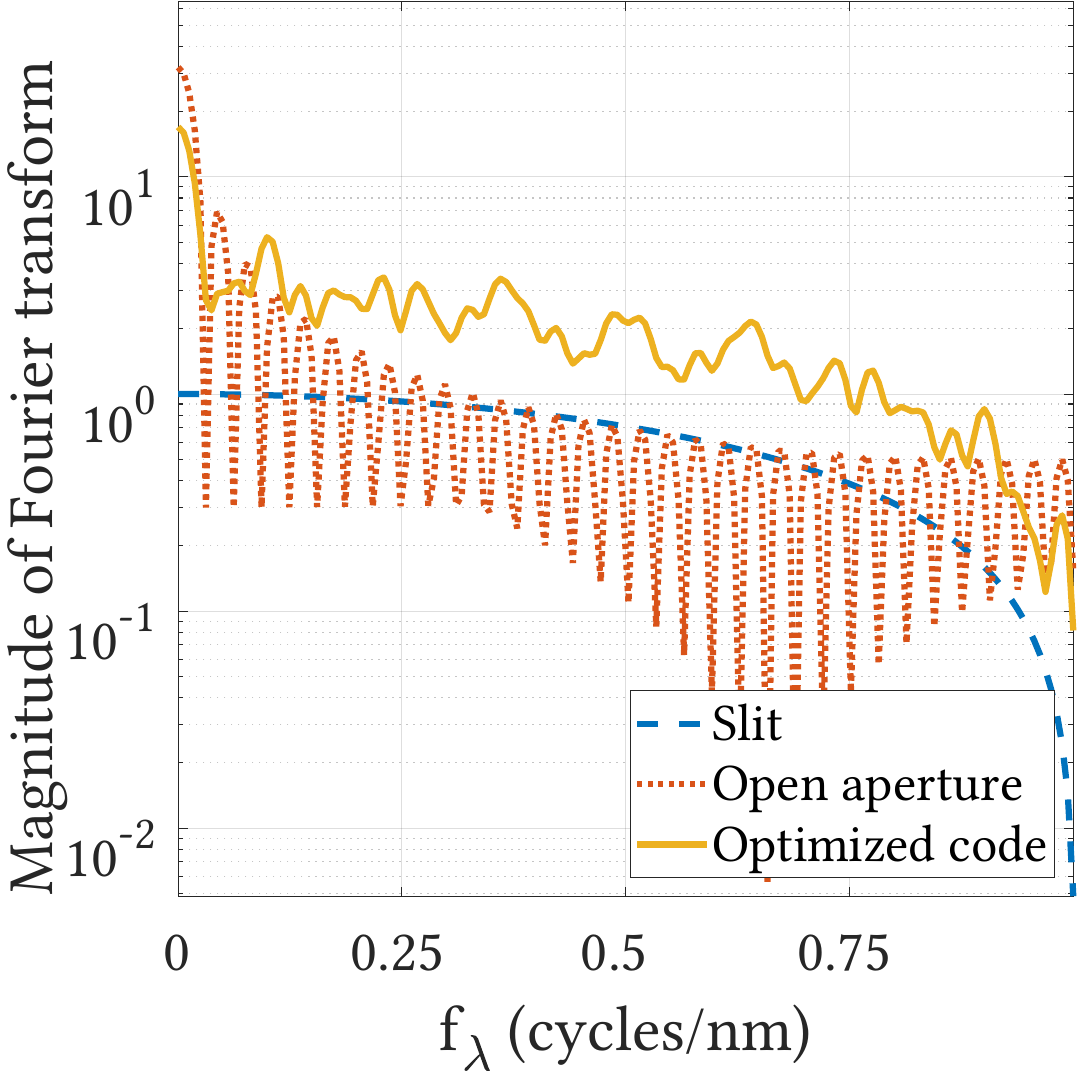}
		\caption{Freq. response of spectral blur}
	\end{subfigure}
	\quad
	\begin{subfigure}[c]{0.475\columnwidth}
		\centering
		\includegraphics[width=\textwidth]{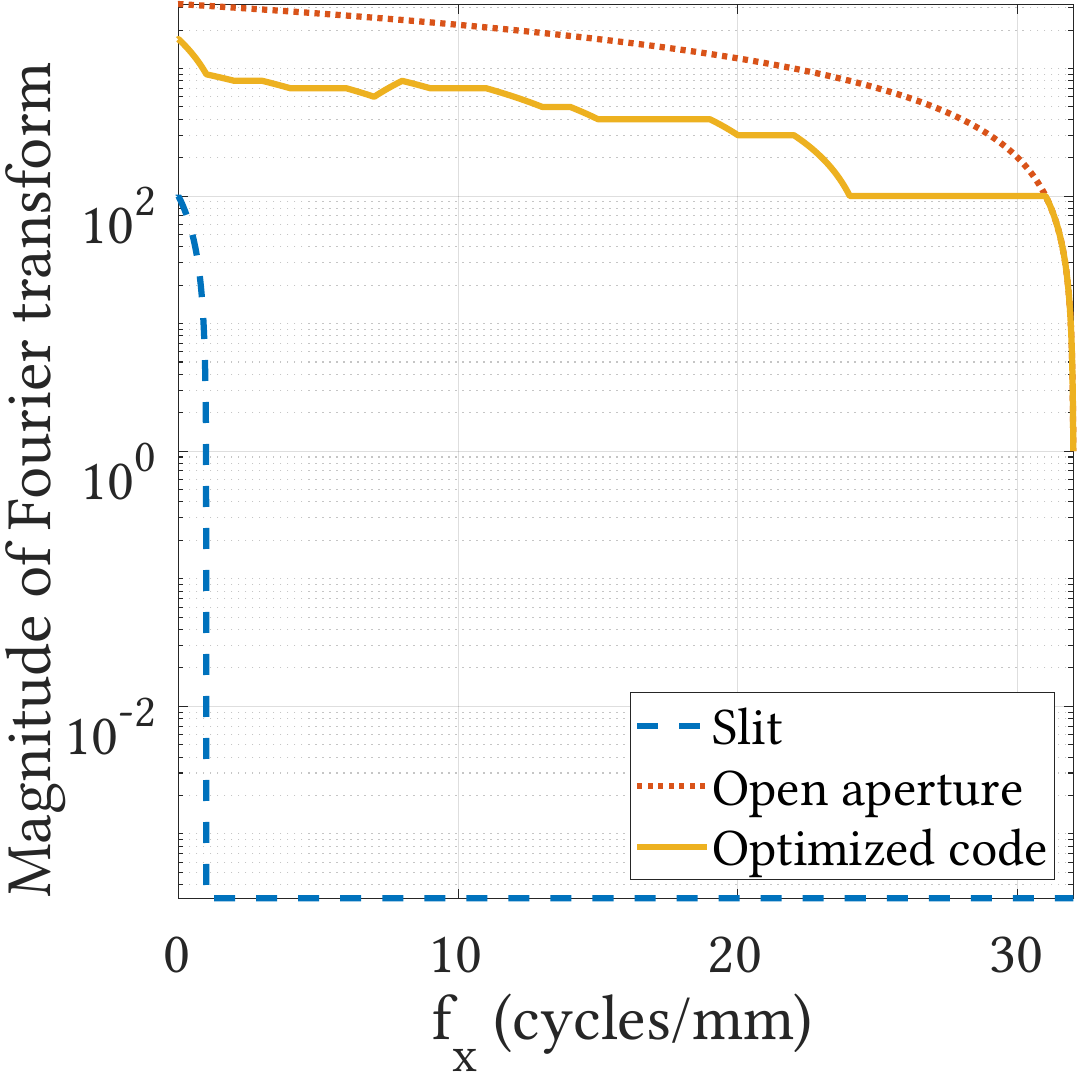}
		\caption{Freq. response of spatial blur}
	\end{subfigure}
	\caption{Frequency response of spatial and spectral blur for various pupil codes. Width of the slit was $100\mu$m, while that of open aperture was $3.2$mm. The length of optimized code is 32-bits, with each bit being $100\mu$m wide, giving a $3.2$mm wide aperture. We assume that a slit can resolve up to 1nm. In the graph, $0.5$ cycles/nm corresponds to a spectral resolution of 1nm, and hence the frequency response of the slit falls off after $0.5$ cycles/nm. Similarly, the maximum spatial resolution is $15\mu m$ and hence $f_x$ is shown till $32$ cycles/mm. For spectral measurements, a slit has a flat frequency response, while an open aperture has several nulls. In contrast, an open aperture has no nulls for spatial measurements, whereas a slit attenuates high frequencies. Optimized codes have a fairly flat frequency response for spectral blur, and no nulls for spatial blur.}
	\label{fig:space_spectrum_ft}
	\vspace{-1em}
\end{figure}

We select an aperture code that leads to invertible blurs for both space and spectrum by solving the following optimization problem:
\begin{equation}
\max_{a_0, \ldots, a_{N-1}} \alpha \min_k \left(| {\bf A}[k] |\right) + (1-\alpha )  \min_k c_k,
\label{eq:obj_invertible}
\end{equation}
under the constraint that the elements of ${\bf a}$ are binary-valued, and $\alpha \in (0, 1)$ is a constant.
For code length $N$ sufficiently small, we can simply solve for the optimal code via exhaustive search of all $2^N-1$ code words. 
We used $N = 32$ and an exhaustive search for the optimal code took over a day.
The resulting code and its performance is shown in Figure \ref{fig:code_compare} and \ref{fig:space_spectrum_ft}; we used $\Delta = 100 \mu$m and $H = 6.4$mm for this result.
A brute force optimization is not scalable for larger codes.
Instead of searching for optimal codes, we can search for approximately optimal codes by iterating over a few candidate solutions. 
This strategy has previously been explored in \cite{raskar2006coded}, where 6 million candidate solutions are searched for a 52-dimensional code.

Figure \ref{fig:space_spectrum_ft} shows the frequency response of both spectral and spatial blurs for the 32-dimensional optimized code.
The advantages of optimized codes are immediately evident --- an open aperture has several nulls in spectral domain, while a slit attenuates all high spatial frequencies. 
The optimized code retains all frequencies in both domains, while increasing light throughput.

\section{Synthetic experiments} \label{sec:synthetic}

\begin{figure*}[!tt]
	\centering
		\begin{subfigure}[c]{0.48\textwidth}
	\begin{subfigure}[t]{0.23\columnwidth}
		\centering
		\includegraphics[width=\columnwidth]{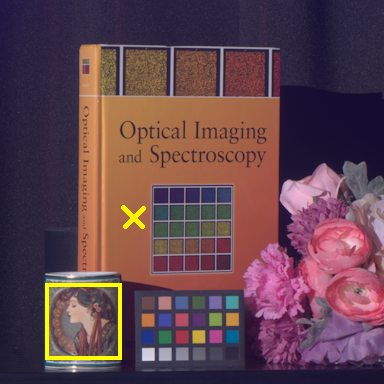}
		\caption{KAIST data \\ ($512\times384\times31$)\\}
	\end{subfigure}
	\hspace{0.1em}
	\begin{subfigure}[t]{0.23\columnwidth}
		\centering
		\includegraphics[width=\columnwidth]{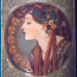}
		\caption{Ground truth}
	\end{subfigure}
	\hspace{0.1em}
	\begin{subfigure}[t]{0.23\columnwidth}
		\centering
		\includegraphics[width=\columnwidth]{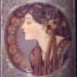}
		\caption{Lin et. al\\$N/M: 29$,\\ PSNR: $29.8$dB}
	\end{subfigure}
	\hspace{0.1em}
	\begin{subfigure}[t]{0.23\columnwidth}
		\centering
		\includegraphics[width=\columnwidth]{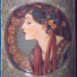}		\caption{Choi et al.\\$N/M: 29$,\\ PSNR: $31.9$dB}
	\end{subfigure}
	\\
	\begin{subfigure}[t]{0.23\columnwidth}
		\centering
		\includegraphics[width=\columnwidth]{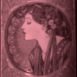}
		\caption{KRISM $K: 1$\\$N/M: 10$,\\ PSNR: $21.8$dB}
	\end{subfigure}
	\hspace{0.1em}
	\begin{subfigure}[t]{0.23\columnwidth}
		\centering
		\includegraphics[width=\columnwidth]{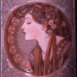}
		\caption{KRISM $K: 2$\\$N/M: 8$,\\ PSNR: $30.2$dB}
	\end{subfigure}
	\hspace{0.1em}
	\begin{subfigure}[t]{0.23\columnwidth}
		\centering
		\includegraphics[width=\columnwidth]{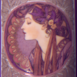}
		\caption{KRISM $K: 3$\\$N/M: 6$,\\ PSNR: $33.0$dB}
	\end{subfigure}
	\hspace{0.1em}
	\begin{subfigure}[t]{0.23\columnwidth}
		\centering
		\includegraphics[width=\columnwidth]{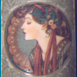}
		\caption{KRISM $K: 4$\\$N/M: 5$,\\ PSNR: $38.6$dB}
	\end{subfigure}
	\\
	\begin{subfigure}[t]{\columnwidth}
		\centering
		\includegraphics[width=\columnwidth]{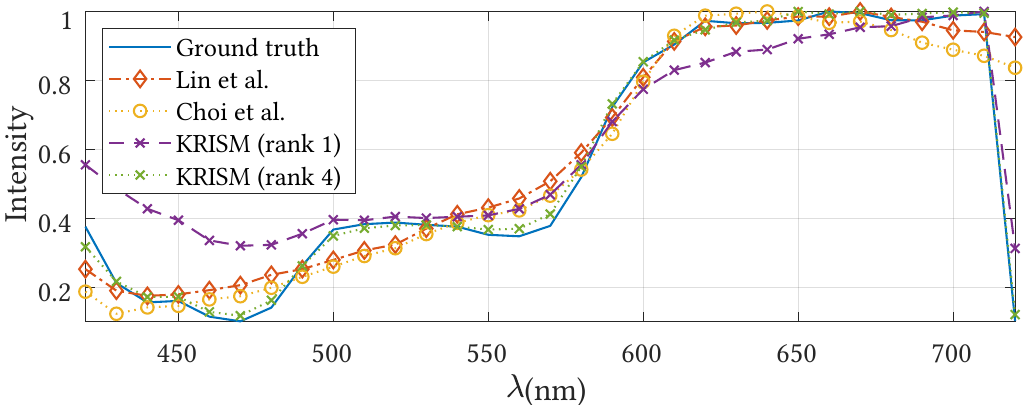}
	\end{subfigure}
	\end{subfigure}
	\hspace{0.4em}	
	\begin{subfigure}[c]{0.48\textwidth}
	\begin{subfigure}[t]{0.23\columnwidth}
		\centering
		\includegraphics[width=\columnwidth]{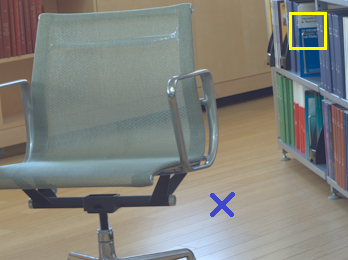}
		\caption{Harvard data \\ ($696\times520\times31$)\\}
	\end{subfigure}
	\hspace{0.1em}
	\begin{subfigure}[t]{0.23\columnwidth}
		\centering
		\includegraphics[width=\columnwidth]{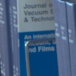}
		\caption{Ground truth}
	\end{subfigure}
	\hspace{0.1em}
	\begin{subfigure}[t]{0.23\columnwidth}
		\centering
		\includegraphics[width=\columnwidth]{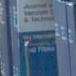}
		\caption{Lin et. al\\$N/M: 29$,\\ PSNR: $26.6$dB}
	\end{subfigure}
	\hspace{0.1em}
	\begin{subfigure}[t]{0.23\columnwidth}
		\centering
		\includegraphics[width=\columnwidth]{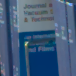}		\caption{Choi et al.\\$N/M: 29$,\\ PSNR: $33.1$dB}
	\end{subfigure}
	\\
	\begin{subfigure}[t]{0.23\columnwidth}
		\centering
		\includegraphics[width=\columnwidth]{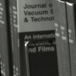}
		\caption{KRISM $K: 1$\\$N/M: 10$,\\ PSNR: $28.7$dB}
	\end{subfigure}
	\hspace{0.1em}
	\begin{subfigure}[t]{0.23\columnwidth}
		\centering
		\includegraphics[width=\columnwidth]{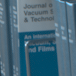}
		\caption{KRISM $K: 2$\\$N/M: 8$,\\ PSNR: $37.4$dB}
	\end{subfigure}
	\hspace{0.1em}
	\begin{subfigure}[t]{0.23\columnwidth}
		\centering
		\includegraphics[width=\columnwidth]{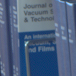}
		\caption{KRISM $K: 3$\\$N/M: 6$,\\ PSNR: $41.2$dB}
	\end{subfigure}
	\hspace{0.1em}
	\begin{subfigure}[t]{0.23\columnwidth}
		\centering
		\includegraphics[width=\columnwidth]{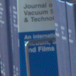}
		\caption{KRISM $K: 4$\\$N/M: 5$,\\ PSNR: $42.7$dB}
	\end{subfigure}
	\\
	\begin{subfigure}[t]{\columnwidth}
		\centering
		\includegraphics[width=\columnwidth]{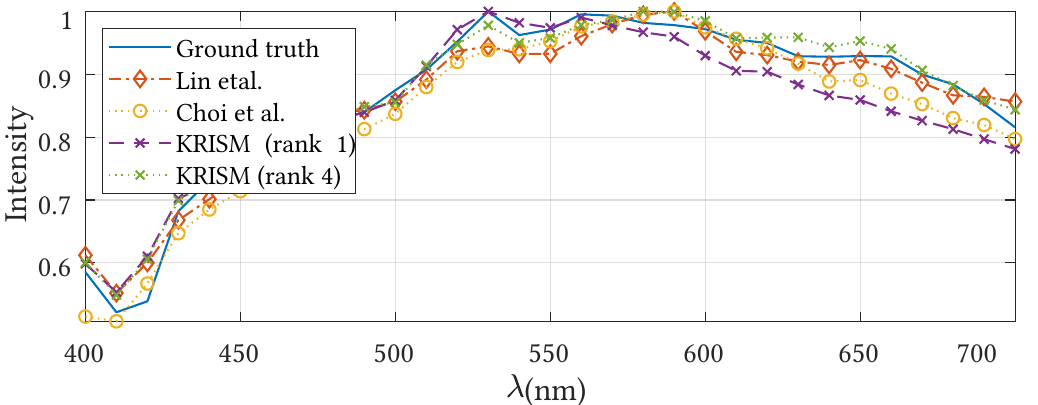}
	\end{subfigure}
	\end{subfigure}
	\caption{Evaluation against snapshot techniques. We compare KRISM with varying rank against results from \cite{lin2014spatial} and \cite{deepcassi2017} in terms of compression as well as accuracy. We show zoomed in image patches for each method and spectrum at pixel marked by a cross. At similar compression rates ($K=1$), KRISM has lower accuracy than snapshot techniques. However, snapshot techniques require solving a complex optimization problem that can be time consuming. In contrast, KRISM requires practically no reconstruction time as the dominant singular vectors are captured directly.}
	\label{fig:lowres_sim}
\end{figure*}
We tested KRISM via simulations on three different datasets, listed in Table \ref{tab:dataset}, and compared against existing approaches. 
For all methods, we simulated both photon and readout noise respectively as Poisson and Gaussian random variables.
All KRISM simulations were done with diffraction effects due to coded aperture.

We quantify performance through compression in measurements $N/M$ which is ratio of number of unknowns to measurements and peak signal to noise ratio (PSNR).
Given a HSI matrix $\bfx$ and its reconstruction $\widehat{\bfx}$, we define peak SNR as \[ \textrm{PNSR} = 20\log_{10}\left( \frac{\|\bfx\|_\infty}{\text{RMSE}(\bfx, \widehat{\bfx})} \right),\] where $\text{RMSE}$ is the root mean squared error defined as
\begin{align}
	\text{RMSE}(\bfx, \widehat{\bfx}) = \sqrt{\frac{1}{N}\sum\limits_{n=1}^{N}(\bfx_n - \widehat{\bfx}_n)^2}.
\end{align}

\begin{table}[!thh]
	\centering
	\includegraphics[width=\columnwidth]{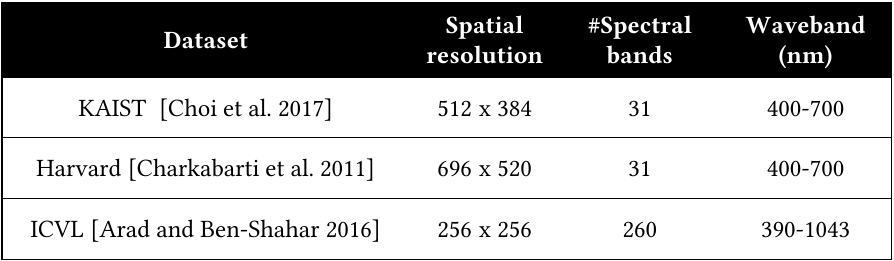}
	\caption{Datasets used for simulations. The spatial resolution for KAIST and Hardvard datasets, and spectral resolution for ICVL dataset was reduced to keep computation tractable with competing methods.}
	\label{tab:dataset}
	\vspace{-2.5em}
\end{table}

\paragraph{Comparison with snapshot techniques.}
Snapshot techniques such as CASSI \cite{wagadarikar2008single} and spatial-spectral encoded CS \cite{lin2014spatial} recover HSI from a single image and hence are appropriate for video-rate hyperspectral imaging. 
In contrast, KRISM is \emph{not} a snapshot technique since, at the very least it requires the measurement of  an image and a spectral profile.
Nevertheless, we compare KRISM against snapshot techniques by varying the number of KRISM iterations.
Figure \ref{fig:lowres_sim} shows performance of these methods with varying number of measurements on KAIST and Harvard datasets.
We observe that in the setting closest to snapshot mode, \citet{deepcassi2017} and \citet{lin2014spatial} do outperform KRISM; this is to be expected since after a single iteration, KRISM provides only a rank-1 approximation.
As the number of KRISM iterations are increased (which allows approximations of higher ranks), KRISM performance improves.
KRISM enjoys advantages when we look at computational cost for reconstruction.
The reconstruction time for \citet{deepcassi2017} is more than 10 minutes\footnote{We used code, dataset and model from \url{https://github.com/KAIST-VCLAB/deepcassi}} even with multiple GPUs, while it runs to several hours for \citet{lin2014spatial}\footnote{We used code, dataset and overcomplete dictionary from the paper itself.}.
In contrast, KRISM requires practically no reconstruction time for recovering the HSI as we directly measure the singular vectors.
\begin{figure}[!tt]
	\centering
	\begin{subfigure}[t]{0.31\columnwidth}
		\centering
		\includegraphics[width=\columnwidth]{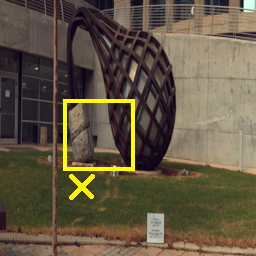}
		\caption{ICVL dataset \\ ($256\times256\times260$)\\}
	\end{subfigure}
	\hspace{0.1em}
	\begin{subfigure}[t]{0.31\columnwidth}
		\centering
		\includegraphics[width=\columnwidth]{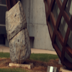}
		\caption{Ground truth}
	\end{subfigure}
	\hspace{0.1em}
	\begin{subfigure}[t]{0.31\columnwidth}
		\centering
		\includegraphics[width=\columnwidth]{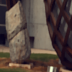}
		\caption{KRISM, $N/M: 43$,\\ PSNR: $47.7$dB}
	\end{subfigure}
	\\
	\begin{subfigure}[t]{0.31\columnwidth}
		\centering
		\includegraphics[width=\columnwidth]{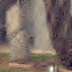}
		\caption{Sun and Kelly $N/M: 5$,\\ PSNR: $41.1$dB}
	\end{subfigure}
	\hspace{0.1em}
	\begin{subfigure}[t]{0.31\columnwidth}
		\centering
		\includegraphics[width=\columnwidth]{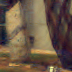}		\caption{Kittle et al.\\$N/M: 5$,\\ PSNR: $43.2$dB}
	\end{subfigure}
	\hspace{0.1em}
	\begin{subfigure}[t]{0.31\columnwidth}
		\centering
		\includegraphics[width=\columnwidth]{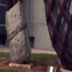}
		\caption{Fazel et al.\\$N/M: 43$,\\ PSNR: $40.7$dB}
	\end{subfigure}
	\\
	\begin{subfigure}[t]{\columnwidth}
		\centering
		\includegraphics[width=\columnwidth]{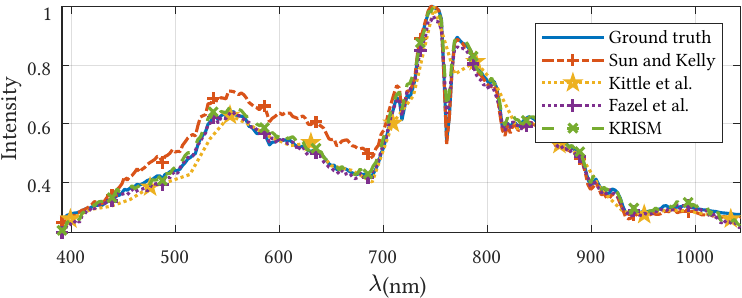}
	\end{subfigure}
	\caption{Evaluation with multi-frame techniques.  We compare KRISM against spatially-multiplexed HSI  \cite{sun2009compressive}, multi-frame version of CASSI \cite{kittle2010multiframe}, and row/column CS \cite{fazel2008compressed}. We show zoomed in image patches for each method and spectrum at pixel marked by a cross. Across the board, KRISM has highest accuracy with fewest measurements.}
	\label{fig:highres_sim}
\end{figure}

\paragraph{Comparison with multi-frame techniques.} 
Since KRISM is essentially a multi-frame technique, we compare against multi-frame version of CASSI \cite{kittle2010multiframe}, and spatially-multiplexed hyperspectral imager \cite{sun2009compressive}.
We simulate spatially-multiplexed HSI imager via randomly permuted Hadamard multiplexed spectra and recover using sparsity of individual bands in wavelet domain.
Note that the compression ratio is lower for \citet{kittle2010multiframe} and \citet{sun2009compressive} since the results were inaccurate for higher compressions\footnote{Please see supplementary for further details.}.
Figure \ref{fig:highres_sim} shows a comparison of recovered spatial and spectral images for ICVL dataset.
The poor performance of \citet{kittle2010multiframe} is due to usage of a translational mask to get multiple measurements.
On the other hand, \citet{sun2009compressive} performs poorly as multiplexing is done only in the spatial domain.
Performance can be improved if we multiplex in the spectral domain  as well; the resulting method is the low-rank CS approach proposed by \citet{fazel2008compressed}.
This results in an increase in accuracy with fewer measurements, as seen in Figure \ref{fig:highres_sim} (f).
Note that CS-based techniques are based on random projections and are not adapted to the scene.
In contrast, KRISM adaptively computes a low-rank approximation leading to an increase in accuracy with the same number of measurements as \citet{fazel2008compressed}.

Based on these simulations, we conclude that KRISM is indeed a compelling methodology when spatial and spectral resolution are  high --- a desirable operating point in many applications.
When the number of spectral bands are smaller, the gains are modest, but nevertheless present.
In the next section, we provide an optical schematic for implementing KRISM.

\section{The KRISM Optical setup} \label{section:optical}
\begin{figure*}[!ttt]
	\centering
	\includegraphics[width=\textwidth]{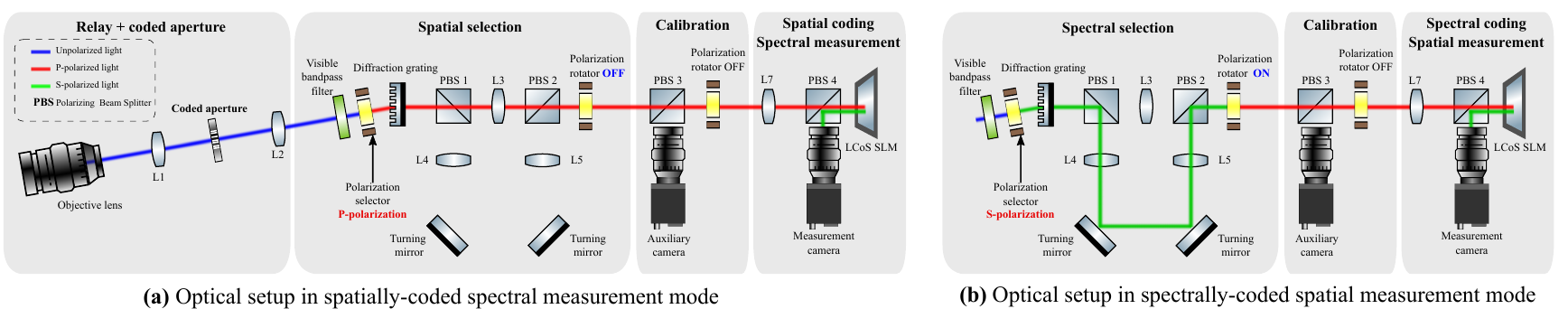}
	\caption{Proposed optical setup in spectral coding (a) and spatial coding (b) mode. The optical method relies on polarization to switch between the two types of coding. When the input light is S-polarized, the LC rotator is switched off, enabling spectrally coded spatial measurements. When the input light instead is P-polarized, the LC rotator is turned on, which enables spatially coded spectral measurements. The input light polarization is controlled by a second LC rotator placed before the grating. With a novel use of LC rotators, our optical setup enables dual coding of hyperspectral scenes with a single camera-SLM pair.}
	\label{fig:optical_setup}
\end{figure*}

We now present an optical design for implementing the two operators presented in Section \ref{section:overview} and analyzed in Section \ref{section:spacespectrum}.
For efficiencies in implementation, we propose a novel design that combines both operators into one compact setup.
Figure \ref{fig:optical_setup} shows a schematic that uses polarization to achieve both operators with a single SLM and a single camera.
First, in Figure \ref{fig:optical_setup}(a), an SLM is placed 2$f$  away from the grating, and an image sensor 2$f$ away from the SLM, implementing spectrally coded spatial measurement operator $\mathcal{I}$.
In Figure \ref{fig:optical_setup}(b), light follows an alternate path where in the SLM is 4$f$ away from the grating; the camera is still 2$f$ away from the SLM. This light path allows us to achieve the spatial-coded spectral measurement operator $\mathcal{S}$.
The two light pathways are combined using a  combination of polarizing beam splitters (PBS) and liquid crystal rotators (LC).
The input light is pre-polarized to be either S-polarized or P-polarized.
When the light is P-polarized, the SLM is effectively 2$f$ units away from the grating leading to implementation of $\mathcal{I}$, the spectrally-coded imager.
When the light is S-polarized, the SLM is 4$f$ units away, provided the polarizing beamsplitter, PBS 3 was absent.
To counter this, an LC rotator is placed before PBS 3 that rotates S-polarization to P-polarization when switched on. 
Hence, when S-light is input in conjugation with the rotator being switched on, we achieve the operator $\mathcal{S}$, a spatially-coded spectrometer.
By simultaneously controlling the polarization of input light and the LC rotator, we can implement both $\mathcal{I}$ and $\mathcal{S}$ operators with a single camera and SLM pair.

\begin{figure*}
	\centering
	\includegraphics[width=\textwidth]{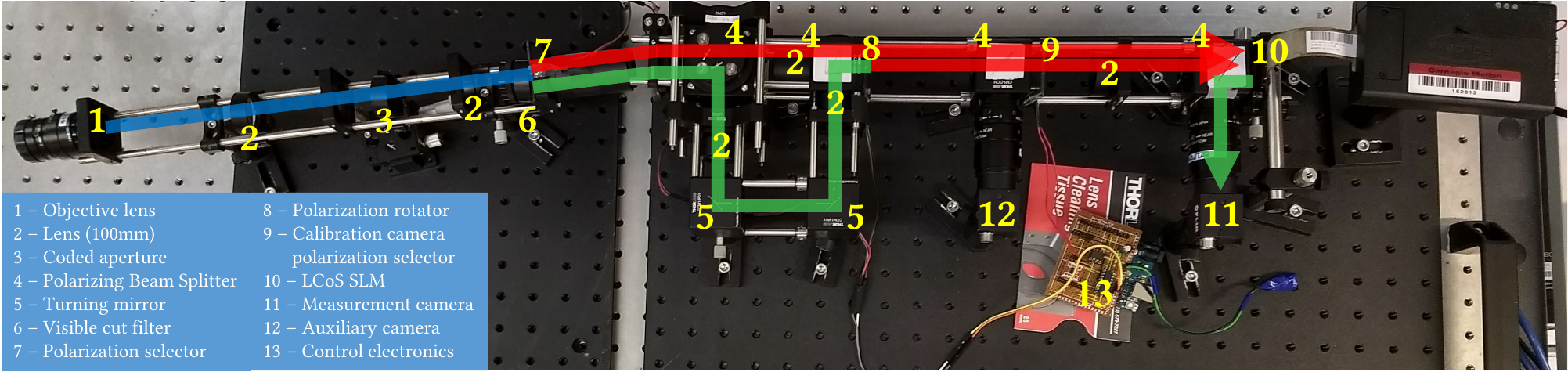}
	\caption{Photograph of our lab prototype. The optical paths for spectral as well as spatial coding shown in Fig. \ref{fig:optical_setup} have been overlaid for easy understanding. Components have been marked, grouped and labeled for convenience.  All other relevant information is available in supplementary material.}
	\label{fig:real_setup}
\end{figure*}

Figure \ref{fig:real_setup} shows our lab prototype with the entire light pathway including the coded aperture placed in the relay system between the objective lens and diffraction grating.
The input polarization is controlled by using a second LC rotator with a polarizer, placed before the diffraction grating.
Finally, an auxiliary camera is used to image the pattern displayed on the SLM. This camera is used purely for alignment of the pattern displayed on the SLM.
A detailed list of components can be found in the supplemental material.
\paragraph{Calibration.} Our optical setup requires three calibration processes.
The first one is camera to SLM calibration. We used an auxiliary camera (Component 12 in Figure \ref{fig:real_setup}) that is directly focused on the SLM for this purpose.
The second one is calibration of wavelengths. We used several narrowband filters to identify the location of wavelengths.
Finally, the third one is radiometric calibration. We used a calibrated Tungsten-Halogen light source to estimate the spectral response of the setup.
A detailed description of the calibration procedure  can be found in the supplementary material.

\paragraph{System characterization.} 
Spectral resolution (FWHM) of the setup was computed using several 1nm narrowband filters across visible wavelengths. Our optical setup provided an FWHM of 2.9nm.
Spatial resolution was computed by capturing photo of a Siemens star, and then deconvolving with a point-spread function obtained by capturing image of a $10\mu m$ pinhole. 
The frequency that achieved 30\% of the modulation transfer function, MTF30,  was found to be nearly 0.4 line pairs/pixel.
All computation details, as well as relevant figures, can be found in the supplementary material. 

\paragraph{Diffraction due to LCoS pattern.} Since the SLM is placed $2f$ away from spectral or spatial measurements, the displayed pattern introduces diffraction blur, creating a non-linear measurement system.
To counter this, we add a constant offset to both positive and negative patterns, which makes the diffraction blur compact enough that the non-linearities can be neglected. 
\paragraph{Spectral deconvolution.}
Measurements by our optical system return spectra at each point, convolved by the aperture code. 
To get the true spectrum, we deconvolved the $k^\textrm{th}$ measured singular vector using a smoothness prior. The specific objective function we used:
\begin{align}
	\min_{\bfv_k} \quad \frac{1}{2}\|\bfy_k - a \ast \bfv_k\|^2 + \eta \| \nabla \bfv_k\|^2,
	\label{eq:spec_deconv}
\end{align}
where $\bfv_k$ is the true spectrum, $\bfy_k$ is the measured spectrum, $a$ is the aperture code, $\nabla \bfx$ is the first order difference of $\bfx$, and $\eta$ is weight of penalty term.
Solution to (\ref{eq:spec_deconv}) was computed using conjugate gradient descent.
Higher $\eta$ favors smoother spectra, and hence is preferred for illuminants with smooth spectra, such as tungsten-halogen bulb or white LED.
For peaky spectra such as CFL, a lower value of $\eta$ is preferred. 
In our experiments, we found $\eta = 1$ to be appropriate for peaky spectra, whereas, $\eta = 10^{3}$ was appropriate for experiments with tungsten-halogen illumination. 
We compare performance of various deconvolution algorithms in the supplementary section.

\paragraph{Spatial deconvolution.} Equation (\ref{eq:spatialblur}) suggests that the spatial blur kernel varies across different spectral bands. 
More specifically, the blur kernels at two different spectra are scaled versions of each other. 
However, we observed that the variations in blur kernels were not significant when we image over a small waveband --- for example, the visible waveband of $420-680$nm.
Given this, we approximate the spatial blur as being spectrally independent, which leads to the following expression:
\begin{align}
	I_S(x, y) \propto \left[ \int_\lambda H(x, y, \lambda) c(\lambda)  d\lambda \right] \ast  p(x, y),
\end{align}
where $p(x, y)$ is the spatial blur.
We estimated the spatial blur kernel by imaging a pinhole and subsequently deconvolved the spatial singular vectors. 
We used a TV prior based deconvolution using the technique in \cite{bioucas2007new} using the image of a pinhole as the PSF.
Details of the deconvolution procedure are in the supplementary section.
\begin{figure*}[!ttt]
	\centering
	\includegraphics[width=\textwidth]{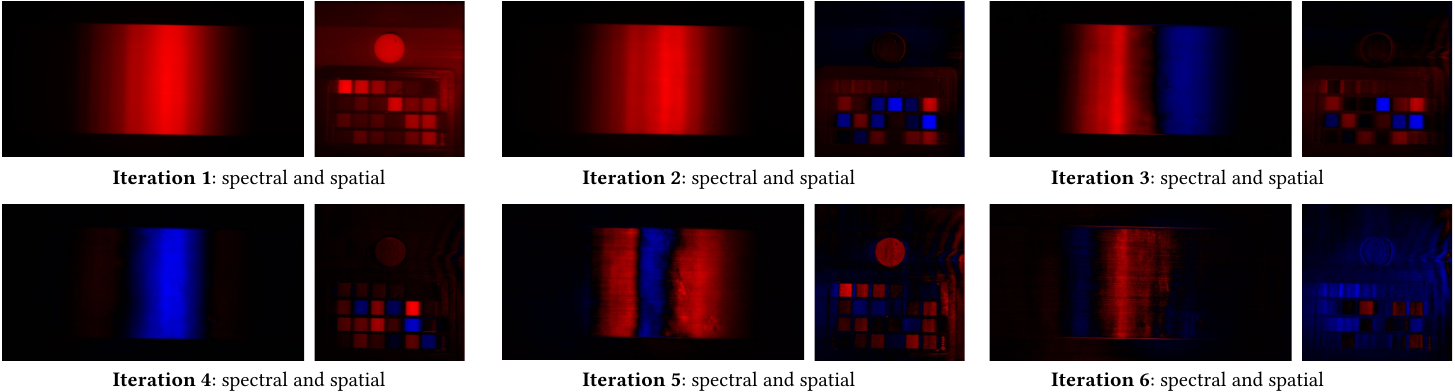}\caption{Data captured during measurement process for a rank-4 approximation of the ``Color checker" scene for six iterations. A picture of the scene is shown in Figure \ref{fig:macbeth_master}. Positive part of data is shown in red and the negative part is shown in blue. KRISM alternates between acquiring spectral and spatial measurements to compute both spatial and spectral singular vectors. The first four iterations involve capturing the dominant wavelengths that includes yellow and green colors, since they have the highest magnitude. The next set of iterations capture the blue and red wavelengths.}
	\label{fig:macbeth_ritz}
\end{figure*}

\section{Real Experiments} \label{section:real}
\begin{figure}[!ttt]
	\centering
	\begin{subfigure}[t]{0.381\columnwidth}
		\centering
		\includegraphics[width=\textwidth]{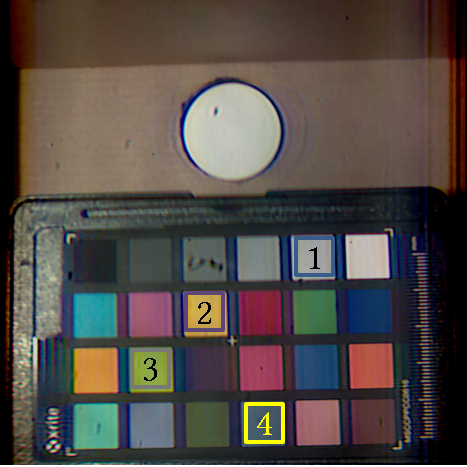}
		\caption{RGB image}
	\end{subfigure}
	\quad
	\begin{subfigure}[t]{0.56\columnwidth}
		\centering
		\includegraphics[width=\textwidth]{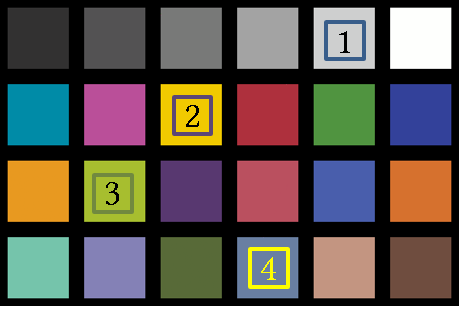}
		\caption{Reference RGB image}
	\end{subfigure}
	\quad
	\begin{subfigure}[t]{\columnwidth}
		\centering
		\includegraphics[width=\textwidth]{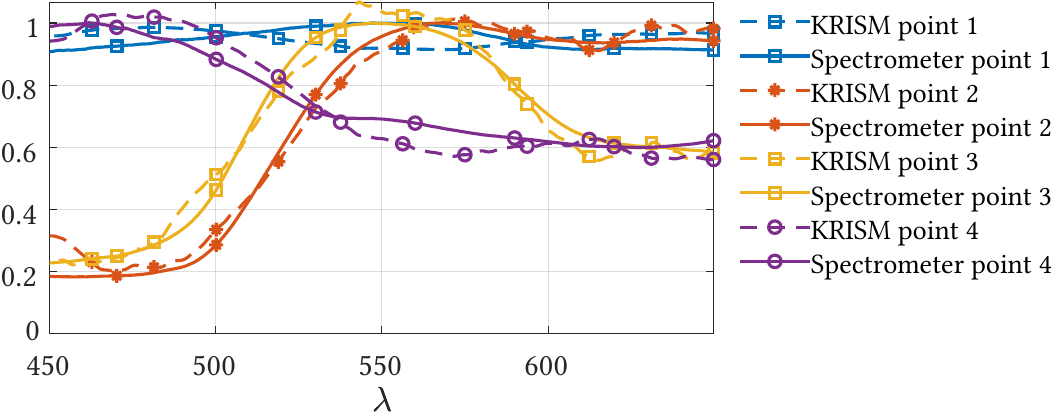}
	\end{subfigure}
	\caption{Macbeth color chart. Spectra is shown at four locations and compared with spectrometer readings. The PSNR is 25dB or higher and the SAM between KRISM spectra and spectrometer readings is less than $6^\circ$.}
	\label{fig:macbeth_master}
\end{figure}

We present several results from real experiments which show the effectiveness of KRISM.
We evaluate the ability to measure singular vectors with high accuracy, and high spatial and spectral resolution capabilities.
Unless specified, experiments involved a capture of a rank-4 approximation of the HSI, with 6 spectral and 6 spatial measurements. 
Lanczos iterations were  initialized with all-ones spatial image to speed up convergence.
HSIs were acquired with a spatial resolution of $560\times550$ pixels and a spectral resolution of $256$ bands between 400nm to 700nm, with 3 nm FWHM.
For verifying spectroradiometric capabilities, we obtained spectral measurements at a small set of spatial points using an Ocean Optics FLAME spectrometer.
We use spectral angular mapper (SAM) \cite{yuhas1992discrimination} similarity and PSNR between spectra measured by our optical setup and that measured with a spectrometer.
SAM between two vectors $\bfx$ and $\widehat{\bfx}$ is defined as $SAM = \cos^{-1}\left( \frac{\bfx^\top \widehat{\bfx}}{\| \bfx \| \|\widehat{\bfx}\|} \right)$.
\paragraph{Visualization of Lanczos iterations}
Figure \ref{fig:macbeth_ritz} shows iterations for the ``Color checker" scene in Figure \ref{fig:macbeth_master}.
The algorithm initially captures brightest parts of the image, corresponding to the spectralon, and the white and yellow patches. 
Consequently, by iteration 5, the blue and red parts of the image are isolated.
The iterations are representative of the signal energy in various wavelengths.
Maximum energy is concentrated in yellow wavelengths, due to tungsten-halogen illuminant and spectral response of the camera.
This is then followed by the red wavelengths, and finally the blue wavelengths.

\begin{figure*}[!ttt]
	\centering
	\begin{subfigure}[c]{0.23\textwidth}
		\centering
		\includegraphics[width=\textwidth]{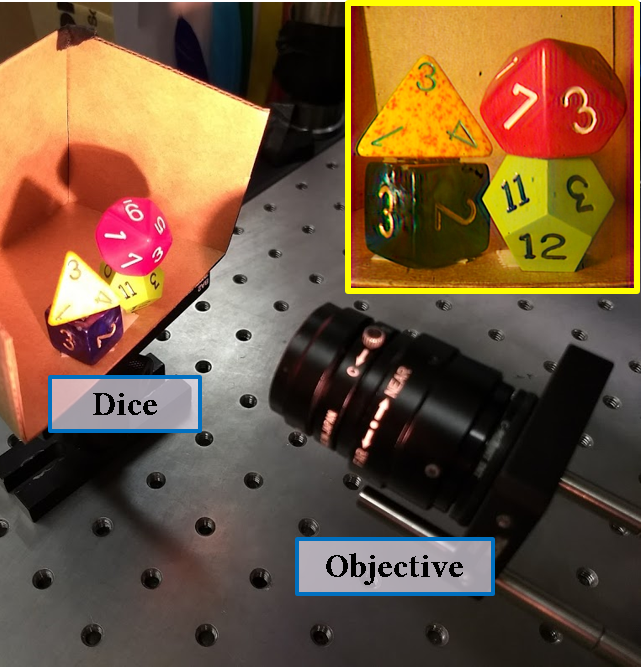}
		\caption{Setup with RGB image in inset}
	\end{subfigure}
	\quad
	\begin{subfigure}[c]{0.23\textwidth}
		\centering
			\includegraphics[width=\textwidth]{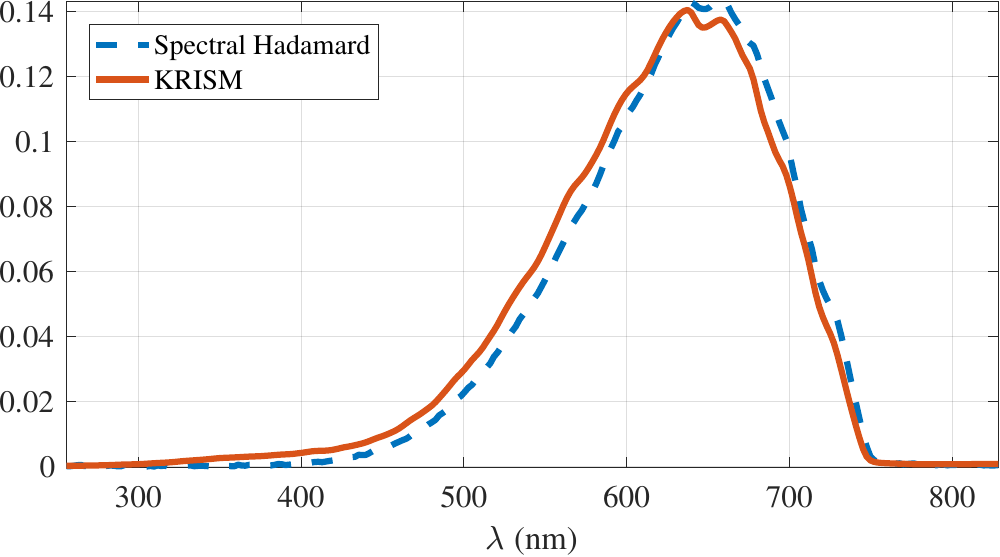}
			\includegraphics[width=0.45\textwidth]{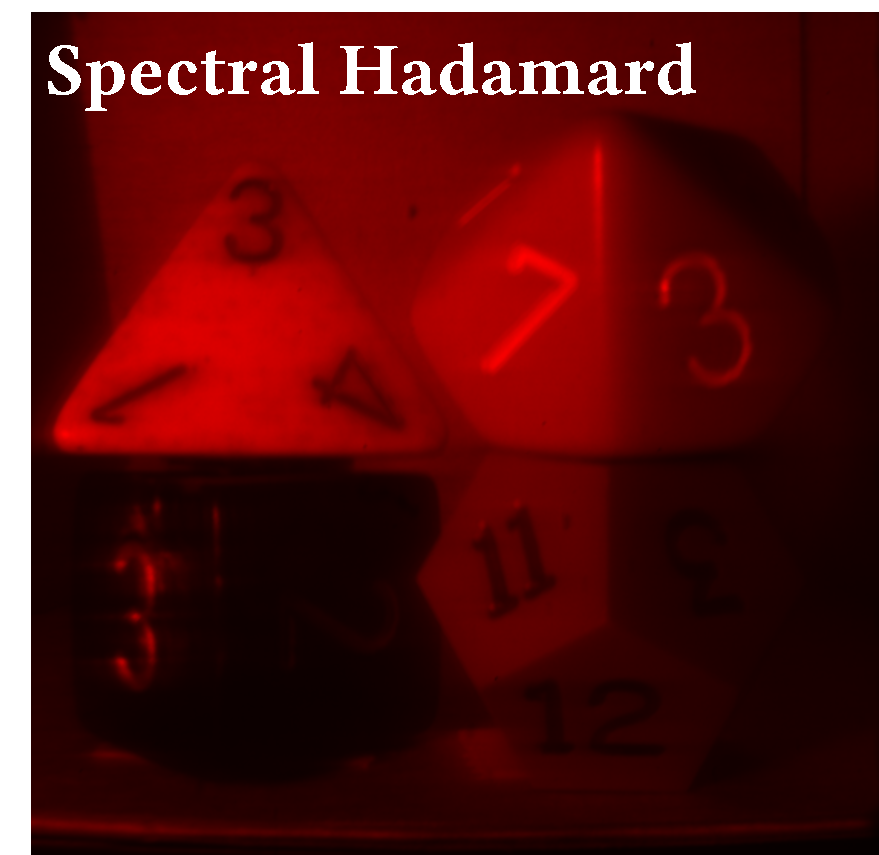}
			\quad
			\includegraphics[width=0.45\textwidth]{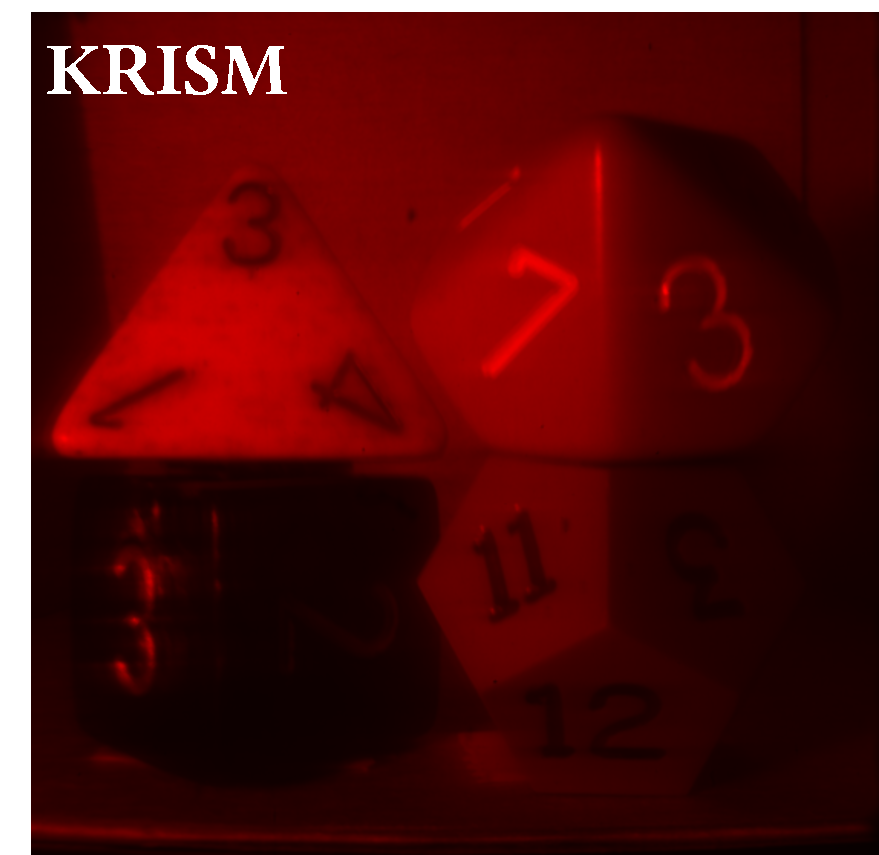}
			\caption{First singular vector}
	\end{subfigure}
	\quad
	\begin{subfigure}[c]{0.23\textwidth}
		\centering
		\includegraphics[width=\textwidth]{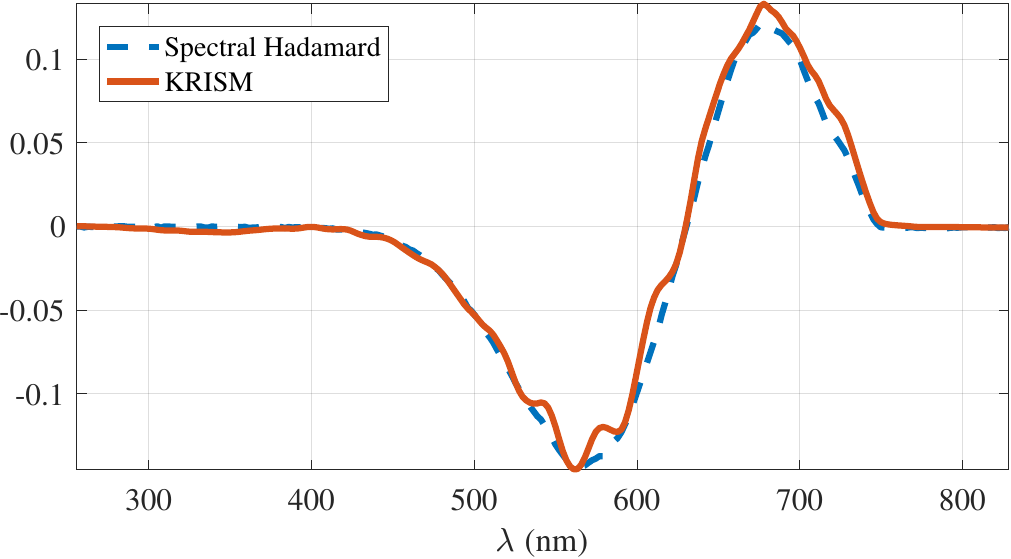}
		\includegraphics[width=0.45\textwidth]{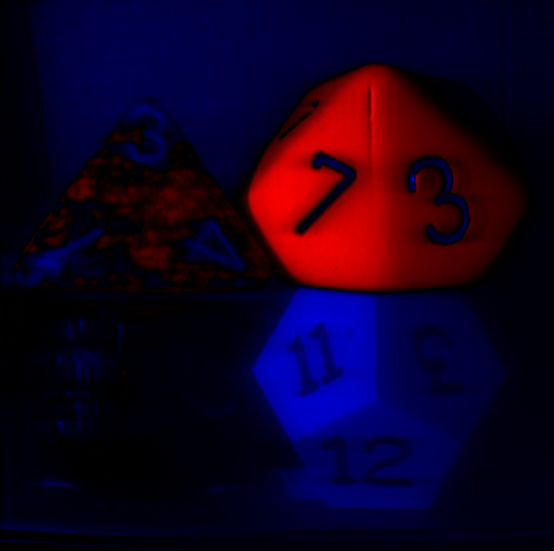}
		\quad
		\includegraphics[width=0.45\textwidth]{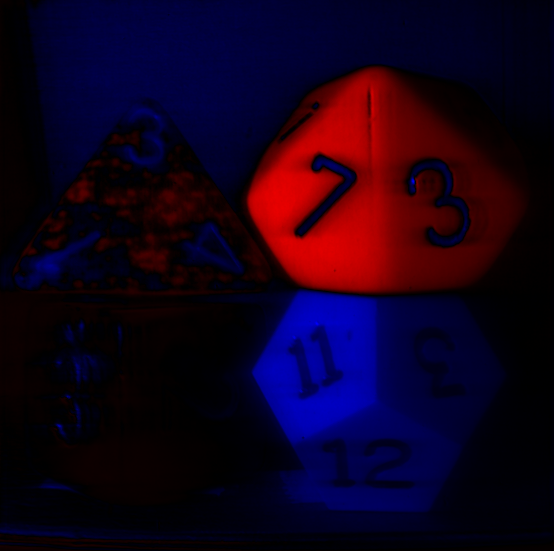}
		\caption{Second singular vector}
	\end{subfigure}
	\quad
	\begin{subfigure}[c]{0.23\textwidth}
		\centering
		\includegraphics[width=\textwidth]{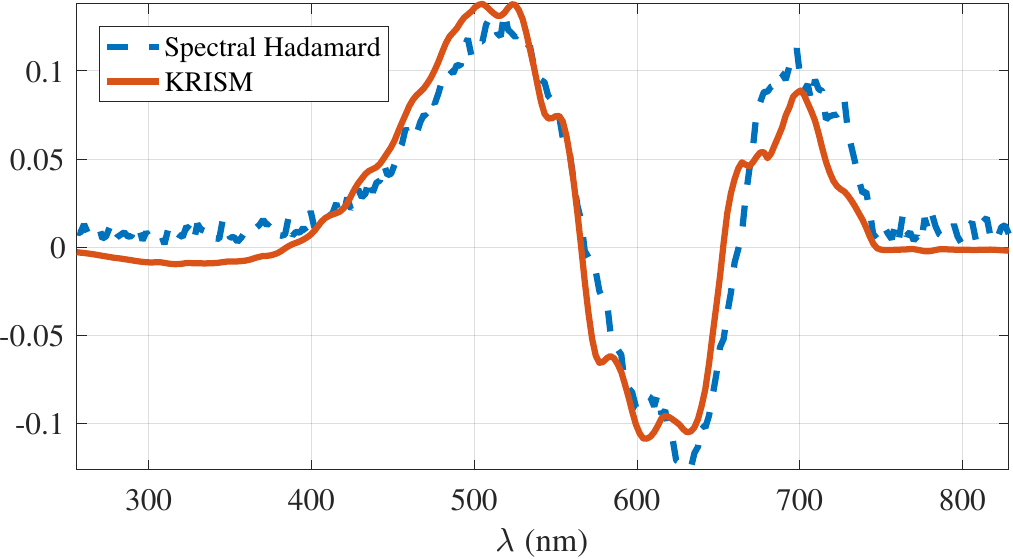}
		\includegraphics[width=0.45\textwidth]{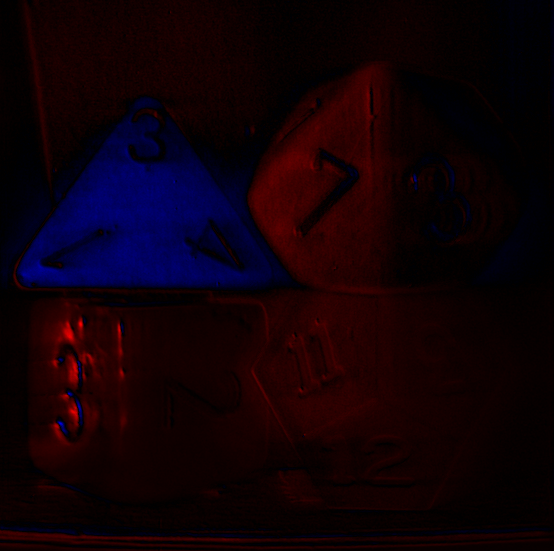}
		\quad
		\includegraphics[width=0.45\textwidth]{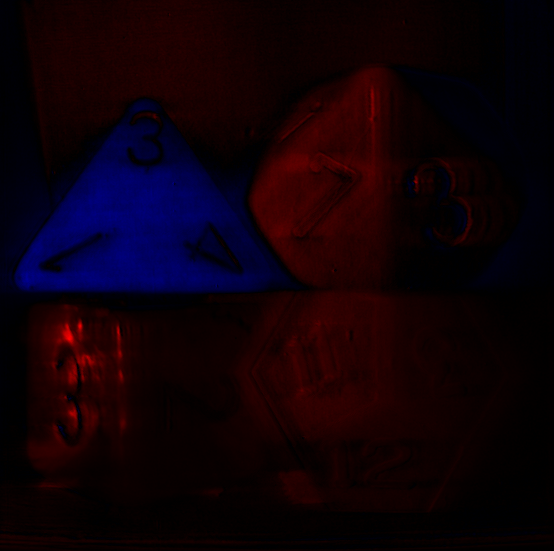}
		\caption{Third singular vector}
	\end{subfigure}
	\caption{Comparison of singular vectors captured via spectrally Hadamard-multiplexed sensing and KRISM for the dice scene. the left image singular vector is from Hadamard multiplexed data and the right one is from KRISM. Blue represents negative values and red represents positive values. KRISM required a total of 6 spectral and 6 spatial measurements to construct 4 singular vectors. While spectral Hadamard sampling method took a total of 49 minutes, KRISM took under 2 minutes. The SAM value between the singular vectors was less than $20^\circ$.}
	\label{fig:dice_eig}
\end{figure*}

\paragraph{Comparison of measured  singular vectors.}
We obtain the complete hyperspectral image through a permuted Hadamard multiplexed sampling in the spectral domain for comparison with ground-truth singular vectors.
We chose a scene with four colored dice for this purpose, shown in Figure \ref{fig:dice_eig} (\textbf{a}).
We then computed 4 singular vectors of spectrally Hadamard-multiplexed data.
Figure \ref{fig:dice_eig} shows a comparison of the spatial and spectral singular vectors.
The singular vectors obtained via Krylov subspace technique are close to the ones obtained through Hadamard sampling. 
On an average, the reconstruction accuracy between KRISM and Hadamard multiplexing was found to be greater than 30dB, while the angle between the singular vectors was no worse than $20^\circ$, with the top three singular vector having an error smaller than $8^\circ$.
Hadamard sampling  took  49 minutes while KRISM took under 2 minutes for 6 spatial and 6 spectral measurements, thus offering a speedup of $20 \times$.
\paragraph{Peaky spectrum illumination}
We imaged a small toy figurine of ``Chopper", placed under CFL, which has a  peaky spectrum, to test high spatio-spectral resolving capability.
Figure \ref{fig:chopper} shows the rendered RGB image and spectra at a representative location.
Spectra at a selected spatial point, as measured by KRISM, and a spectrometer are  shown as well.
The SAM between spectrum measured by KRISM and that measured by spectrometer was found to be $14.7^\circ$.
Notice that the location of the peaks, as well as the heights match accurately. 
Indeed, the chopper example establishes the high spatio-spectral resolution capabilities of KRISM.

\begin{figure*}[!tt]
	\centering
	\begin{subfigure}[t]{\textwidth}
		\centering
		\begin{subfigure}[t]{0.02\textwidth}
			\centering
			\includegraphics[width=\textwidth]{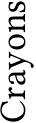}
		\end{subfigure}
		\quad
		\begin{subfigure}[t]{0.203\textwidth}
			\centering
			\includegraphics[width=\textwidth]{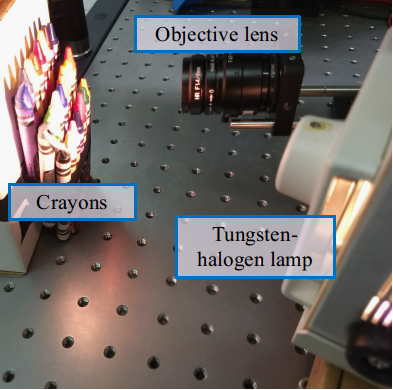}
		\end{subfigure}
		\quad
		\begin{subfigure}[t]{0.32\textwidth}
			\centering
			\includegraphics[width=\textwidth]{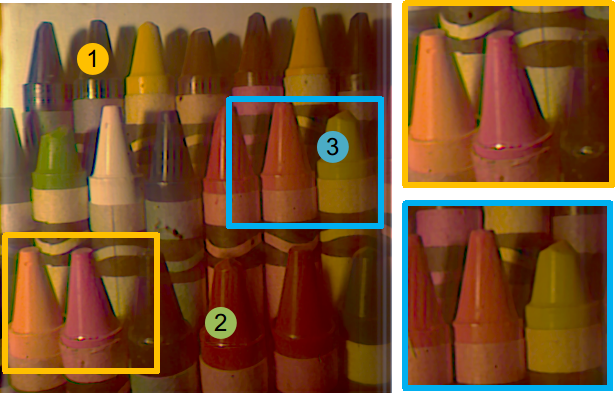}
		\end{subfigure}
		\quad
		\begin{subfigure}[t]{0.378\textwidth}
			\centering
			\includegraphics[width=\textwidth]{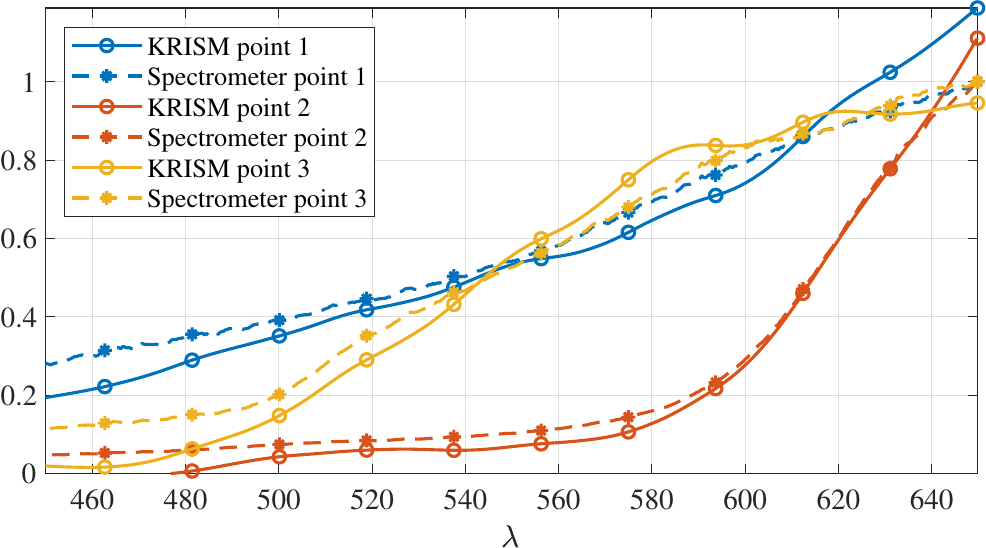}
		\end{subfigure}
	\end{subfigure}
	
	\vspace{0.2cm}
	
	\begin{subfigure}[t]{\textwidth}
		\centering
		\begin{subfigure}[t]{0.02\textwidth}
			\centering
			\includegraphics[width=\textwidth]{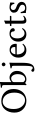}
		\end{subfigure}
		\quad
		\begin{subfigure}[t]{0.203\textwidth}
			\centering
			\includegraphics[width=\textwidth]{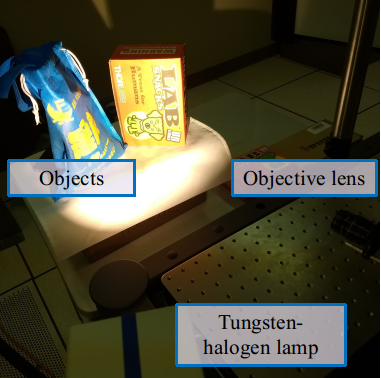}
		\end{subfigure}
		\quad
		\begin{subfigure}[t]{0.32\textwidth}
			\centering
			\includegraphics[width=\textwidth]{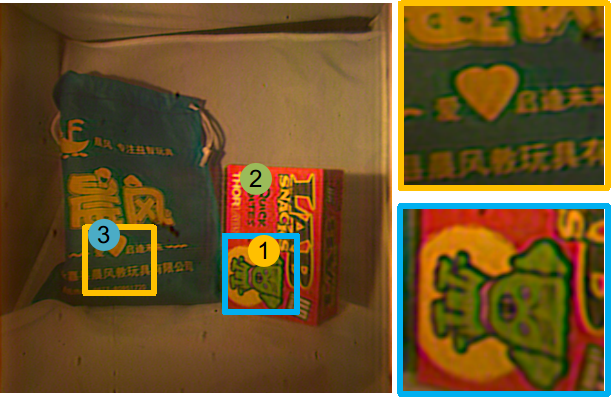}
		\end{subfigure}
		\quad
		\begin{subfigure}[t]{0.378\textwidth}
			\centering
			\includegraphics[width=\textwidth]{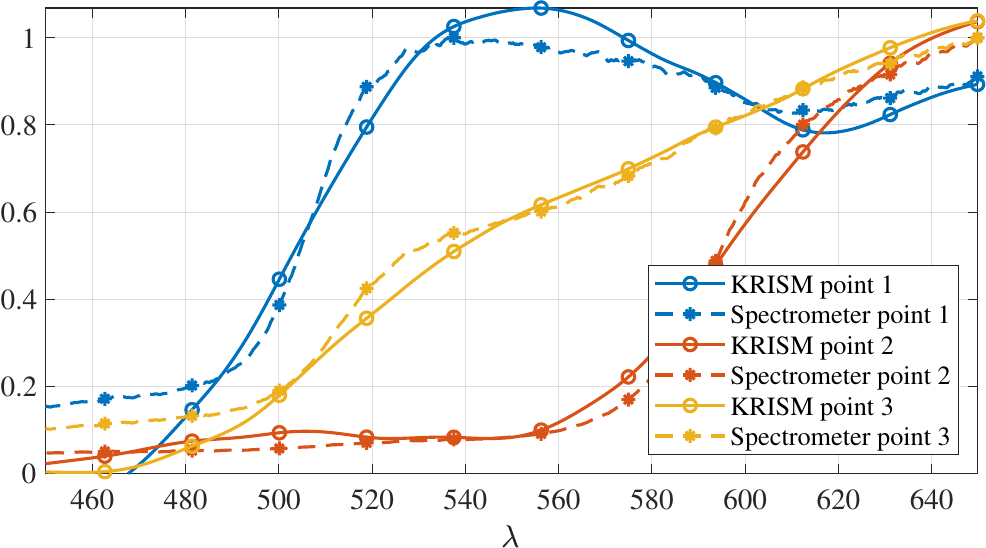}
		\end{subfigure}
	\end{subfigure}

	\vspace{0.2cm}

	\begin{subfigure}[t]{\textwidth}
		\centering
		\begin{subfigure}[t]{0.02\textwidth}
			\centering
			\includegraphics[width=\textwidth]{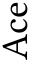}
		\end{subfigure}
		\quad
		\begin{subfigure}[t]{0.203\textwidth}
			\centering
			\includegraphics[width=\textwidth]{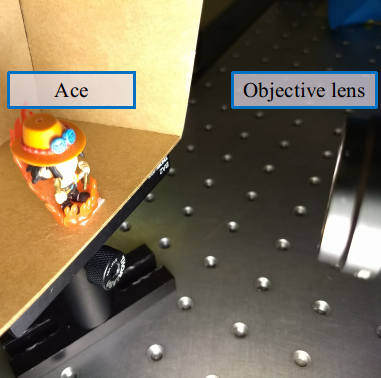}
		\end{subfigure}
		\quad
		\begin{subfigure}[t]{0.32\textwidth}
			\centering
			\includegraphics[width=\textwidth]{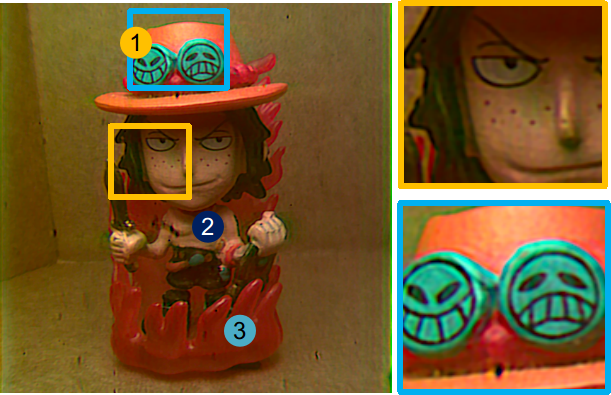}
		\end{subfigure}
		\quad
		\begin{subfigure}[t]{0.378\textwidth}
			\centering
			\includegraphics[width=\textwidth]{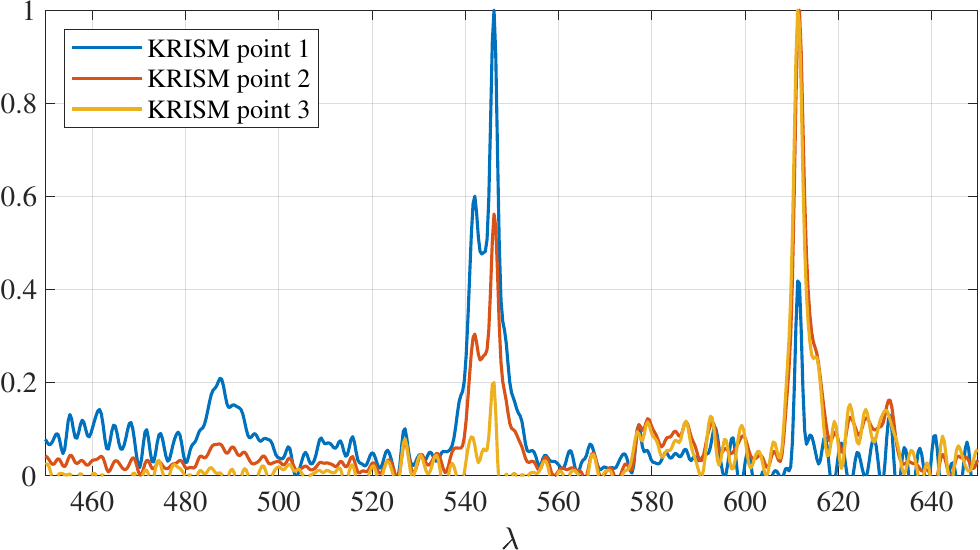}
		\end{subfigure}
	\end{subfigure}

	\vspace{0.2cm}
	
	\begin{subfigure}[t]{\textwidth}
		\centering
		\begin{subfigure}[t]{0.02\textwidth}
			\centering
			\includegraphics[width=\textwidth]{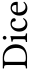}
		\end{subfigure}
		\quad
		\begin{subfigure}[t]{0.202\textwidth}
			\centering
			\includegraphics[width=\textwidth]{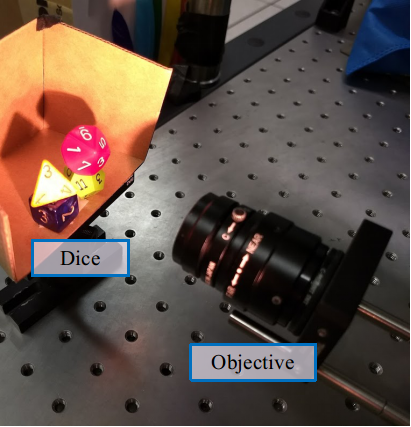}
		\end{subfigure}
		\quad
		\begin{subfigure}[t]{0.32\textwidth}
			\centering
			\includegraphics[width=\textwidth]{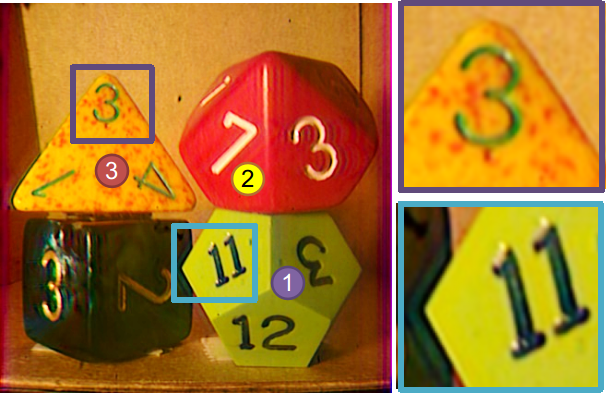}
		\end{subfigure}
		\quad
		\begin{subfigure}[t]{0.378\textwidth}
			\centering
			\includegraphics[width=\textwidth]{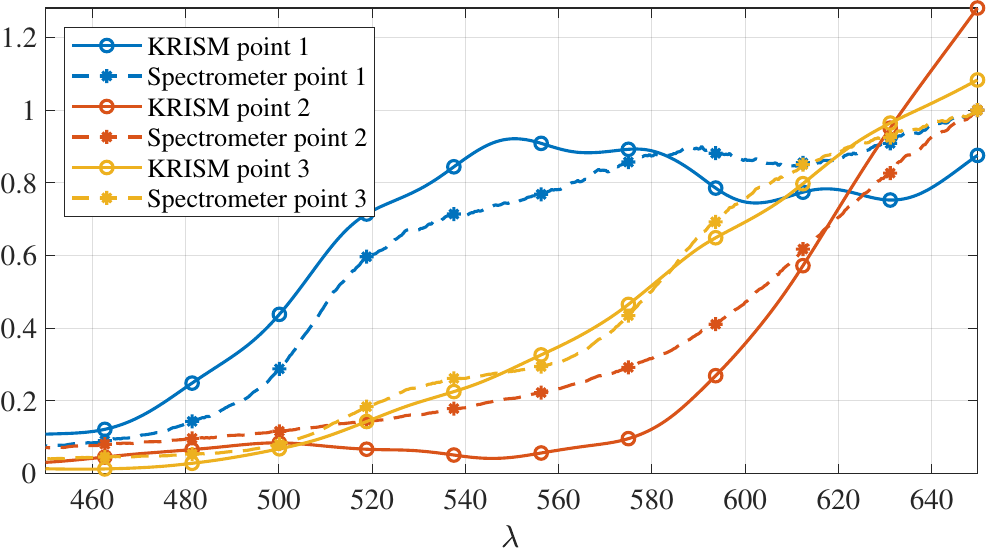}
		\end{subfigure}
	\end{subfigure}	
	
	\vspace{0.2cm}
	
	\begin{subfigure}[t]{\textwidth}
		\centering
		\begin{subfigure}[t]{0.02\textwidth}
			\centering
			\includegraphics[width=\textwidth]{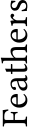}
		\end{subfigure}
		\quad
		\begin{subfigure}[t]{0.203\textwidth}
			\centering
			\includegraphics[width=\textwidth]{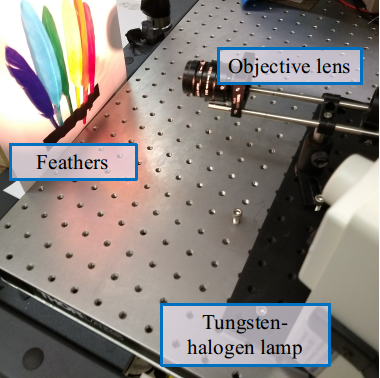}
		\end{subfigure}
		\quad
		\begin{subfigure}[t]{0.32\textwidth}
			\centering
			\includegraphics[width=\textwidth]{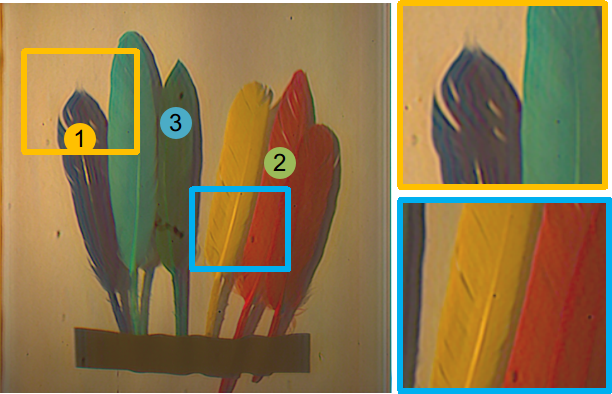}
		\end{subfigure}
		\quad
		\begin{subfigure}[t]{0.378\textwidth}
			\centering
			\includegraphics[width=\textwidth]{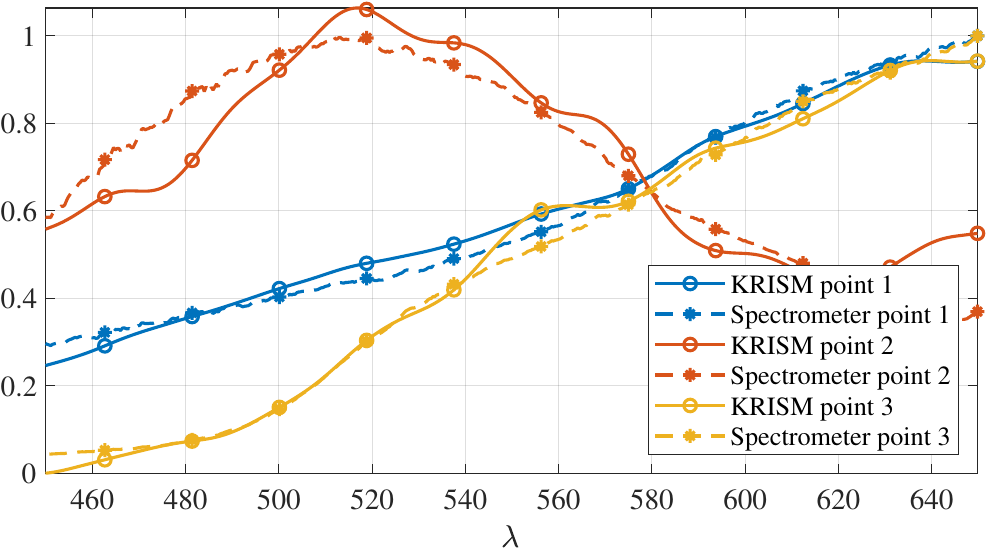}
		\end{subfigure}
	\end{subfigure}
	\caption{Real data captured with our optical setup. We show the physical setup used for capturing the data, rendered RGB image with some interesting patches zoomed in, and spectra at some points, compared with a spectrometer. The results are promising, as the spectra is very close to spectrometer readings (PSNR > 20dB), and the spatial images are captured in high resolution.}
	\label{fig:real}
\end{figure*}

\paragraph{Diverse real experiments.}
Figure \ref{fig:real} shows several real world examples captured with our optical setup, with a diverse set of objects.
For verification with ground truth, we captured spectral profiles at select spatial locations.
The ``Dice" and ``Objects" scene captures several more colorful objects with high texture.
The zoomed-in pictures show the spatial resolution, while the comparison of spectra highlights the fidelity of our system as a spectral measurement tool.
``Ace" scene was captured by placing the toy figurine under CFL illuminant, which is peaky.
We could not obtain ground truth with a spectrometer, as the toy was too small to reliably probe with a spectrometer.
The peaks are located within 2nm of ground truth peaks, and the relative heights of the peaks match the underlying color.
``Crayons" scene consists of numerous colorful wax crayons illuminated with a tungsten-halogen lamp. 
The closeness of spectra with respect to spectrometer readings shows the spectral performance of our setup.
Finally, ``Feathers" consists of several colorful feathers illuminated by tungsten-halogen lamp.
The fine structure of feathers is well captured by our setup.

\paragraph{Color checker.} Since our setup is optimized for viewing in 400nm-700nm, we evaluated our system on the 24-color Macbeth color chart.
The Macbeth color chart consists of a wide gamut of colors in visible spectrum that are spectrally well separated, and forms a good test bench for visible spectrometry.
We placed the ``Color passport" and spectralon plug in front of our camera and illuminated it with a tungsten-halogen DC light source.
The spectralon has a spectrally flat response, and hence helps estimate the spectral response of the illuminant+spectrometer system. 
This enables measurement of true radiance of the color swatches.
Since the spectra is smooth, we used least squares recovery of the spectrum, with $\ell_2$ penalty on the first difference of spectral singular vectors.
The captured data was then normalized by dividing spectrum of all points with the spectrum of the spectralon.
Figure \ref{fig:macbeth_master} shows the captured image against reference color chart along with spectra at select locations plotted along with ground truth spectra. 
On an average, the PSNR between spectra measured by KRISM and that measured by spectrometer is greater than 25dB, while the SAM is less than $6^\circ$. 

\section{Discussion and conclusion} \label{section:discussion}
We presented a novel hyperspectral imaging methodology called KRSIM, and provided an associated novel optical system for enabling optical computation of hyperspectral scenes to acquire the top few singular vectors in a fast and efficient manner.
Through several real experiments, we establish the strength of KRISM in three important aspects: 1) the ability to capture singular vectors of the hyperspectral image with high fidelity, 2) the ability to capture an approximation of the hyperspectral image with $20\times$ or faster acquisition rate compared to Nyquist sampling, and 3)  the ability to measure simultaneously at high spatial and spectral resolution.
We believe that our setup will trigger several new experiments in adaptive imaging for fast and high resolution hyperspectral imaging.
\begin{figure}[!tt]
	\centering
	\begin{subfigure}[c]{0.31\columnwidth}
		\centering
		\includegraphics[width=\textwidth]{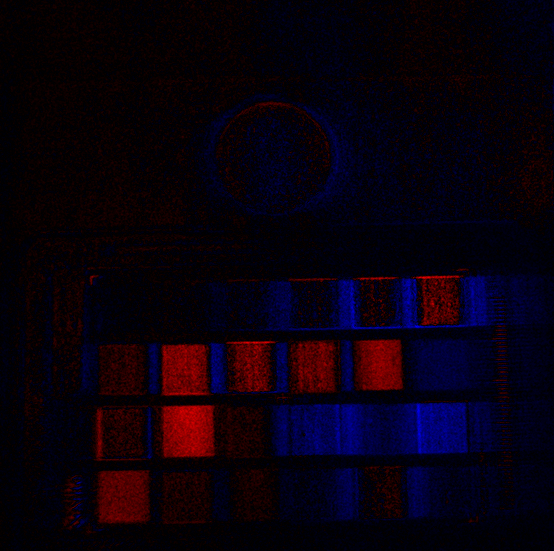}
		\caption{Spectral Hadamard}
	\end{subfigure}
	\hspace{0.1em}
	\begin{subfigure}[c]{0.31\columnwidth}
		\centering
		\includegraphics[width=\textwidth]{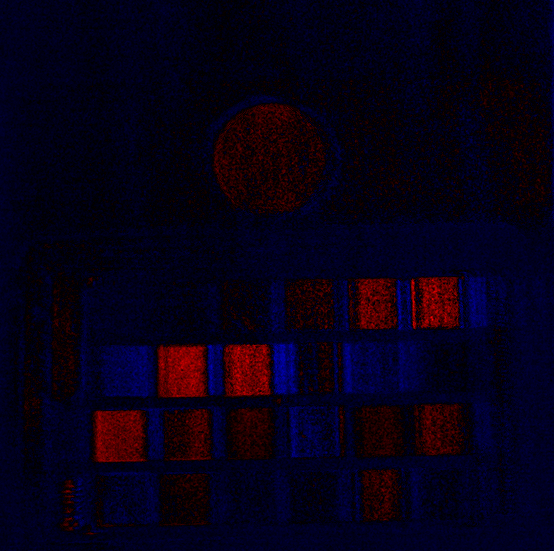}
		\caption{KRISM}
	\end{subfigure}
	\hspace{0.1em}
	\begin{subfigure}[c]{0.31\columnwidth}
		\centering
		\includegraphics[width=\textwidth]{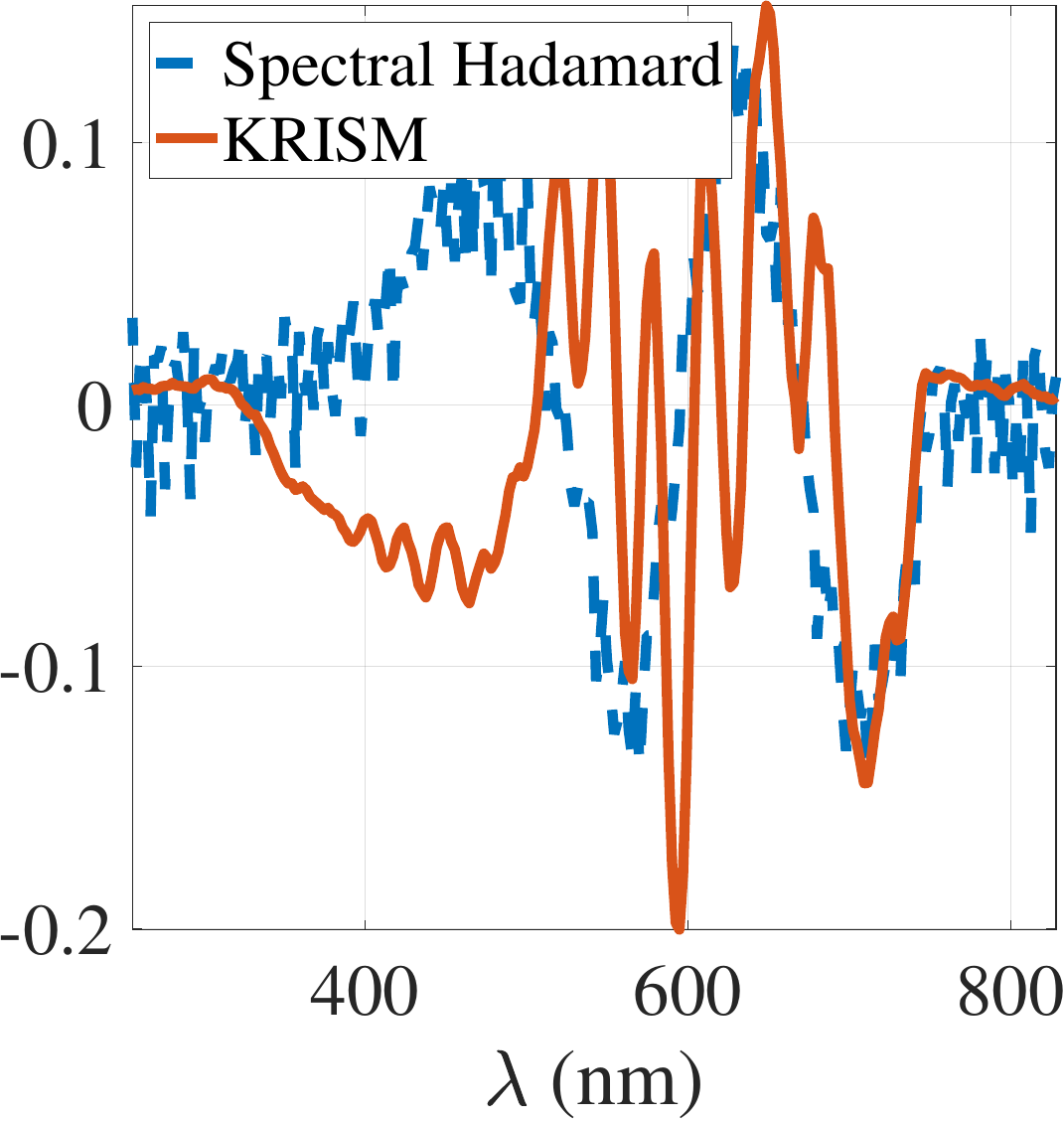}
		\caption{Spectra}
	\end{subfigure}
	\caption{Capturing higher singular vectors. Since KRISM computes higher singular vectors by progressively blocking more light, photon noise dominates measurements after some iterations resulting in noisy estimates of singular vectors. The above example shows inaccurately estimated fifth singular vectors measured for ``color checker" scene with our lab prototype.}
	\label{fig:failure}
\end{figure}

\paragraph{Added advantages.} 
There  are two additional advantages to KRISM. One, since we capture the top few singular vectors directly, there is a data compression from the acquisition  itself.
Two, the only recovery time involves deconvolution of a few spatial and spectral singular vectors, which is significantly less than the time required for recovery of hyperspectral images from CS measurements. 

\paragraph{Beyond low-rank volumes.} Key to our paper is the assumption that the underlying HSI is low-rank. 
Sensing a high rank HSI will require several measurements which negates the benefits of KRISM.
However, there are several other matrix sampling techniques that rely on row or column sensing \cite{havsan2007matrix,ou2011lightslice} to capture information about high rank matrices in an efficient manner. 
Since the proposed setup is capable of computing arbitrary matrix-vector products, such matrix sampling techniques can be implemented efficiently.

\paragraph{Effect of photon noise} Although Krylov subspace based methods are very robust to noise \cite{simoncini2003theory}, the quality of the singular vectors degrade as the rank of acquisition is increased (see Figure \ref{fig:failure}).
This is primarily due to photon noise, as we progressively block most of the energy contained in initial singular vectors.
This can be mitigated by increasing the exposure time of measurements for higher singular vectors.
All said, the problem of noisy higher singular vectors exists with any kind of sampling scheme and hence needs separate attention via a good noise model. 

\section{Acknowledgement}
The authors thank Prof. Ioannis Gkioulekas (Robotics Institute, CMU) for  valuable feedback, and Ms. Yi Hua (ECE Department, CMU) for help with making the figures.
The authors acknowledge support via the NSF CAREER grant CCF-1652569, the  National Geospatial-Intelligence Agency’s Academic Research Program (Award No. HM0476-17-1-2000), and the Intel ISRA on compressive sensing.
Vishwanath Saragadam also gratefully acknowledges support via the Prabhu and Poonam Goel fellowship.

\bibliographystyle{ACM-Reference-Format}
\bibliography{refs}

\begin{appendices}
	\section{Supplementary Material}
	This article supplements the main paper with several simulations and real world experiments.
The remainder of this article is organized as follows.
\begin{itemize}[leftmargin=*]
\item \textbf{Section \ref{section:sup_spacespectrum} --- Derivation of spatio-spectral blur.} We provide an in-depth derivation of the spatial and spectral blur relationship due to a coded aperture mentioned in Section 4 of the main paper.
\item \textbf{Sections \ref{section:sup_deconv} --- Code design.} We provide details on the design of coded aperture, comparisons with alternate codes such as M-sequences, and specifications of the deconvolution technique used for both space and spectrum.
\item \textbf{Section \ref{section:sup_components} --- Details of optical implementation.} We provide a comprehensive list of  components used for building the setup.
\item \textbf{Section \ref{section:sup_design} ---Explanation of design choices.} We discuss some of the design considerations for implementing our optical setup. 
\item \textbf{Section \ref{section:sup_calibration} --- Calibration.} This section serves as a guide for calibrating the proposed optical setup.

\item \textbf{Section \ref{section:sup_real} --- Real results.} We provide additional visualization of the real data shown in the main paper.
\item \textbf{Section \ref{section:sup_synthetic} --- Additional simulation results.} We present some more simulations with emphasis on performance across a diverse set of datasets. 
\end{itemize}
	
	\section{Coded apertures for simultaneous sensing of space and spectrum} \label{section:sup_spacespectrum}
	\begin{figure*}[!tt]
	\centering
	\includegraphics[width=\textwidth]{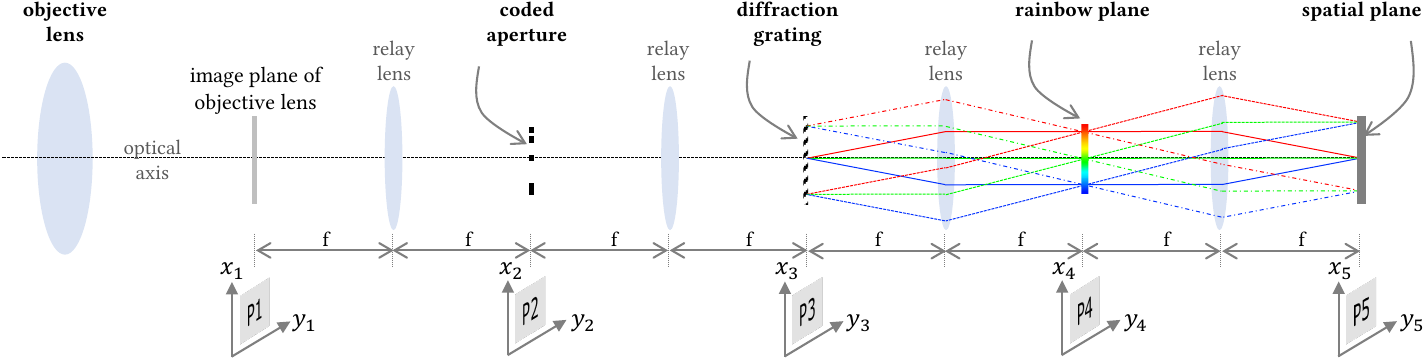}
	\caption{Schematic of an optical setup that simultaneously captures spatial images and spectral profiles. P1 is the image plane of the objective lens, P2 contains spatial frequencies of the image. We place a coded aperture, $a(x, y)$ at this plane. P3 contains the image plane blurred by the coded aperture. We place a diffraction grating at this plane to disperse light into different wavelengths. P4 contains the resultant spectrum and P5 contains the spatial image that is a copy of P3. We use this schematic to explain the derivation of measurements on planes P4 and P5.}
	\label{fig:schematic}
\end{figure*}
The main paper provided a brief derivation of the effect of coded aperture on spatial and spectral blur. We do a more rigorous proof here.
Figure \ref{fig:schematic} explains the optical setup that we will consider for all our derivations. In particular, we place a coded aperture at plane P2, which introduces diffraction blur in both spatial and spectral measurements.
We obtain spectral measurements on P4 and spatial measurements on P5. The goal of this section is to derive the relationship between measurements on P4 and P5 to the coded aperture and the scene's HSI.

\subsection{Assumptions}
The first assumption is that the image is spatially \textit{incoherent}, an assumption that is realistic for most real world settings. 
This implies that the spatial frequencies add up in intensities and not amplitudes.
Next, the diffraction grating is assumed to disperse light along x-axis, which implies no dispersion along the y-axis.
Finally, we assume an ideal thin lens model for all the lenses. This implies that the Fourier transform property of ideal thin lenses holds for all computations.

\subsection{Basics}
The derivation in the sequel relies on the so called Fourier transform property of lenses \cite{goodman2005introduction}.
Suppose that the complex-valued phase field at the plane $z= 0$ is given as $i_{0}(x, y, \lambda)$, where $(x, y)$ denote spatial coordinates on the plane and $\lambda$ denotes the wavelength of light.
Lets place an ideal thin lens with focal length $f$ at $z = f$ and whose optical axis is aligned along the $z-$axis.
The Fourier transforming property states that the complex phasor field that is formed at the plane $z = 2f$ is given as
\[ i_{2f}(x', y', \lambda) = \frac{1}{j\lambda f} I_{0} \left( \frac{x'}{\lambda f}, \frac{y'}{\lambda f}, \lambda \right), \]
where $I_0(u, v, \lambda)$ is the 2D Fourier transform of $i_0(x, y, \lambda)$ along the first two dimensions.

We can compute the 2D Fourier transform of $i_{2f}$ using the scaling property,
\[ I_{2f}(u', v', \lambda) = \frac{1}{j \lambda f} (\lambda f)^2 i_0(-\lambda f u', -\lambda f v', \lambda). \]
The negative signs in the argument of $i_0(\cdot)$ comes from the Fourier transform being the Hermitian of the inverse Fourier transform.
If we now placed a second ideal thin lens of focal length $f$ at $z= 3f$, then the field at $z= 4f$ can be computed as
\begin{align*}
i_{4f}(x'', y'', \lambda) &= \frac{1}{j\lambda f} I_{2f} \left( \frac{x''}{\lambda f}, \frac{y''}{\lambda f}, \lambda \right), \\
&= \frac{1}{(j \lambda f)^2} (\lambda f)^2 i_0(-x'', -y'',  \lambda) \\
&= - i_0(-x'', -y'', \lambda)
\end{align*}
The assembly above, with two lenses at $z= f$ and $z= 3f$ is referred to as a $4f$ system.
As we see above, the $4f$ system replicates the field at $z=0$ at $z=4f$, barring a flip of the coordinate axis; this property is useful for the following discussion.

\subsection{Propagation of signal}
We will use Figure \ref{fig:schematic} as a guide for the derivation.
An objective lens focuses a scene onto its image plane, denoted as P1.
Assuming that all light is incoherent, let the complex phasor at P1 be denoted as $h(x_1, y_1, \lambda)$; note that intensity of this complex field is the hyperspectral image $H(x_1, y_1, \lambda)$  that we seek to measure, i.e., $H(x_1, y_1, \lambda) = | h(x_1, y_1, \lambda)|^2.$

Since we assume an incoherent model, we analyze the system for a point light source and then extend it to a generic image by adding up only intensities.

\paragraph{Field at Plane P1.}
Consider a point light source at $(x_0, y_0)$ with complex amplitude $h(x_0, y_0, \lambda)$.
The overall phasor field at P1 is given as $$i_1(x, y, \lambda) = \delta(x- x_0, y-y_0)h(x_0, y_0, \lambda).$$

\paragraph{Field at Plane P2.}
Using Fourier transform property of lens, we get the field on plane P2 to be,
\begin{align}
	\widehat{i}_2(x_2, y_2, \lambda) &= \frac{1}{j\lambda f}I_1\left(\frac{x_2}{\lambda f}, \frac{y_2}{\lambda f}, \lambda\right)\nonumber\\
					   &= \frac{1}{j\lambda f} h(x_0, y_0, \lambda) e^{-\frac{2\pi j}{\lambda f}(x_0 x_2 + y_0 y_2)},
\end{align}
where $I_1(u, v, \lambda)$ is the continuous 2D Fourier transform of $i_1(x, y, \lambda)$ along the first two dimensions.
The field just after the coded aperture is given by,
\begin{align}
	i_2(x_2, y_2, \lambda) &= a(x_2, y_2)\widehat{i}_2(x_2, y_2, \lambda)\nonumber\\
			&= \frac{1}{j\lambda f}a(x_2, y_2)h(x_0, y_0, \lambda)e^{-\frac{2\pi j}{\lambda f}(x_0 x_2 + y_0 y_2)}
\end{align}

\paragraph{Field at Plane P3.}
Using Fourier transform property of lens a second time, the field just before the diffraction grating is
\begin{align}
	\widehat{i}_3(x_3, y_3, \lambda) &= \frac{1}{j\lambda f}I_2\left(\frac{x_3}{\lambda f}, \frac{y_3}{\lambda f}, \lambda\right)\nonumber\\
			&= \frac{1}{(j\lambda f)^2}h(x_0, y_0, \lambda) A\left(\frac{x_3 + x_0}{\lambda f}, \frac{y_3 + y_0}{\lambda f}\right),
\end{align}
where $A(u, v)$ is the continuous 2D FT of $a(x, y)$.
Since the diffraction grating is assumed to disperse along x-axis, we model it as a series of infinite slits, given by,
\begin{align}
	d(x, y) = \frac{1}{\nu_0}\sum\limits_{k=-\infty}^{\infty} \delta\left(x - \frac{k}{\nu_0}\right),
\end{align}
where $\nu_0$ is the grove density of the diffraction grating, measured in grooves per unit length. 
The $\frac{1}{\nu_0}$ factor ensures that light does not get amplified as it propagates through the setup.
The field just after the diffraction grating is hence given by,
\begin{align}
	i_3(x_3, y_3, \lambda) &= \widehat{i}_3(x_3, y_3, \lambda) d(x_3, y_3) \nonumber\\
						   &= \frac{h(x_0, y_0, \lambda)}{(j\lambda f)^2} A\left(\frac{x_0+x_3}{\lambda f}, \frac{y_0+y_3}{\lambda f}\right)\frac{1}{\nu_0}\sum\limits_{k=-\infty}^{\infty} \delta\left(x_3 - \frac{k}{\nu_0}\right).
\end{align}

\paragraph{Field at the Rainbow Plane P4.} To calculate the field at P4, we first need an expression for $D(u, v)$, the 2D Fourier transform of $d(x, y)$.
\begin{align}
	D(u, v) &= \delta(v) \sum\limits_{k=-\infty}^{\infty} \delta(u - k\nu_0)
\end{align}
The field on plane P4 is given as:
\begin{align}
	 &= \frac{1}{(\lambda f)^2}\frac{1}{j\lambda f}\widehat{I}_3\left(\frac{x_4}{\lambda f}, \frac{y_4}{\lambda f}, \lambda\right) \ast D\left(\frac{x_4}{\lambda f}, \frac{y_4}{\lambda f}\right)\nonumber\\
	 &=\frac{1}{(\lambda f)^2} \frac{1}{j\lambda f}\widehat{I}_3\left(\frac{x_4}{\lambda f}, \frac{y_4}{\lambda f}, \lambda\right) \ast \left( \delta\left(\frac{y_4}{\lambda f}\right)\sum\limits_{k=-\infty}^{\infty}\delta\left(\frac{x_4}{\lambda f} - k\nu_0\right) \right)\nonumber\\
	 &= -\frac{1}{j\lambda f} h(x_0, y_0, \lambda)\cdot \nonumber\\ &\sum\limits_{k=-\infty}^{\infty}a(-(x_4 - k\nu_0 \lambda f), -y_4) e^{j\frac{2\pi}{\lambda f}(x_0(x_4 - k\nu_0 \lambda f)+y_0 y_4)},
\end{align}	
where "$\ast$" represents continuous 2D convolution. Since we are only interested in the first order of diffraction, we set k = 1, giving us,
\begin{align}
	& i_4(x_4, y_4, \lambda) = \cdots \nonumber\\
	& \cdots -\frac{1}{j\lambda f} h(x_0, y_0, \lambda )a(-(x_4 - \nu_0 \lambda f), -y_3) e^{j\frac{2\pi}{\lambda f}(x_0(x_4 - \nu_0 \lambda f)+y_0 y_4)}.
\end{align}

\paragraph{Field at the Spatial Plane P5.}
Finally, the field on plane P5 is,
\begin{align}
	i_5(x_5, y_5, \lambda) &= \frac{1}{j\lambda f}I_4\left(\frac{x_5}{\lambda f}, \frac{y_5}{\lambda f}\right)\nonumber\\
			&= \frac{1}{(j\lambda f)^2} h(x_0, y_0, \lambda) e^{-j2\pi x_5 \nu_0}A\left(-\frac{x_5-x_0}{\lambda f}, -\frac{y_5 - y_0}{\lambda f}\right)
\end{align}

\subsection{Measurement by camera}
A camera can only measure intensity of the field. Assuming a camera with spectral response $c(\lambda)$, the measurement on plane P4 is,
\begin{align}
	M_4(x, y) &= \int\limits_{\lambda}|i_4(x, y, \lambda)|^2c(\lambda)d\lambda\nonumber\\
			  &= \int\limits_{\lambda} \frac{1}{\lambda^2 f^2}| h(x_0, y_0, \lambda )|^2 a^2(-x + \nu_0 \lambda f, -y) c(\lambda ) d\lambda\nonumber\\
			  &= \widehat{H}_{x_0, y_0} \left( \frac{x}{f \nu_0}\right) \ast a^2(-x, -y), 
\end{align}
where $\ast$ is a 2D convolution and
\[  \widehat{H}_{x_0, y_0} \left( \frac{x}{f \nu_0}\right)  = H(x_0, y_0, \lambda) \frac{c(\lambda)}{\lambda^2 f^2}.\]
Extending to a generic incoherent image case, we get the following expression,
\begin{align}
	M_4(x, y) &= \int\limits_{x_0} \int\limits_{y_0} \widehat{H}_{x_0, y_0} \left( \frac{x}{f \nu_0}\right) \ast a^2(-x, -y) dx_0 dy_0 \nonumber \\
	&=  \left [ \int\limits_{x_0} \int\limits_{y_0} \widehat{H}_{x_0, y_0} \left( \frac{x}{f \nu_0}\right)  dx_0 dy_0  \right] \ast a^2(-x, -y) \nonumber \\
			 &= \left( \widetilde{c}\left(\frac{x}{f\nu_0}\right) S\left(\frac{x}{f \nu_0}\right)\right)  \ast a^2(-x, -y),
			 \label{eq:specblur}
\end{align}
where
\begin{align*}
S(\lambda) = \int_{x_0} \int_{y_0} {H}(x_0, y_0, \lambda) dx_0 dy_0, \text{ and, }
\widetilde{c}(\lambda) = \frac{c(\lambda)}{\lambda^2 f^2}.
\end{align*}
We can observe from (\ref{eq:specblur}) that the image formed at the rainbow plane P4 is a convolution of the scene's spectrum $S(\lambda)$  modified with the camera response as well as $1/(\lambda f)^2$ with the square of the aperture code.

\begin{figure}[!tt]
	\centering
	\includegraphics[width=\columnwidth]{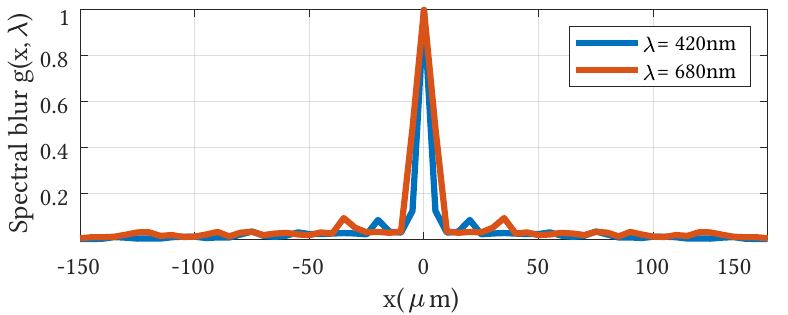}
	\caption{Simulation of spatial blur at $420$nm and $680$nm for the optimized code. Simulation was done with a lens of focal length $100$mm, and a camera pixel width of $5\mu m$. Our system was designed only for a narrow range of $260$nm, over which the blur was almost the same. Hence we assumed the spatial blur to be spectrally invariant.}
	\label{fig:psf_vs_lambda}
\end{figure}

Similarly, the intensity on plane P5 is given by,
\begin{align}
	M_5(x, y) &= \int\limits_{\lambda}|i_5(x, y, \lambda)|^2c(\lambda)d\lambda\nonumber\\
			  &=  \int\limits_{\lambda}  |h(x_0, y_0, \lambda )|^2 \left|\frac{1}{\lambda^2 f^2}A\left(-\frac{x-x_0}{\lambda f}, -\frac{y-y_0}{\lambda f}\right)\right|^2 c(\lambda ) d\lambda\nonumber\\
			  &=  \int\limits_{\lambda}  |h(x_0, y_0, \lambda )|^2 \frac{c(\lambda)}{\lambda^2 f^2}\left|\frac{1}{\lambda f}A\left(-\frac{x-x_0}{\lambda f}, -\frac{y-y_0}{\lambda f}\right)\right|^2  d\lambda\nonumber\\
			  &= \int\limits_{\lambda}  \widehat{H}(x_0, y_0, \lambda ) p(x-x_0, y - y_0, \lambda) d\lambda, 
\end{align}
where $p(x, y, \lambda) = \left|\frac{1}{\lambda f} A\left(-\frac{x}{\lambda f}, -\frac{y}{\lambda f}\right)\right|^2$ can be seen as the spatial PSF. 
Extending to a generic incoherent image case, we get the following expression,
\begin{align}
	M_5(x, y) &= \int\limits_{x_0} \int\limits_{y_0} \int\limits_{\lambda} \widehat{H}(x_0, y_0, \lambda ) p(x-x_0, y - y_0, \lambda) d\lambda\label{eq:spatial_meas} 
\end{align}
\begin{figure*}[!ttt]
	\centering
	\begin{subfigure}[t]{0.32\textwidth}
		\centering
		\includegraphics[width=0.48\textwidth]{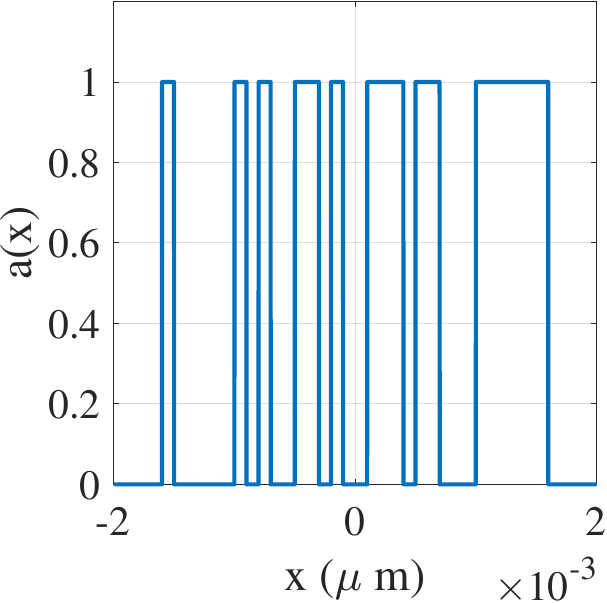}
		\includegraphics[width=0.35\textwidth]{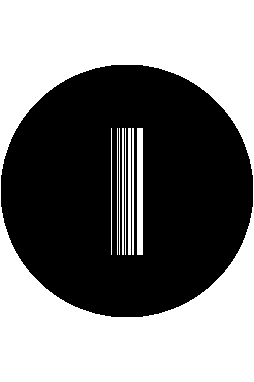}
		\caption{Optimized code and aperture for $N=32$.}
	\end{subfigure}
	\begin{subfigure}[t]{0.32\textwidth}
		\centering
		\includegraphics[width=0.45\textwidth]{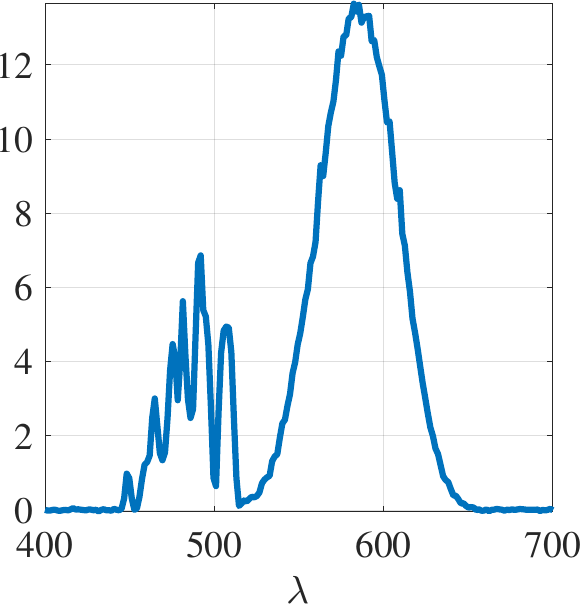}
		\quad
		\includegraphics[width=0.45\textwidth]{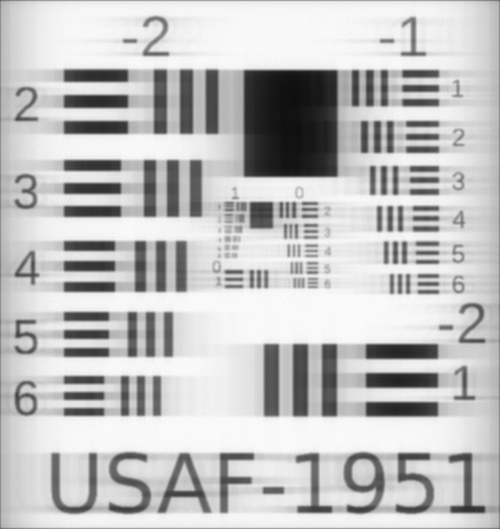}
		\caption{Raw measurements.}
	\end{subfigure}
	\begin{subfigure}[t]{0.32\textwidth}
		\centering
		\includegraphics[width=0.45\textwidth]{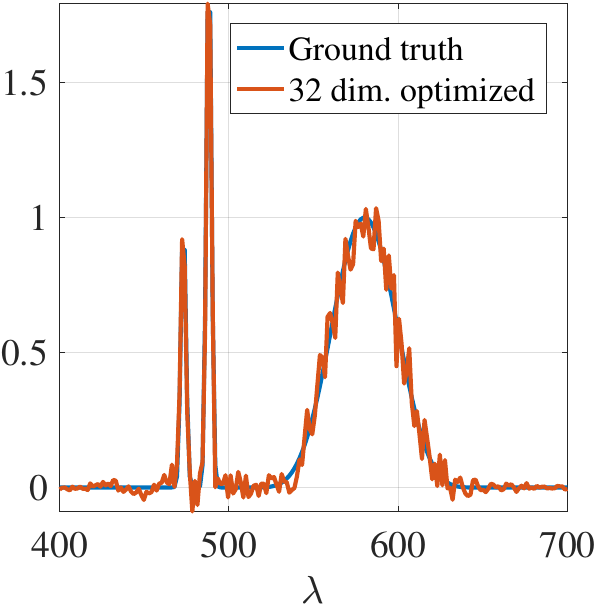}
		\quad
		\includegraphics[width=0.45\textwidth]{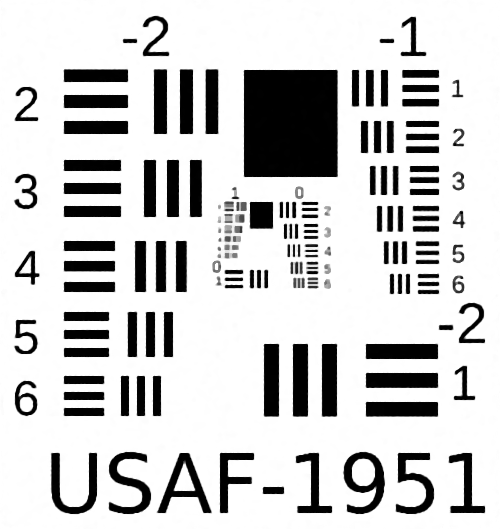}
		\caption{Deconvolved spectrum and image respectively\\(Spectrum PSNR: 31.9dB; Spatial PSNR: 27.2dB)}
	\end{subfigure}
	\\
	\begin{subfigure}[t]{0.32\textwidth}
		\centering
		\includegraphics[width=0.48\textwidth]{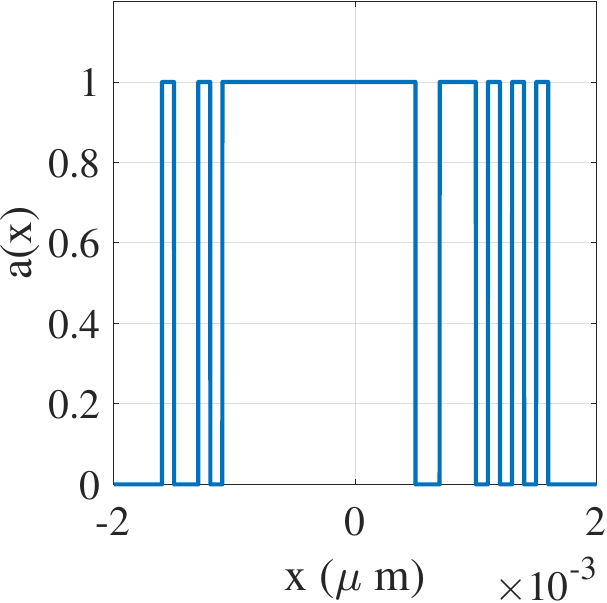}
		\includegraphics[width=0.35\textwidth]{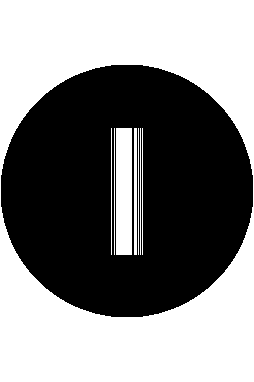}
		\caption{Spatially-compact code and aperture for $N=32$.}
	\end{subfigure}
	\begin{subfigure}[t]{0.32\textwidth}
		\centering
		\includegraphics[width=0.45\textwidth]{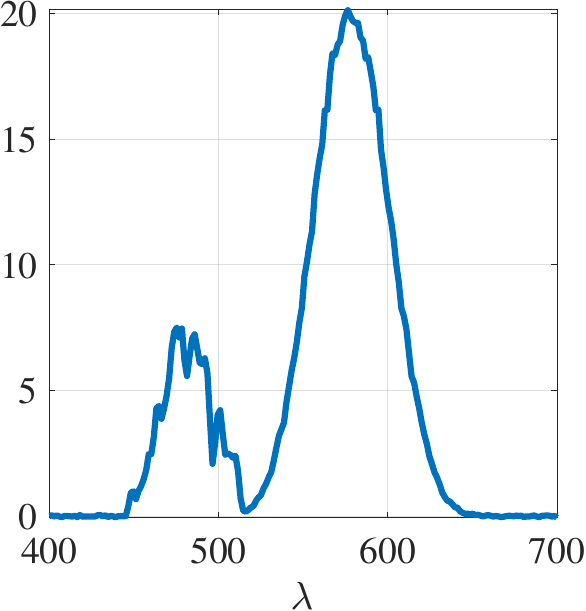}
		\quad
		\includegraphics[width=0.45\textwidth]{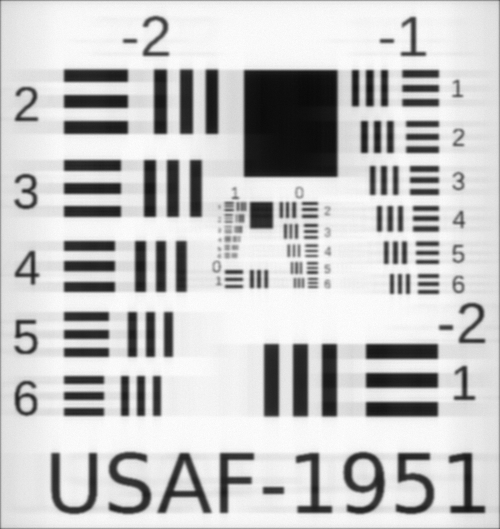}
		\caption{Raw measurements.}
	\end{subfigure}
	\begin{subfigure}[t]{0.32\textwidth}
		\centering
		\includegraphics[width=0.45\textwidth]{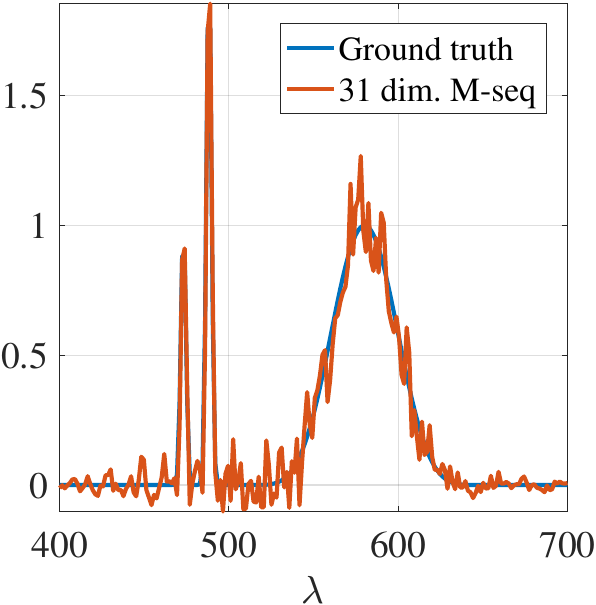}
		\quad
		\includegraphics[width=0.45\textwidth]{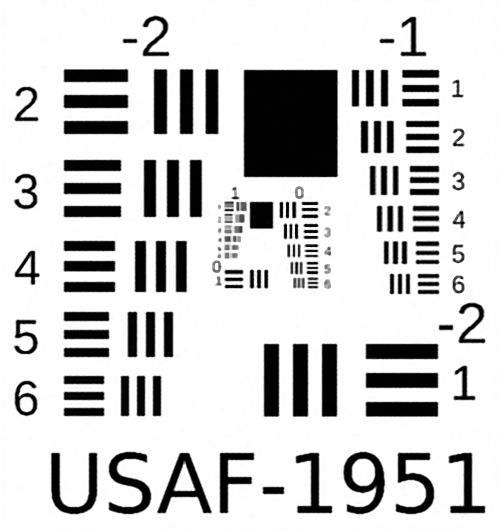}
		\caption{Deconvolved spectrum and image respectively\\(Spectrum PSNR: 29.0dB; Spatial PSNR: 26.4dB)}
	\end{subfigure}
	\\
	\begin{subfigure}[t]{0.32\textwidth}
		\centering
		\includegraphics[width=0.48\textwidth]{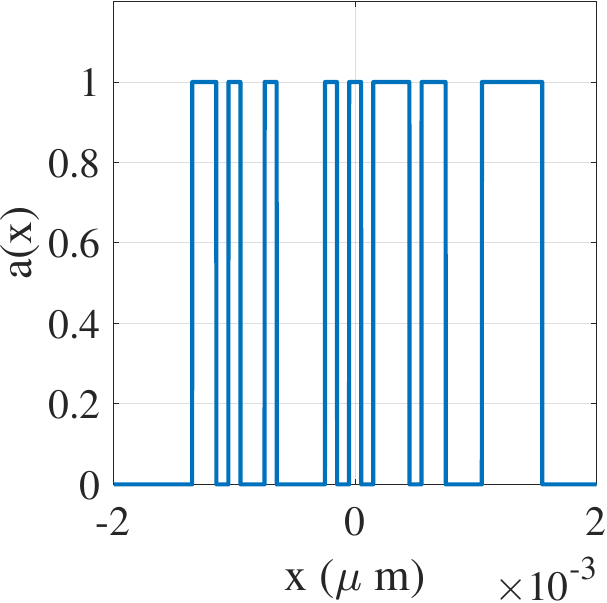}
		\includegraphics[width=0.35\textwidth]{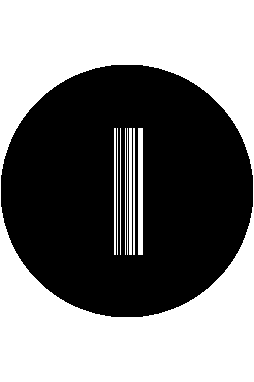}
		\caption{M-sequence code and aperture for $N=31$.}
	\end{subfigure}
	\begin{subfigure}[t]{0.32\textwidth}
		\centering
		\includegraphics[width=0.45\textwidth]{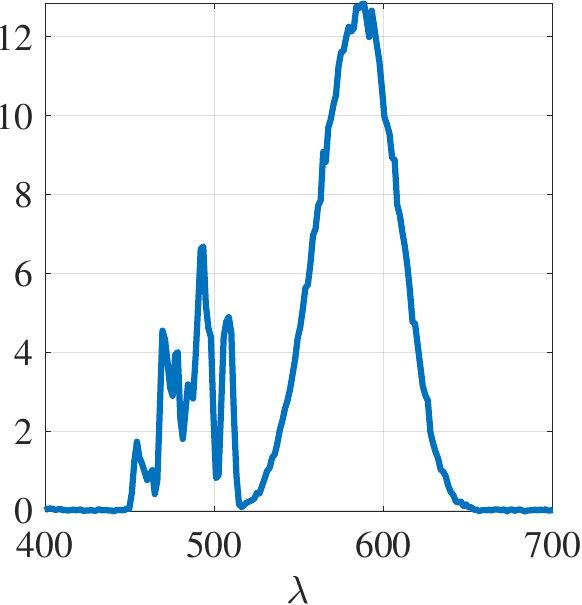}
		\quad
		\includegraphics[width=0.45\textwidth]{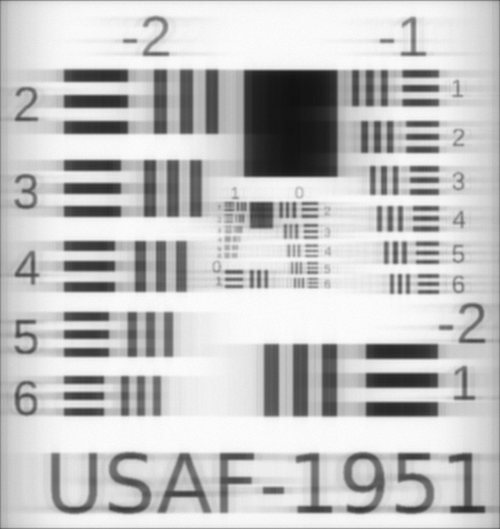}
		\caption{Raw measurements.}
	\end{subfigure}
	\begin{subfigure}[t]{0.32\textwidth}
		\centering
		\includegraphics[width=0.45\textwidth]{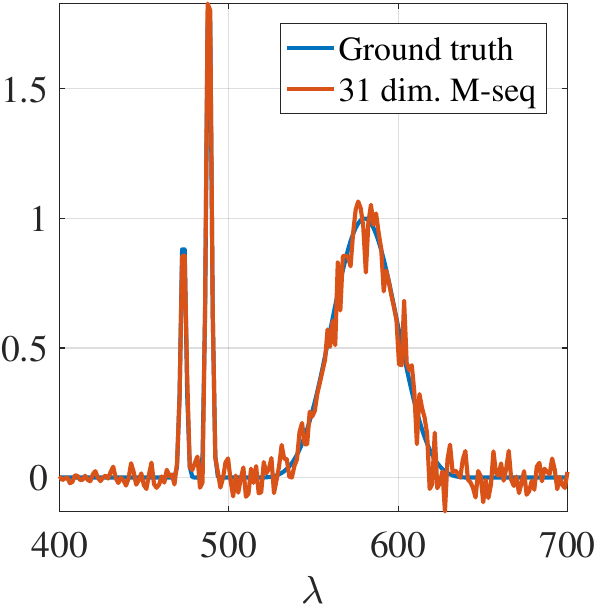}
		\quad
		\includegraphics[width=0.45\textwidth]{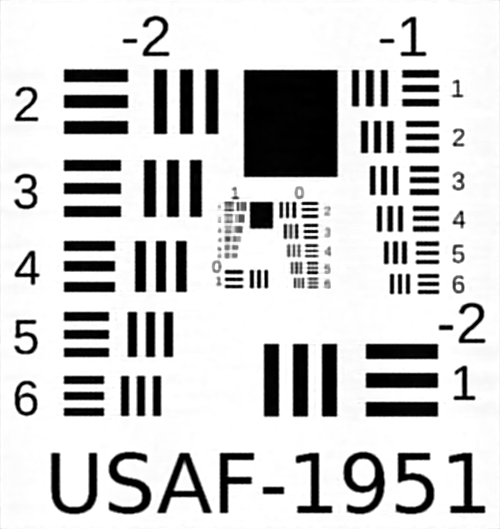}
		\caption{Deconvolved spectrum and image respectively\\(Spectrum PSNR: 29.4dB; Spatial PSNR: 26.4dB)}
	\end{subfigure}
	\caption{Comparison of performance of various codes. We compare optimized codes, spatially compact codes and M-sequences. Simulations were performed with added readout and poisson noise. Spectral deconvolution was done with Wiener deconvolution, while spatial deconvolution was done with TV-prior. While spatially-compact codes (row 2) offer better performance in spatial images, spectral deconvolution accuracy is very low. M-sequences (row 3) perform moderately well for both spectral and spatial deconvolution. However, optimized codes perform the best overall (row 1).}
	\label{fig:code_compare}
\end{figure*}
The expression above suggests that the image associated with each spectral channel is convolved with a different blur kernel; further, the blur kernel for different wavelengths are simple scaled versions of each other.
This implies that we need to design codes and deconvolve them for each channel separately.
However, this can be avoided for the following two reasons.
First, since the kernels are scaled versions of each other, if one of them is invertible then so are the rest.
Second, since our optical setup imaged within a narrow spectral band of $420-680$nm, the variance in spatial blur is not significant.
Since the blur of coded aperture is compact, the pixellation makes the differences in blur sizes insignificant.
Figure \ref{fig:psf_vs_lambda} shows a comparison of spatial blur of optimized code at two representative wavelengths of $420$nm and $680$nm as seen by a camera with $5 \mu m$ pixel width. 
As is evident, the blur size is largely invariant to wavelength, and hence we assumed that the spatial blur is spectrally invariant, giving us,
\[ M_5(x, y) \approx  \left[ \int_\lambda \widehat{H}(x_0, y_0, \lambda ) d\lambda \right] \ast p(x, y, \lambda_0). \]
For optimizing the spatial blur kernel, we chose a design wavelength of $\lambda = 500$nm.
However, if we were to image over a larger span of wavelengths, such as $300-1100$nm, spectral bands have to be deconvolved individually.

	\section{Code selection} \label{section:sup_deconv}
	
We now discuss alternate designs for the coded aperture and the specific algorithm we used for deconvolving the spatial and spectral measurements.

\subsection{Choice of codes}
We discussed a way of obtaining optimized codes that promote invertibility in both spectrum and space in section 3.1 of the main paper. 
In this section, we show other choices for coded apertures and compare their performance.
\subsubsection{Spatially compact codes}
Instead of pursuing spatially invertible codes, we can optimize for codes which introduce compact spatial blur.
In such a case, the goal would be to suppress side lobes of the PSF of spatial PSF.
Let $P_a(x)$ be the spatial PSF created by $a(x)$. If $\eta_1, \eta_2$ be the first and second maximum peak heights of $P_a(x)$, then maximizing the ratio $\eta_2/\eta_1$ leads to spatially imperceptible blur.
Combined with an invertible spectral blur, we formulate the overall objective function as:
\begin{equation}
\max_{a_1, \ldots, a_N} \alpha \min_k \left(| {\bf A}[k] |\right) + (1-\alpha )\left(\frac{\eta_2}{\eta_1}\right), 
\label{eq:obj_imperceptible}
\end{equation}
where $\alpha \in (0, 1)$ is a constant. As with optimized codes, we brute forced the optimal solution.
\subsubsection{M-sequences}
Maximal length sequences, or M-sequences for short, are optimal codes when using circular convolution. Their PSD is flat and hence is desirable as blur functions.
However, since our convolution is linear, M-sequences are not necessarily the optimal choice.
\begin{figure}[!ttt]
	\centering
	\begin{subfigure}[t]{\columnwidth}
		\centering
		\includegraphics[width=\textwidth]{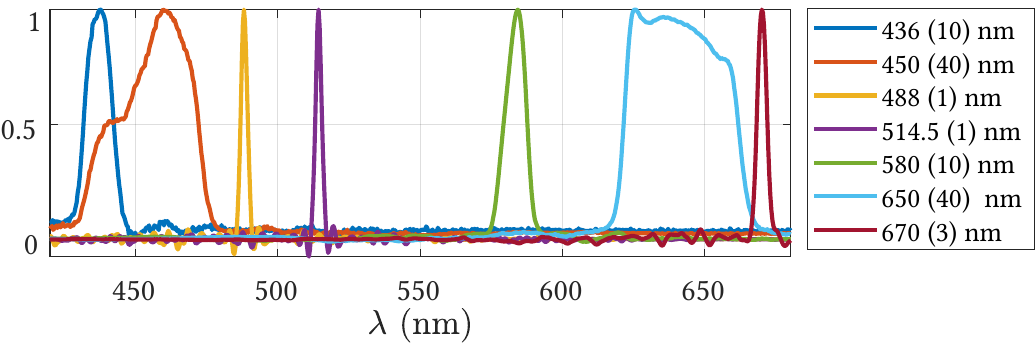}
		\caption{Spectra of narrowband filters.}
	\end{subfigure}
	\begin{subfigure}[t]{\columnwidth}
		\centering
		\includegraphics[width=\textwidth]{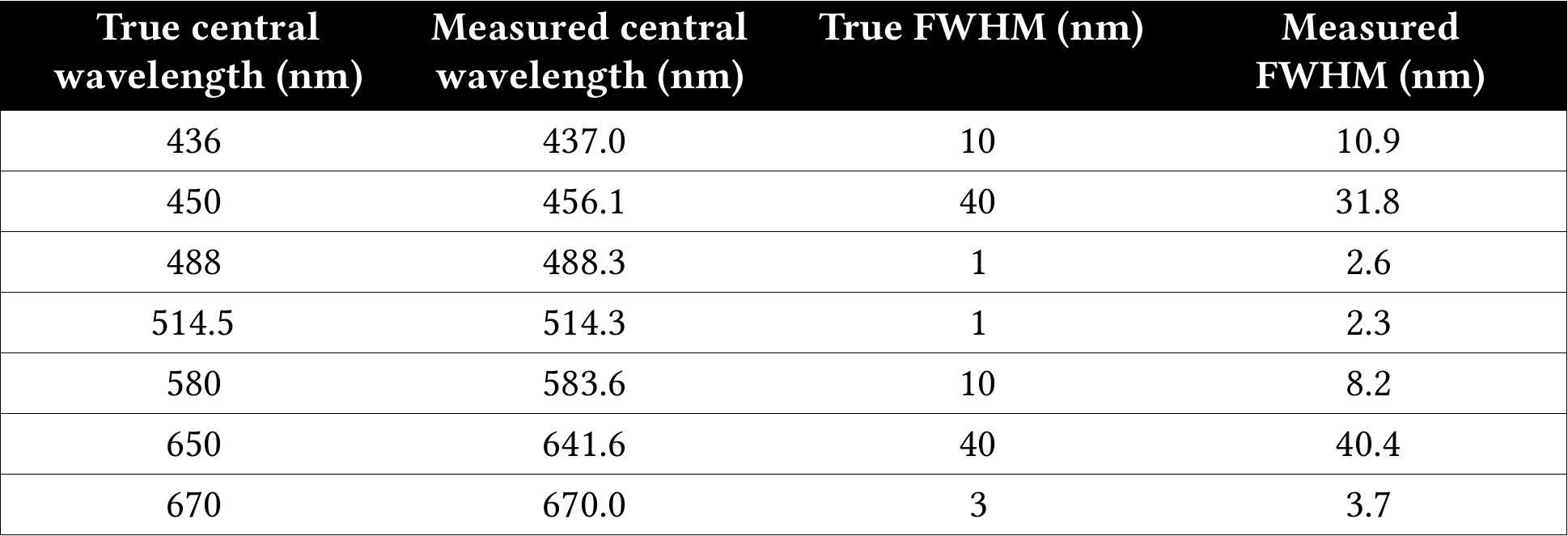}
		\caption{Central wavelength and FWHM of the filters}
	\end{subfigure}
	\caption{Spectra of some narrowband filters with datasheet central wavelength and FWHM (in parenthesis) provided in legend. We used Weiner deconvolution to obtain the true spectra. (b) tabulates the estimated central wavelength and FWHM, along with datasheet values. The accuracy of central wavelength and FWHM establishes the accuracy as well as high-resolution capabilities of our optical setup.}
	\label{fig:spec_narrowband}
\end{figure}
\begin{figure}[!ttt]
	\centering
	\begin{subfigure}[t]{\columnwidth}
		\centering
		\includegraphics[width=\textwidth]{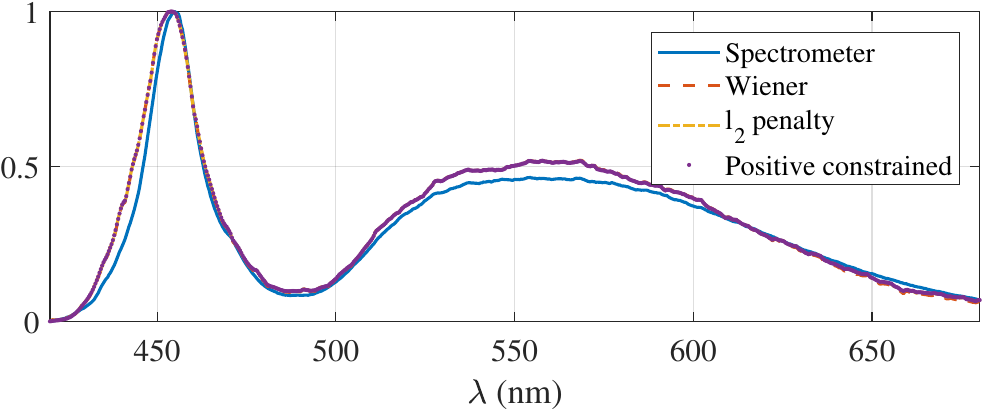}
		\caption{White Light Emitting Diode light source}
	\end{subfigure}
	\begin{subfigure}[t]{\columnwidth}
		\centering
		\includegraphics[width=\textwidth]{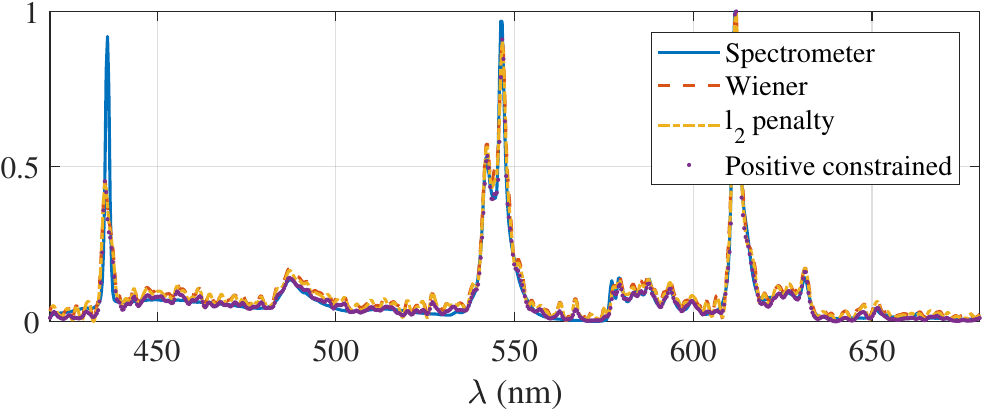}
		\caption{White Compact Fluorescent Lamp}
	\end{subfigure}
	\begin{subfigure}[t]{\columnwidth}
		\centering
		\includegraphics[width=\textwidth]{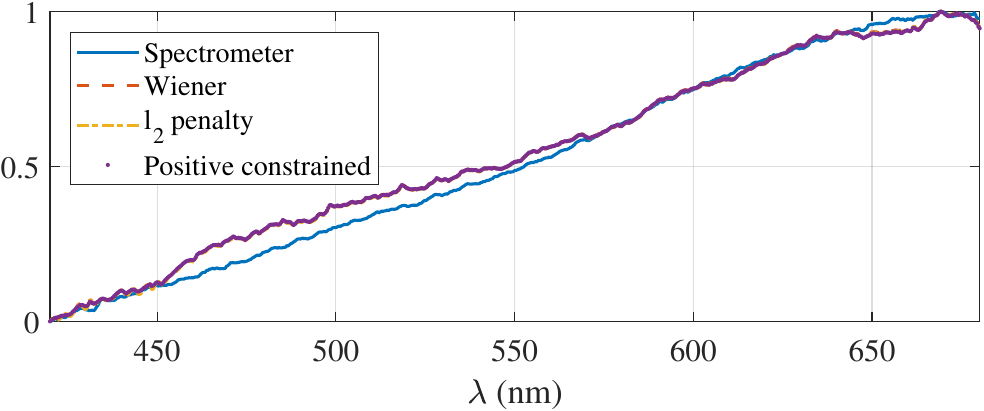}
		\caption{Tungsten-halogen light source..}
	\end{subfigure}
	\caption{Spectra of commonly found light sources: (a) Tungsten-halogen, (b) Light Emitting Diode (LED), and (c) Compact Fluorescent Lamp (CFL). Results are shown for Wiener deconvolution, $\ell_2$ penalized deconvolution and positive-constrained deconvolution. CFL shows some error in blue wavelengths, as the machine vision camera we used was not reliable for deep blue wavelengths. The deconvolved results are robust to choice of algorithm, as the aperture code was designed to be invertible.}
	\label{fig:spec_deconv}
\end{figure}
\begin{figure}[!ttt]
	\centering
	\begin{subfigure}[t]{0.47\columnwidth}
		\centering
		\includegraphics[width=\textwidth]{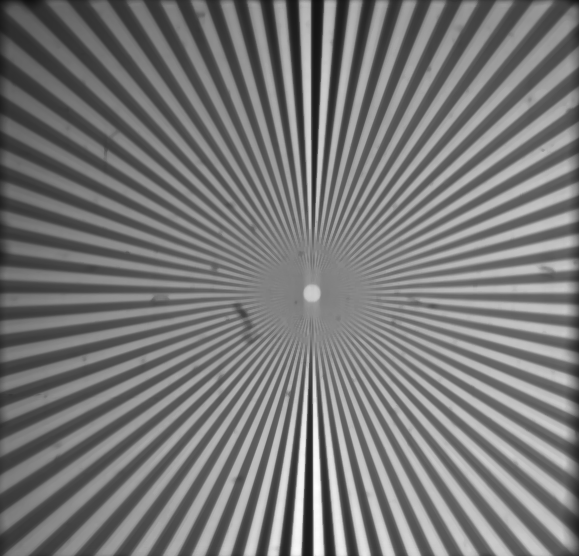}
		\caption{Before deconvolution}
	\end{subfigure}
	\quad
	\begin{subfigure}[t]{0.47\columnwidth}
		\centering
		\includegraphics[width=\textwidth]{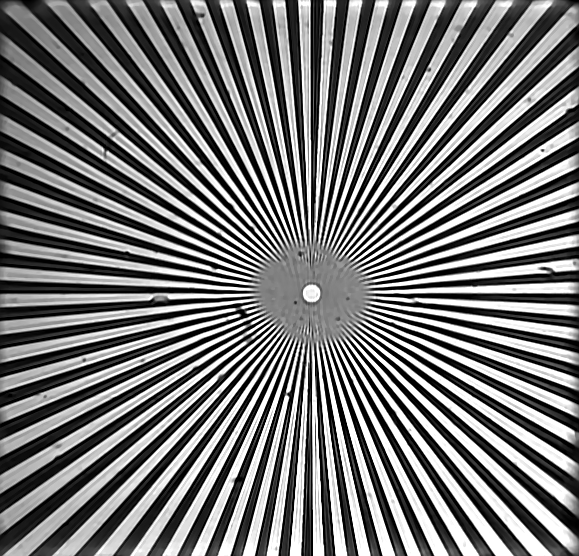}
		\caption{TV-prior deconvolution}
	\end{subfigure}
	\quad
	\begin{subfigure}[t]{0.47\columnwidth}
		\centering
		\includegraphics[width=\textwidth]{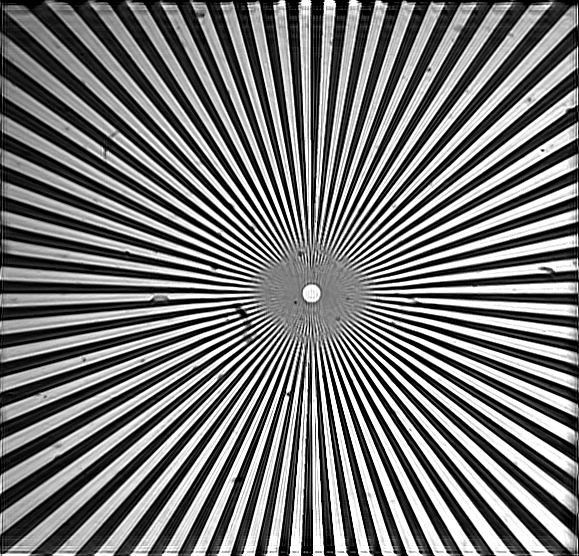}
		\caption{Wiener deconvolution}
	\end{subfigure}
	\quad
	\begin{subfigure}[t]{0.47\columnwidth}
		\centering
		\includegraphics[width=\textwidth]{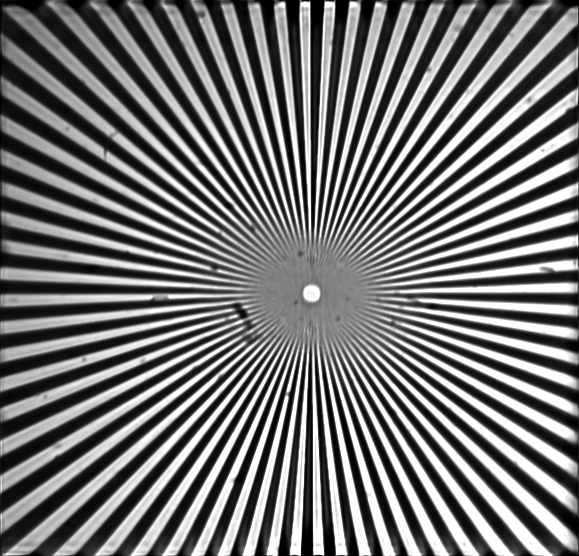}
		\caption{Richardson-Lucy deconvolution}
	\end{subfigure}
	\caption{Deconvolution results of a Siemen star with various deconvolution algorithms. Due to invertible nature of PSF,  deconvolution is robust to choice of method. However, TV-prior gave best results in terms of higher contrast ratio.}
	\label{fig:spatial_deconv}
\end{figure}
\subsection{Performance comparison}
We compare optimized codes, spatially compact codes and M-sequ- ences for their performance in spatial deconvolution and spectral deconvolution.
To test spectral deconvolution, we created a spectrum with two closely spaced narrowband peaks and a broadband peak, and blurred them with various codes.
Readout noise and shot noise were added to adhere to real world measurements.
Finally, deconvolution was done with wiener filter.
To test spatial deconvolution, we used Airforce target and blurred with the scaled PSD of the pupil codes, and added noise.
Deconvolution was done with a TV prior in all cases.

Figure \ref{fig:code_compare} shows a comparison of performance for spectral and spatial deconvolution.
As expected, optimized codes perform the best for spectral deconvolution, while spatially compact codes perform worse.
Spatially compact codes perform the best in this case, while optimized codes come close.
Since hyperspectral imaging requires good spatial as well as spectral resolution, we chose optimized codes.
\subsection{Spectral deconvolution}
Recall that our optical setup measures a blurred version of the true spectrum.
Specifically, if the aperture code is $a(x)$ and the spectrum to be measured is $s(\lambda)$, our optical setup measures $y(\lambda) = a(\lambda)\ast s(\lambda) + n(\lambda)$, where $n(\lambda)$ is additive white gaussian noise.
The addition of noise prevents us from simply dividing in Fourier domain.
Fortunately, since the aperture code was designed to be invertible, it is fairly robust to noise.
A naive solution, such as Wiener deconvolution, hence, works very well. 
If the noise is too high, or the spectra is known to be smooth, we can impose an $\ell_2$ penalty on the difference and solve the following optimization problem:
\begin{align}
	\min_{\bfs} \quad \frac{1}{2}\|\bfy - \bfa \ast \bfs\|^2 + \eta \| \nabla \bfs\|^2,
\end{align}
where $\bfy$ is the measured spectrum, $\bfa$ is the aperture code, $\bfs$ is the spectrum to be recovered, and $\nabla \bfs$ is the first difference of $\bfs$.
Further priors, such as positivity constrains give better results as well.
Figure \ref{fig:spec_deconv} shows a comparison of spectra of various commonly available light sources, as well as a comparison with spectrometric measurements.
We showed results for three forms of deconvolution, namely, Wiener deconvolution, $\ell_2$ regularized deconvolution, and positivity constrained deconvolution.
Figure \ref{fig:spec_narrowband} shows results for some narrowband filters. We computed the central wavelength and Full Width Half Max (FWHM) for each filter and compared it against the numbers provided by the company.
As expected, the FWHM of 1nm filters is between 2nm and 3nm, as the FWHM of our optical setup is 3nm.
FWHM for 10nm filters and 40nm filters is close to the ground truth values. 
\subsection{Spatial deconvolution}
\begin{figure}[!tt]
	\centering
	\includegraphics[width=\columnwidth]{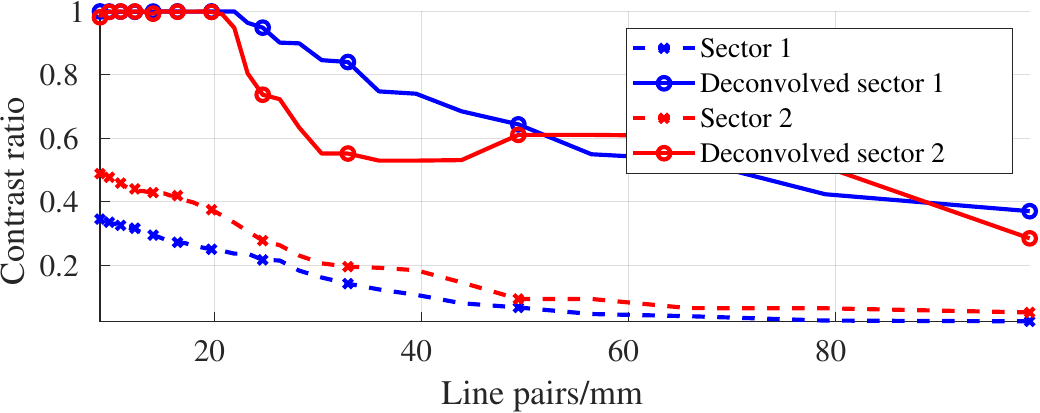}
	\caption{MTF plot before and after deconvolution. PSF was estimated by capturing image of a $10\mu m$ pinhole. Deconvolution was then done using a TV prior on the image gradients. There is a marked improvement in contrast ratio after deconvolution. The MTF30 value for both sectors jumps from 20 line pairs/mm to 90 line pairs/mm.}
	\label{fig:mtf}
\end{figure}

The presence of a coded aperture introduces a blur in spatial domain, which is the scaled power spectral density of the coded aperture.
Our optimization procedure accounts for invertible spectral blur as well as invertible spatial blur.
Hence, deconvolution is stable even in the presence of noise.
While naive deconvolution procedures such as Wiener deconvolution or Richardson-Lucy work well, we imposed total variance (TV) penalty on the edges to get more accurate results.
Figure \ref{fig:spatial_deconv} shows blurred image of Siemen star, and deconvolution with TV-prior, Wiener and Richardson-Lucy algorithms.
As with spectral deconvolution, spatial deconvolution is fairly robust to choice of algorithm.
We chose TV-prior, as it returned the sharpest results.
Figure \ref{fig:mtf} shows MTF before and after deconvolution, with TV-prior.
Deconvolution signficantly improves the MTF30 value, which jumps from 20 line pairs/mm to 90 line pairs/mm. 

	\section{List of components} \label{section:sup_components}
\begin{figure*}[!ttt]
	\centering
	\includegraphics[width=\textwidth]{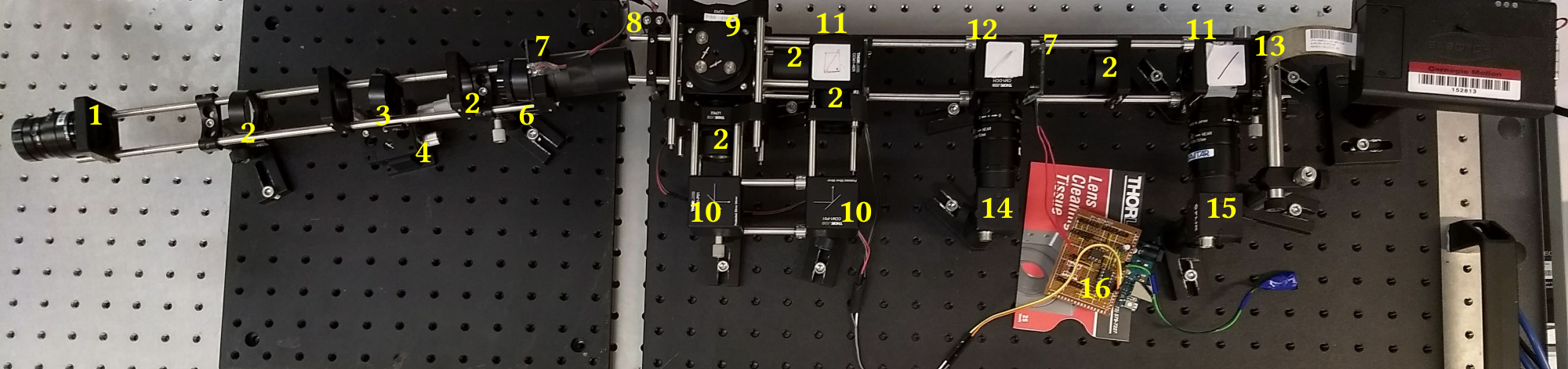}
	\includegraphics[width=\textwidth]{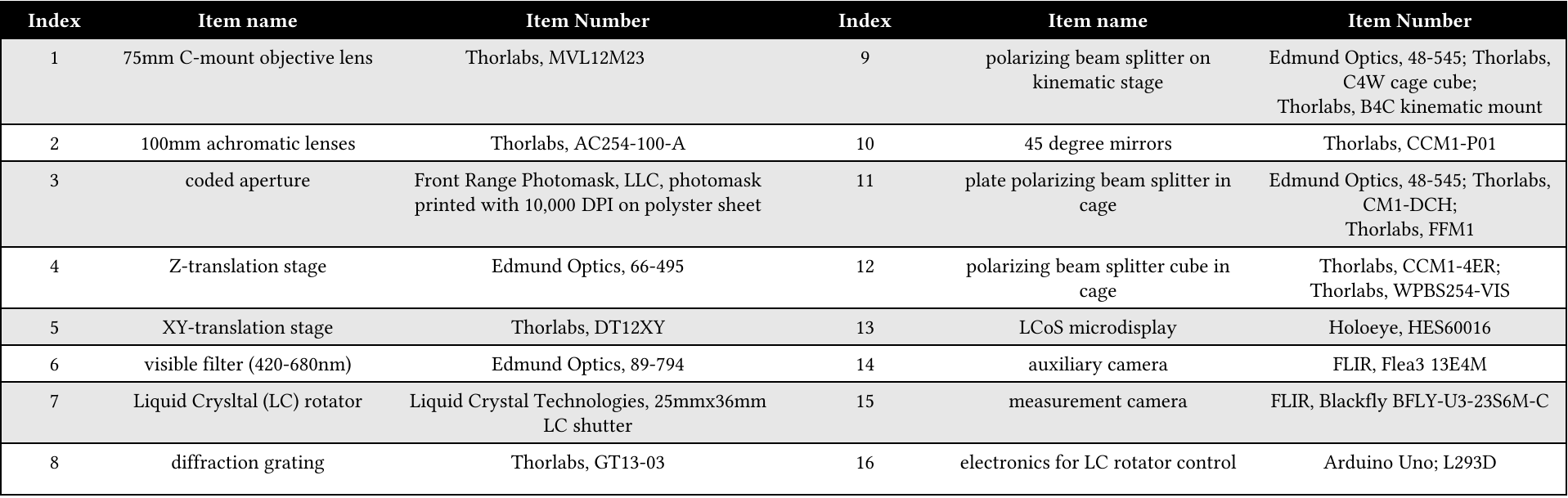}
	\caption{List of components for the KRISM optical setup. All the important components, their company and item number have been listed for reference. Construction component names such as cage plates, rods, posts and breadboards have been omitted for brevity.}
	\label{fig:components}
\end{figure*}

	Figure \ref{fig:components} shows an annotated image of the optical setup we built along with a list of components along with their company and item number.
The system was optimized for a central wavelength of 580nm and hence
the relay arm till the diffraction grating has been tilted at $10^\circ$ with respect to the diffraction grating to correct for schiempflug.
Lenses in the relay arm are tilted by $5^\circ$ with respect to the diffraction grating so that the objective can be aligned with the relay arm without any further tilt.
The first beamsplitter (component 8) and the second turning mirror (component 10) have been placed on a kinematic platform to correct for misalignments in the cage system.
It is of importance that we chose an LCoS instead of a DMD for spatial light modulation. The reasons:
\begin{itemize}[leftmargin=*]
\item Since the output after modulation by DMD is not rectilinear to the DMD plane, it introduces further scheimpflug, which is hard to correct.
\item DMD acts as a diffraction grating with Littrow configuration, as it is formed of extremely small mirror facets. This will introduce artifacts in measurements which are non-linear.
\end{itemize}

Some more design considerations are enumerated below:
\begin{enumerate}[leftmargin=*]
	\item \textit{Lenses.} We used 100mm achromats for all lenses except the last lens before cameras. Achromats were the most compact and economical choice for our optical setup, while offering low spatial and spectral distortion.
	\item \textit{Polarizing beam splitters.} We used wire grid polarizing beamsplitters everywhere to ensure low dependence of spectral distortion on angle of incidence, and increase the contrast ratio.
	\item \textit{Using an objective lens for measurement camera.} Note that a lens is placed between the LCoS and measurement sensor which converts spatially-coded image to spectrum and coded spectrum to spatial image. Instead of using another achromat, we used an objective lens set to focus at infinity. Since objective lenses are free of any distortions, and are optimized to focus at infinity, this significantly improves resolution of measurements.
	\item \textit{Diffraction grating.} We used an off-the-shelf transmissive diffraction grating with 300 groves/mm, which offered most compact spectral dispersion without any overlap with higher orders. This ensured that there would be no spectral vignetting at any point in the setup. Further discussion about the choice of grove density is provided in section \ref{section:sup_design}
	\item \textit{Polarization rotators.} We bought off-the-shelf Liquid Crystal (LC) shutters and peeled off the polarizers on either sides to construct polarization rotators. This is the most economic option, while offering contrast ratios as high as 400:1. The key drawback is that the settling time is 330ms, which prevents their usage at very high rate. A natural workaround is to incorporate binary Ferroelectric shutters which have a low latency rate of 1ms. However, since ours was only a lab prototype, we decided to go with the cheaper option.
\end{enumerate}

	\section{Design considerations} \label{section:sup_design}
	We outline some design choices we made and the rationale behind them in this section.
\begin{figure}[!tt]
	\centering
	\includegraphics[width=\columnwidth]{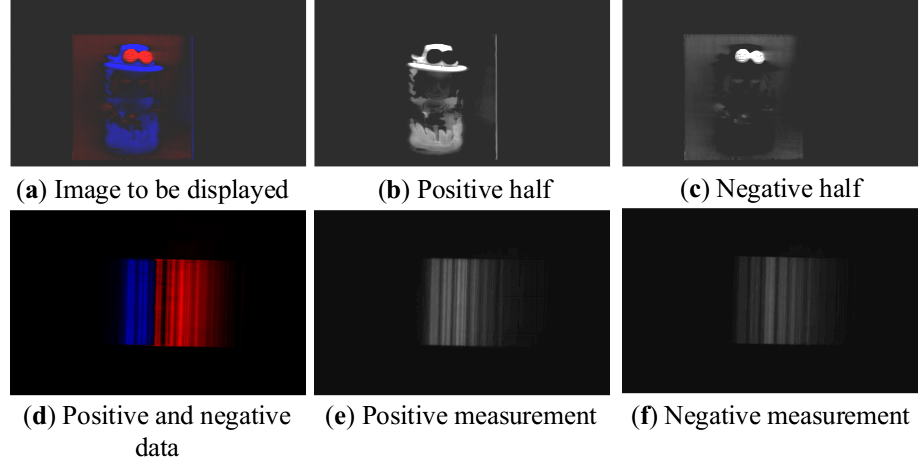}
	\caption{Our optical setup requires measurement of positive and negative data. Since measurements are linear, we split the image to be displayed (\textbf{a}) into positive part in (\textbf{b}), with a maximum value of 0.0026 and negative part in (\textbf{c}) part, with a maximum value of 0.0029. By capturing positive (\textbf{e}) and negative data (\textbf{f}), $\bfy_p$ and $\bfy_n$ respectively, the required measurement is evaluated as $\bfy = 0.0026\bfy_p - 0.0029\bfy_n$.}
	\label{fig:posneg}
\end{figure}
\subsection{Choice of code size}
The pupil code has two free parameters, the length of the code $N$ and the pitch size $\Delta$. 
The two parameters control the invertibility of spectrum and imperceptibility of spatial images.
To understand our design choices, we present constrains and physical dimensions of various measurements.
Let each lens in the optical setup have a focal length $f$ and aperture diameter $a_L$. 
Let pixel pitch of measurement camera be $p$. This implies that the camera can capture all spatial frequencies up to $f_{max} = \frac{1}{2p} m^{-1}$.
Let the size of grating be $a_g$ in each dimension and its grove density be $g\, groves/mm$.

We capture wavelengths from $\lambda_1 = 420 nm$ to $\lambda_2 = 680 nm$. The grating equation is given by, $a \sin(\theta) = m\lambda$, where $a$ is the groves spacing and $m$ is the order of diffraction, 1 in our case. Solving for angular spread of spectrum, we get, $\Delta \theta = \sin^{-1}\left(\frac{\lambda_2}{a}\right)-\sin^{-1}\left(\frac{\lambda_1}{a}\right)$
The size of spectrum then is $f\tan(\Delta \theta) + N\Delta$.
The minimum resolvable wavelength is $\Delta \lambda \approx \frac{\Delta}{af}$. 
To avoid vignetting in a $4F$ system, we require that the pupil plane be no larger than $a_L - a_g mm$, giving us $f\tan(\Delta \theta) + N\Delta \le a_L - a_g$.

Recall that the pupil code is $a(x) = b(x) * \sum_{k=0}^{k=N-1}a[k]\delta(x-k\Delta)$, where $a[k]$ is the binary pupil code and $b(x) = 1 \quad -\Delta/2 \le x \le \Delta/2$. 
Using the formula for PSF of an incoherent system, we know that the Fourier transform of the PSF is $F_{PSF} = C_a (\lambda f u )$, where $C_a(x)$ is the linear autocorrelation of $a(x)$ and $u$ is spatial frequency in $1/m$.
To capture all spatial frequencies, we need $F_{PSF}(u)$ to be non-zero for $u \ge f_{max}$, which gives us $N\Delta \ge \lambda f \frac{1}{2 p}$.

In our optical setup, we have $a_L = 25 mm, a_g = 12.5mm, p = 5 \mu m$, $f = 100 mm$, $\lambda = 500 nm$, and $\Delta = 100 \mu m$, which leaves us $N$, and $g$ as free variables. 
To prevent vignetting, we need $N\Delta$ to be less than $a_L - a_g - f\tan(\Delta \theta)$, which means that N increases as $g$ decreases.
Increasing $N$ increases resolution of images, but the optimization problem for optimal binary code becomes lengthy. 
On the other hand, increasing $\Delta$ can increase spatial resolution, but the spectral resolution reduces. 
Keeping practical considerations in mind, we set $N = 32$, which took close to a day to optimize.
Further, $g = 300 groves/mm$ was the smallest grove density we obtained as off-the-shelf component.
\vspace{-1em}
\subsection{Handling positive/negative data} 
When computing singular vectors, the data to be measured, as well as the data to be displayed on the LCoS contains negative values.
Since our optical devices cannot handle negative data, we make two positive measurements and combine them.
We split the data to be displayed on the LCoS into positive and negative parts.
Then, we capture positive data with positive part on the LCoS, and then repeat the process for negative data.
By taking the difference of the positive and negative data, we obtain the required measurement.
Figure \ref{fig:posneg} shows an example of capture of data with positive/negative data. 
The data in (\textbf{a}) shows the positive/negative image to  be displayed on the LCoS, which is split into positive (\textbf{b}) and negative (\textbf{c}) halves, which are separately displayed on the LCoS, to capture positive (\textbf{e}) and negative (\textbf{f}) data. 
The final required measurement is then obtained by appropriately weighing and subtracting the two measurements.

	\section{Calibration} \label{section:sup_calibration}
	\begin{figure}[!ttt]
	\centering
	\includegraphics[width=\columnwidth]{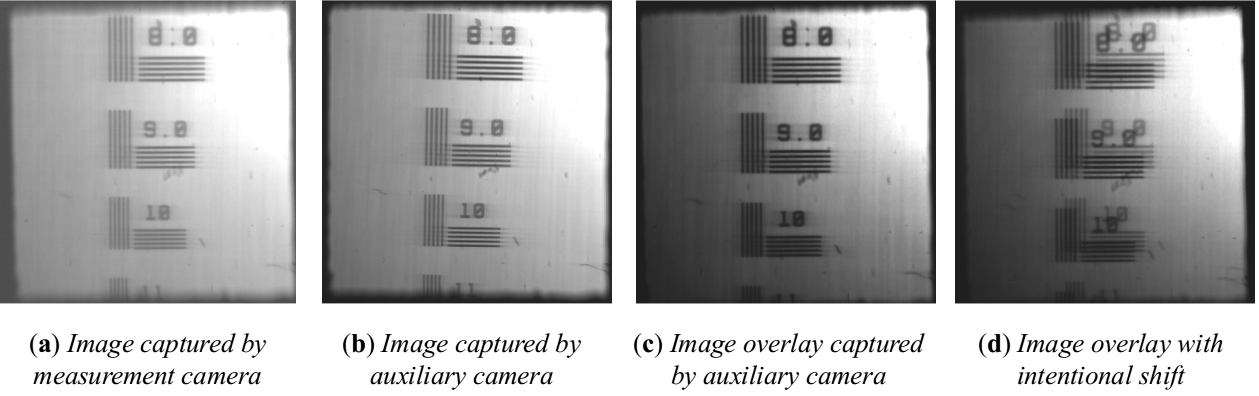}
	\quad
	\includegraphics[width=\columnwidth]{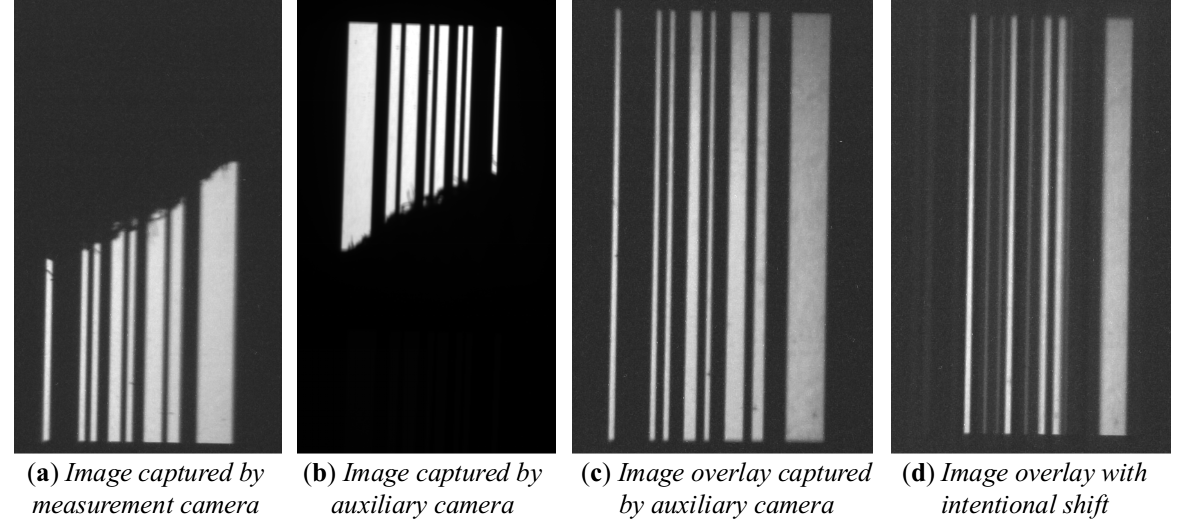}
	\caption{Images showing spatial and spectral calibration accuracy obtained by placing a target image in front of the camera. Spatial registration is done by capturing image of a known target placed in front of the optical setup by the measurement camera (\textbf{a}) and auxiliary camera (\textbf{b}). We then register the two images to obtain a similarity transformation that maps images from the measurement camera to the LCoS. To verify our registration, we keep the target in front of the setup and display the image captured by the measurement camera on LCoS after mapping. (\textbf{c}) shows the image then captured by the auxiliary camera. The images overlap well, implying that the registration process was successful. (\textbf{d}) shows an intentional shift induced in the measured image and displayed back on the LCoS. There is a visible shift in the target image. A similar process is followed for the spectral measurement registration as well. Instead of a target, we place a narrowband filter in front of the optical setup and illuminate it with a broadband light source. Since the pupil code image is vertically symmetric, there will be a 180 degrees ambiguity. We get rid of that by sticking a tape at the bottom and capturing images, shown in (\textbf{e}) and (\textbf{f}). A successful registration process results in (\textbf{g}), with image on LCoS very well overlapping with the mapped measurement image. (\textbf{h}) shows overlay with an intentional shift, resulting in an image that does not look like the pupil code.}
	\label{fig:calib_test}
\end{figure}
We now outline calibration steps for the proposed optical setup.
Firstly, we need a mapping between the captured image and the image displayed on the SLM.
Secondly, we need calibration of wavelengths, and finally, we need spectral response calibration of the system for high-fidelity measurements.
\paragraph{Camera-SLM calibration.}
Recall that the power method for estimating eigen vectors requires the multiplication $\bfx_2 = H \bfx_1$, where $\bfx_1 = H^\top \bfx_0$ is a spatial measurement, displayed on the SLM and $\bfx_2$ is the measurement made by the camera.
Hence, we need a one-to-one mapping between the measured image and the LCoS.
To do this, we added a second, calibration camera, henceforth called the auxiliary camera, which directly sees the image on the LCoS. The calibration steps are:
\begin{enumerate}[leftmargin=*]
	\item Find pixel to pixel correspondence between LCoS and auxiliary camera using gray or binary codes.
	\item Place known target in front of the camera.
	\item Capture the image of the target using the primary camera. Let this image be $I_1$.
	\item Capture the image of the target on the LCoS using second camera. Let this image be $I_2$.
	\item Register $I_1$ and $I_2$ using a similarity transform.
\end{enumerate}

The steps are then repeated for the spectrum. 
Instead of placing a known target image, a narrow band filter is placed.
This creates the coded aperture pattern on both the cameras. 
The image of the coded aperture for the narrow band filters can be used for registering the cameras for spectral measurements.
For robustness, we combined images of two narrow band filters, namely 514.5nm with an FWHM of 1nm and 670nm with an FWHM of 3nm, which helped registration of the camera and LCoS over a larger field of view.

Figure \ref{fig:calib_test} shows spatial and spectral calibration results. (\textbf{a}) shows the images of target captured by auxiliary camera and (\textbf{b}) shows capture by measurement camera. 
The calibration process was verified by displaying the captured target image back on the SLM and then capturing the image of LCoS by auxiliary camera. The result is shown in (\textbf{c}).
(\textbf{d}) shows the result if the registration were not successful, showing ghosting of the two images.
(\textbf{e}) and (\textbf{f}) show image of spectrum of a narrowband filter.
Since the pupil code is vertically symmetric, we stuck a piece of tape at the bottom, creating a trapezoidal shape, which was then easy to register. (\textbf{g}) shows the overlay image captured by the auxiliary camera, for verification.
A good registration results in an image that looks like the aperture code itself.
(\textbf{h}) shows the result of an intentional shift, to show the effect of a bad registration.
In both cases, we used Matlab's built in SURF based automatic image registration technique for estimating a similarity transform between the two captured images.

\begin{figure}[!ttt]
\centering
\begin{subfigure}[t]{0.45\columnwidth}
\centering
\includegraphics[width=\textwidth]{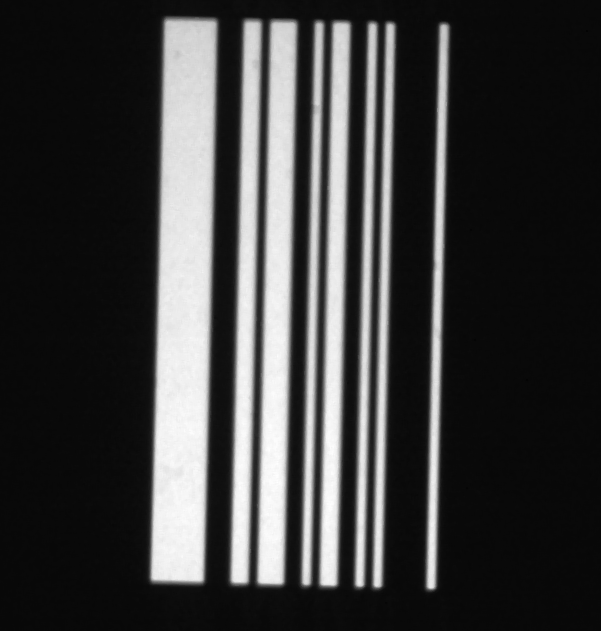}
\vspace{0.2em}
\caption{Captured image}
\end{subfigure}
\quad
\begin{subfigure}[t]{0.45\columnwidth}
\centering
\includegraphics[width=\textwidth]{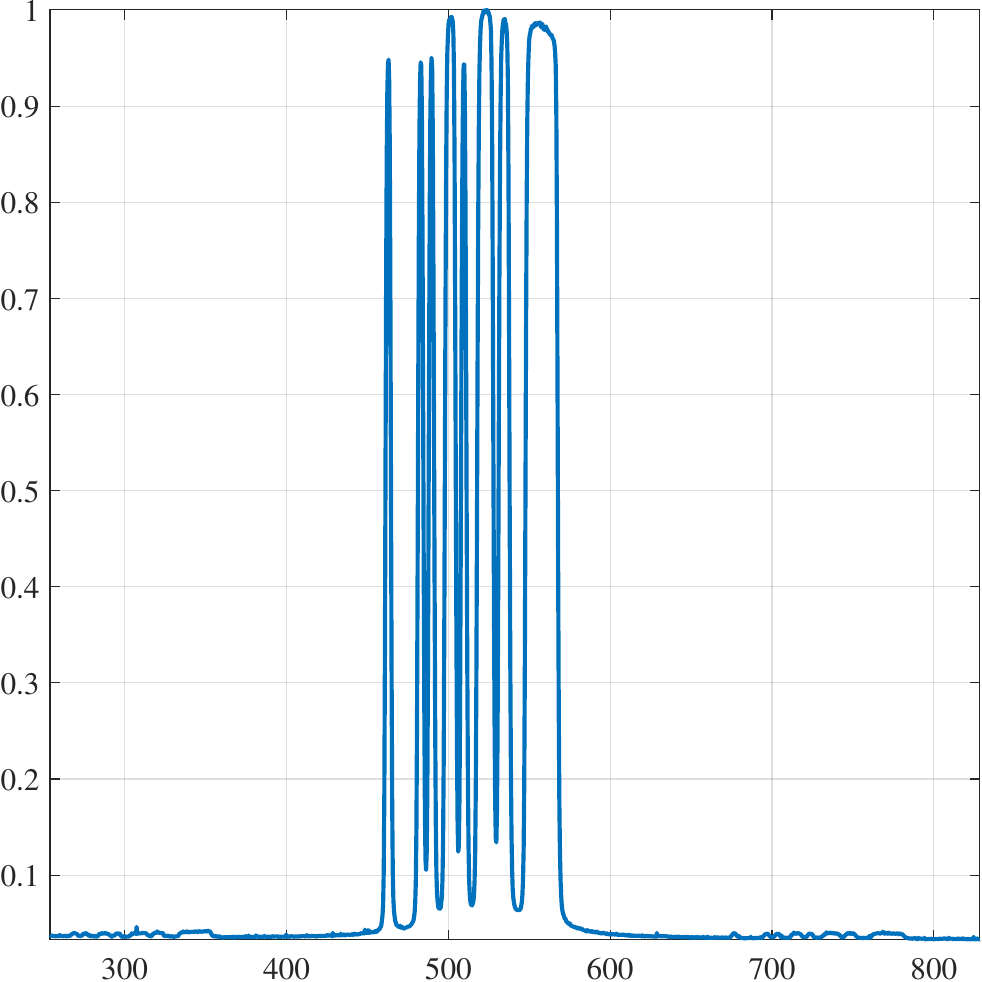}
\caption{Spectrum computed from image}
\end{subfigure}
\quad
\begin{subfigure}[t]{\columnwidth}
\centering
\includegraphics[width=\textwidth]{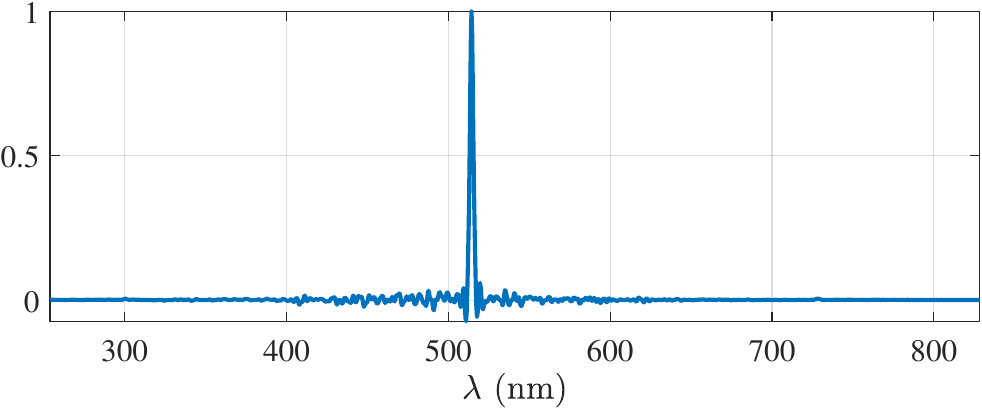}
\caption{Deconvolved spectrum}
\end{subfigure}
\caption{Calibrating of code and wavelengths location. We start with image of spectrum of a narrow band filter (a), 514.4nm in this case. Then the image is corrected for rotation and summed vertically to obtain the spectrum (b). The spectrum is thresholded to get the binary code which is then used to deconvolve the observed spectrum to obtain the spectrum of the narrow band filter in (c). We used Wiener deconvolution for obtaining the sharp spectrum.}
\label{fig:spectrum_deconv}
\end{figure}

\paragraph{Wavelength calibration.} Wavelength calibration requires two steps -- 1) Estimating the binary code of the coded aperture and 2) Estimating locations of wavelengths.
We found thresholding the measured spectrum to be a robust way of estimating the binary code of the coded aperture.
To calibrate wavelength locations, we use three filters of known spectral response. 
Specifically, we use 488nm, 514.5nm and 670nm spectral filters with FWHM of 1nm, 1nm and 3nm respectively.
Since spectral spread is linear, two known wavelengths are sufficient.
However, for robustness, we use a third filter and then linearly interpolate to get the wavelength positions.

\begin{figure}[!ttt]
	\centering
	\begin{subfigure}[t]{0.3\columnwidth}
		\centering
		\includegraphics[width=\textwidth]{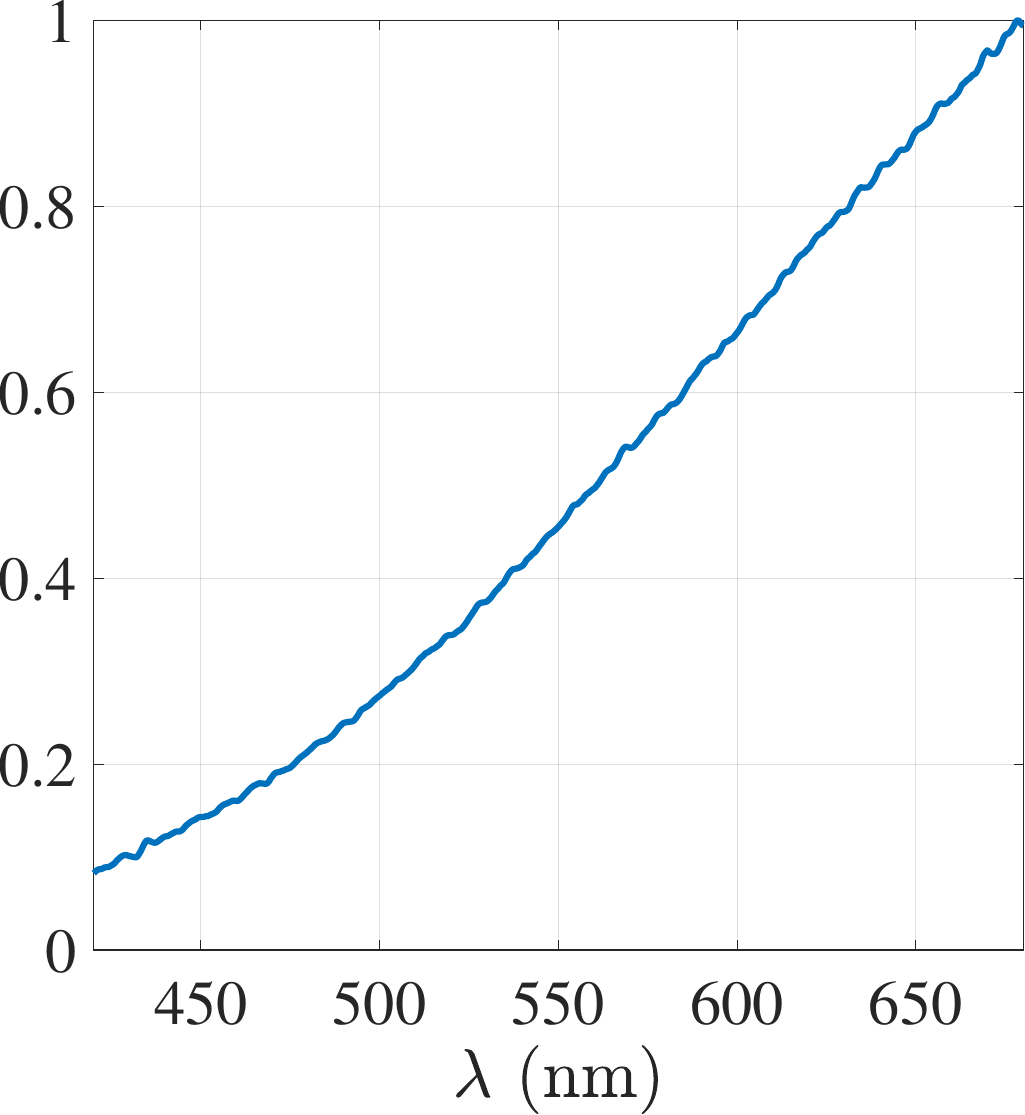}
		\caption{Ground truth spectrum}
	\end{subfigure}
	\begin{subfigure}[t]{0.3\columnwidth}
		\centering
		\includegraphics[width=\textwidth]{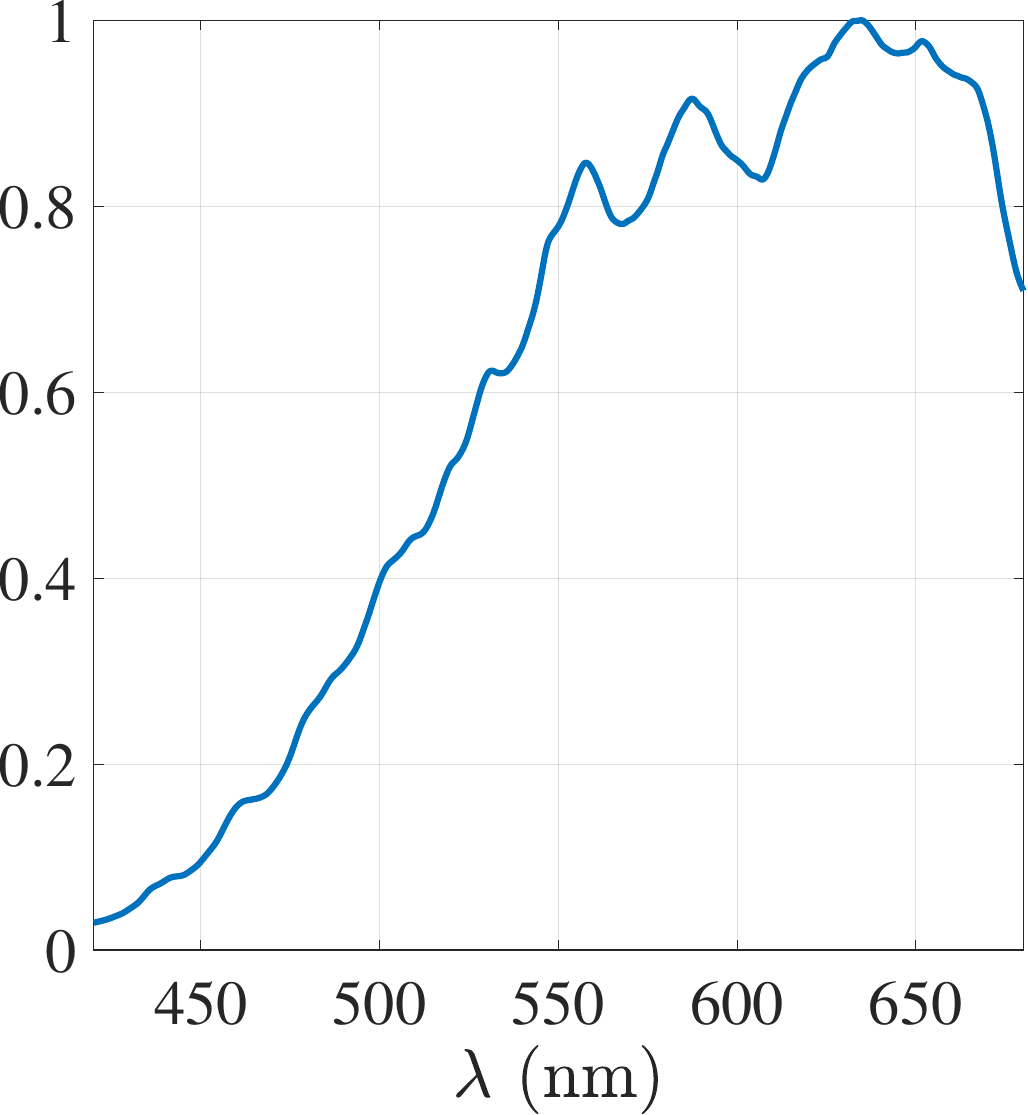}
		\caption{Raw spectrum}
	\end{subfigure}
	\begin{subfigure}[t]{0.3\columnwidth}
		\centering
		\includegraphics[width=\textwidth]{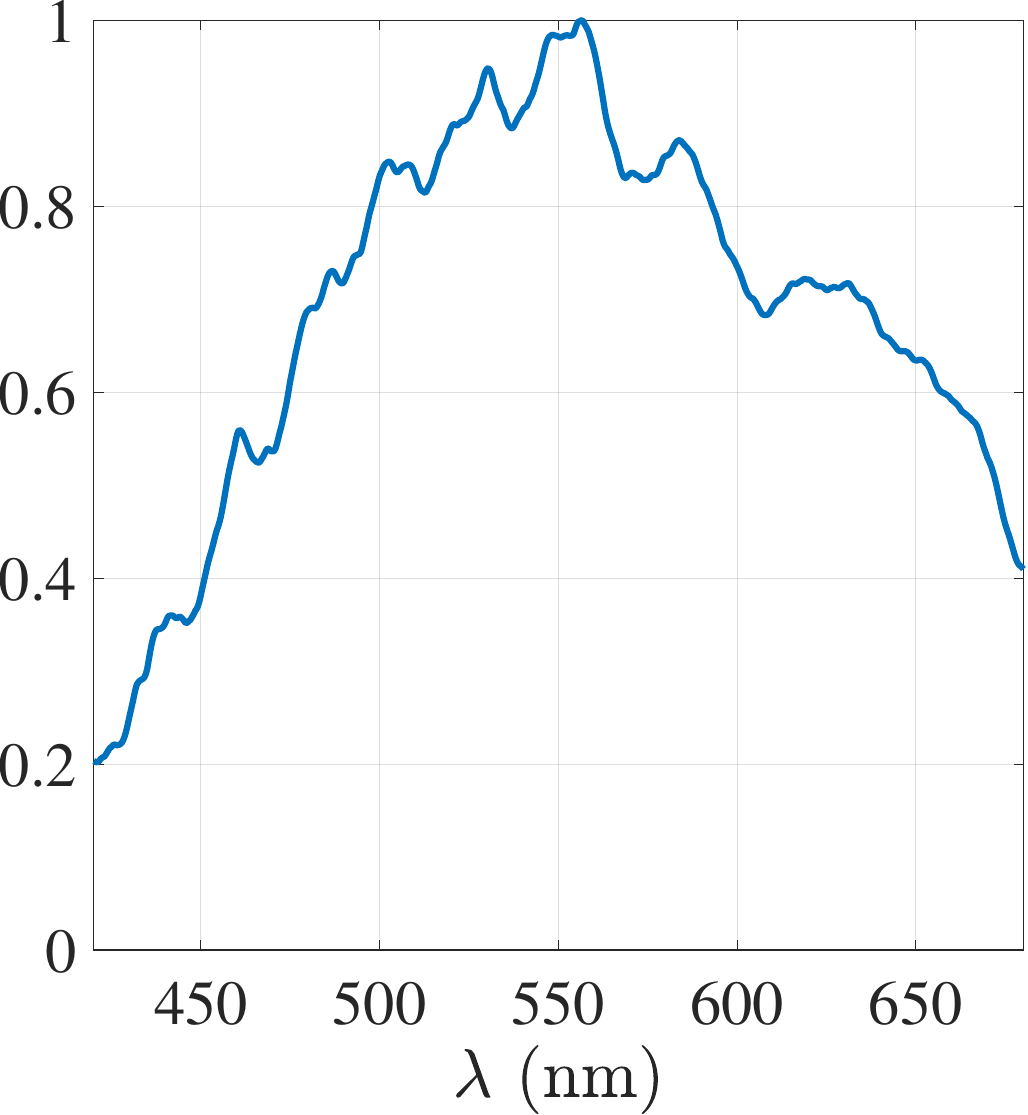}
		\caption{System spectral response}
	\end{subfigure}
	\caption{Calibration process for the optical setup. We used the tungsten-halogen light souce ``SL1-CAL" from Stellarnet. The ground truth spectrum in (a) was provided as part of the light source. We then measured spectrum of the light source by reflecting it off spectralon and deconvolving it with aperture code, shown in (b). Measured spectrum was then divided by ground truth spectrum to obtain system response, shown in (c).}
	\label{fig:response_calib}
\end{figure}

Figure \ref{fig:spectrum_deconv} shows the image for wavelength calibration pipeline. We first obtain image of spectrum of a narrow band filter.
After correcting for rotation, we obtain spectrum by summing the image vertically.
This helps estimate the binary code, which is then used to deconvolve the observed spectrum to get spectrum of the narrow band filter.
The peak of the narrow band filter is used as a known location.
The process is then repeated for 488nm and 670nm filters to get wavelengths.
\paragraph{Spectral response of camera / Radiometric calibration.}
The measured image and specturm on the camera plane is given by
\begin{align*}
	I_S(x, y) &\propto \int\limits_{\lambda} \left(H(x, y, \lambda) \ast \left| \frac{1}{\lambda f} A\left(-\frac{x}{\lambda f}, -\frac{x}{\lambda f}\right)\right|^2\right) c(\lambda) d\lambda\\
	I_R(x, y) &\propto a(x, y) \ast \left( s\left(\frac{x}{\lambda f\nu_0}\right) c\left(\frac{x}{\lambda f\nu_0}\right) \right),
\end{align*}
where $c(\lambda)$ is the spectral response of the camera.
For true spectrometric readings, contribution of $c(\lambda)$ needs to be removed. 
This can be achieved by calibrating the spectrometric measurements with a known light source.
The tungsten-halogen light source, ``SL1-CAL" from Stellarnet was used for this purpose.
To compute $c(\lambda)$, we assumed  that the true spectrum, $c_t(\lambda)$ of the light source is known.
We then measured spectrum of the light source, $c_m(\lambda)$ with our optical setup.
The spectral response of the system was then computed as $c(\lambda) = \frac{c_m(\lambda)}{c_t(\lambda)}$.
This procedure is illustrated in Figure \ref{fig:response_calib}.
	
	\section{Real experiments} \label{section:sup_real}
\begin{figure}[!tt]
	\centering
	\begin{subfigure}[t]{0.48\columnwidth}
		\centering
		\includegraphics[width=\textwidth]{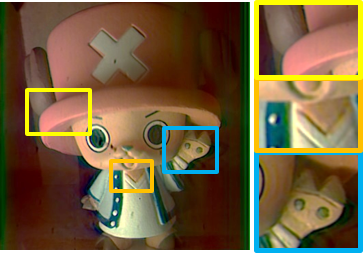}
		\caption{KRISM}
	\end{subfigure}
	\hspace{0.1em}
	\begin{subfigure}[t]{0.48\columnwidth}
		\centering
		\includegraphics[width=\textwidth]{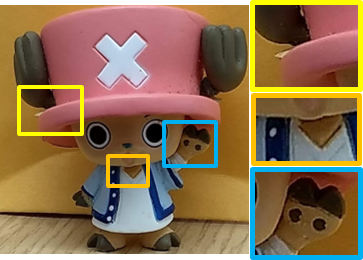}
		\caption{Cellphone}
	\end{subfigure}
	\caption{Comparison of RGB images with our lab prototype and cellphone. Note that the KRISM image is not white balanced whereas the cellphone image is white balanced by default. The zoomed in patches show edges in the toy, establishing high spatial resolution of our lab prototype.}
	\label{fig:chopper_compare}
\end{figure}

We provide visualizations for some of the real experiments presented in the main paper.
Specifically, we compare the captured singular vectors for two scenes with spectrally Hadamard multiplexed measurements.
We also show spectral band images for Macbeth chart and crayons chart, showing the intensity variation of various colors.

\subsection{Visualizing spatial images}
\begin{figure*}[!ttt]
	\centering
	\begin{subfigure}[t]{0.2\textwidth}
		\centering
		\includegraphics[width=\textwidth]{figures/macbeth/rgb.png}
		\caption{``Color checker" scene}
	\end{subfigure}
	\quad
	\begin{subfigure}[t]{0.75\textwidth}
		\centering
		\includegraphics[width=\textwidth]{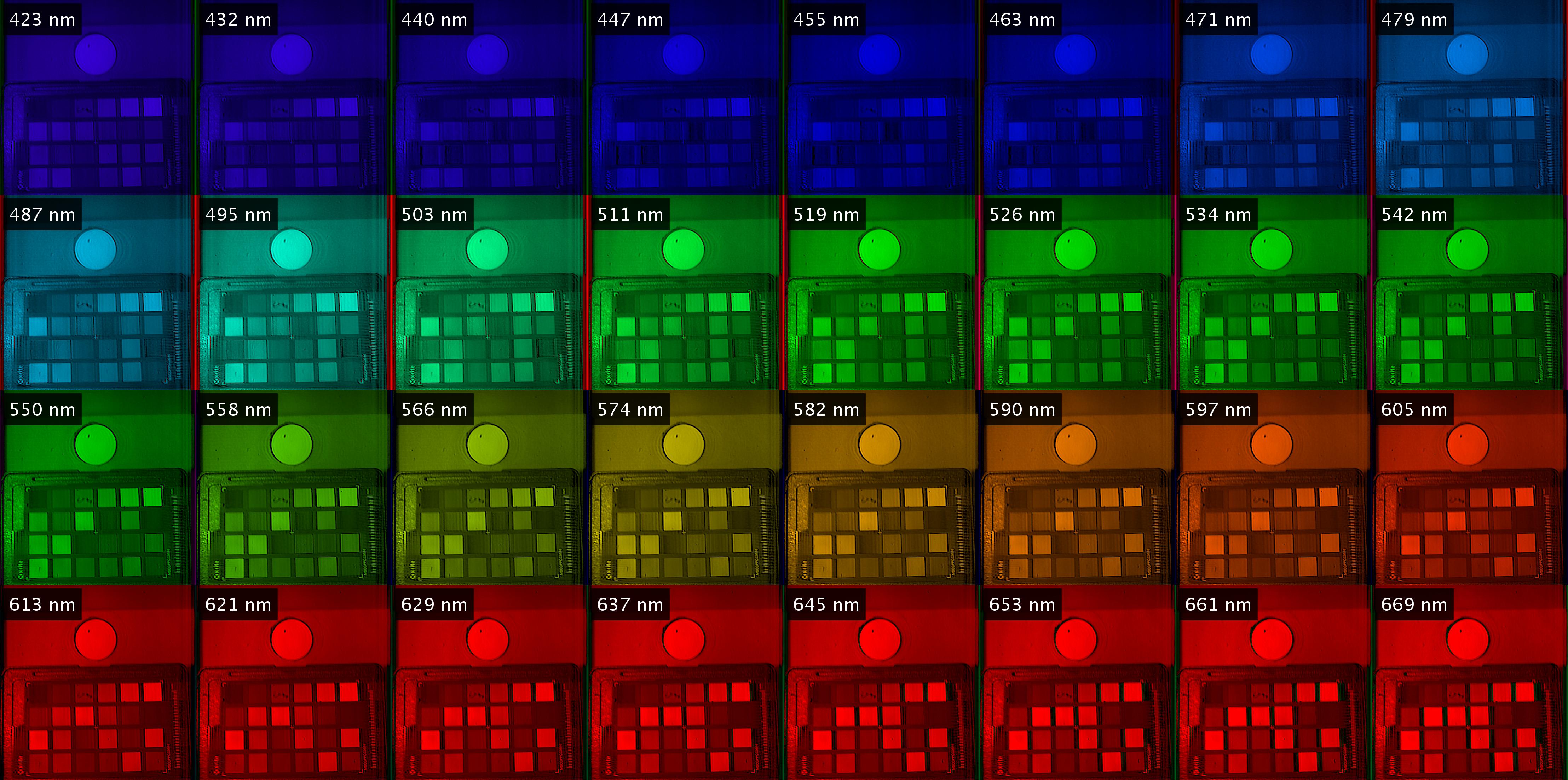}
		\caption{Spectral band images}
	\end{subfigure}
	\\
	\begin{subfigure}[t]{0.2\textwidth}
		\centering
		\includegraphics[width=\textwidth]{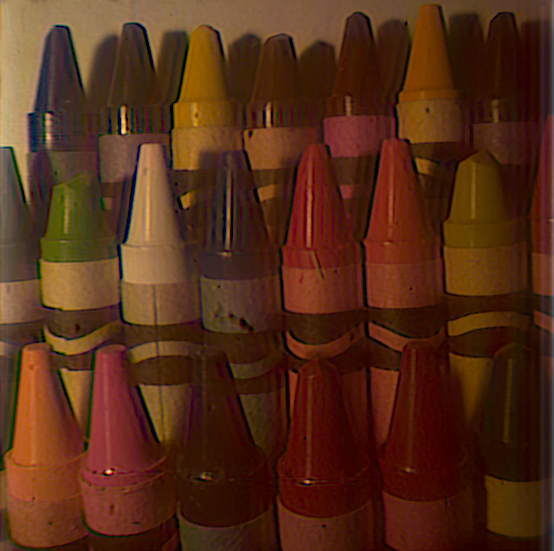}
		\caption{``Crayons" scene}
	\end{subfigure}
	\quad
	\begin{subfigure}[t]{0.75\textwidth}
		\centering
		\includegraphics[width=\textwidth]{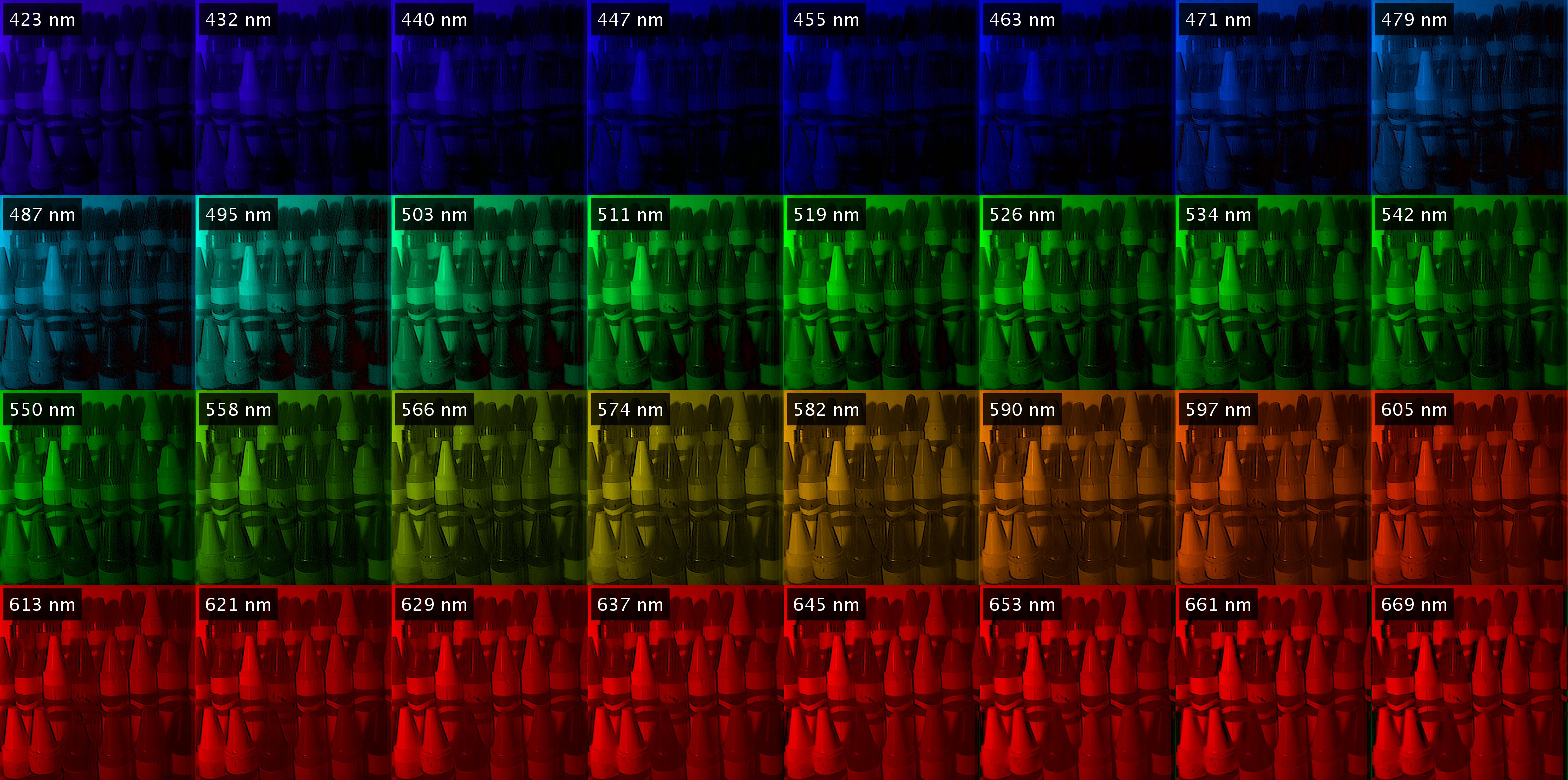}
		\caption{Spectral band images}
	\end{subfigure}
	\caption{Crayons and Macbeth scene images across different wavelengths. Both scenes were captured by illuminating the subjects with a tungsten-halogen bulb and then obtaining a rank-4 approximation with KRISM. The intensity of the crayons across wavelengths correctly reflects the individual color.}
	\label{fig:bands}
\end{figure*}

Figure \ref{fig:bands} shows images across various wavelengths for the ``color checker" scene and ``crayons" scene.
In particular, The images show the variation of intensity of each color swatch/crayon across wavelengths, with blur objects being brighter initially, green objects in the middle and red objects finally.
Figure \ref{fig:chopper_compare} shows RGB image captured by our lab prototype as well as a cellphone camera for the Chopper scene.
The insets shows textured areas which show high spatial resolution of our prototype.
\begin{figure*}[!ttt]
	\centering
	\begin{subfigure}[c]{0.23\textwidth}
		\centering
		\includegraphics[width=\textwidth]{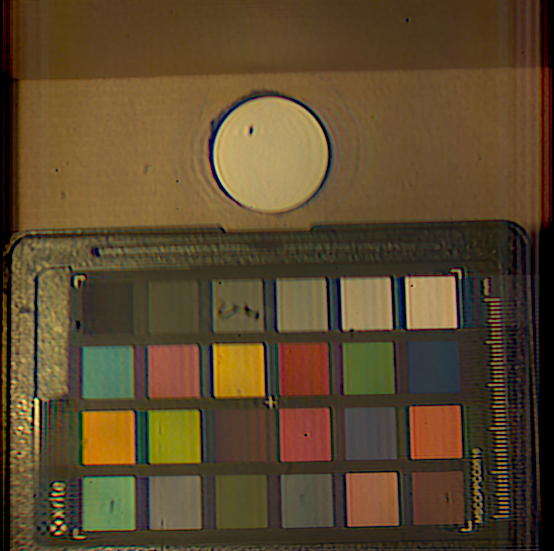}
		\caption{``Color checker" scene}
	\end{subfigure}
	\quad
	\begin{subfigure}[c]{0.23\textwidth}
		\centering
		\includegraphics[width=\textwidth]{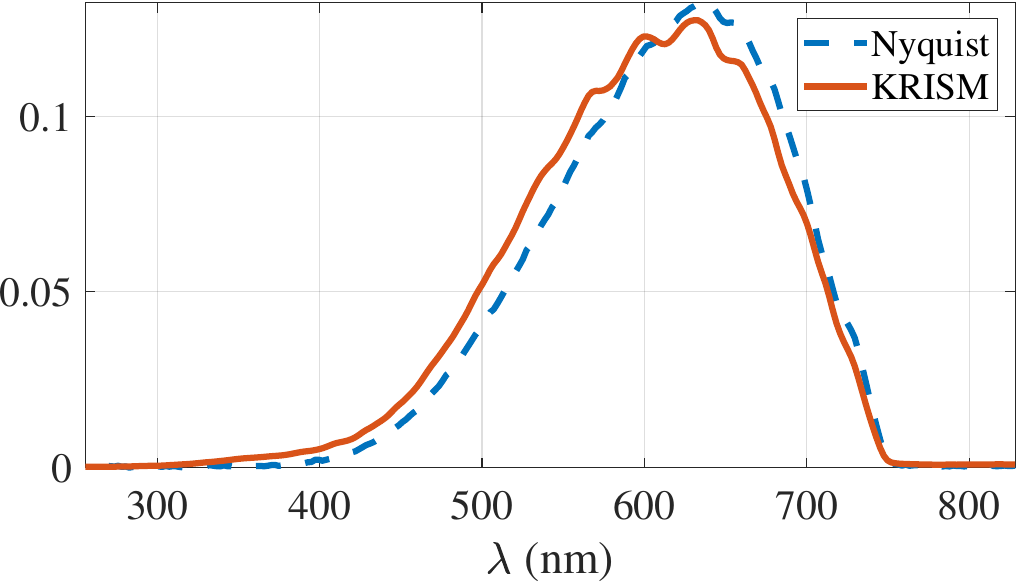}
		\includegraphics[width=0.45\textwidth]{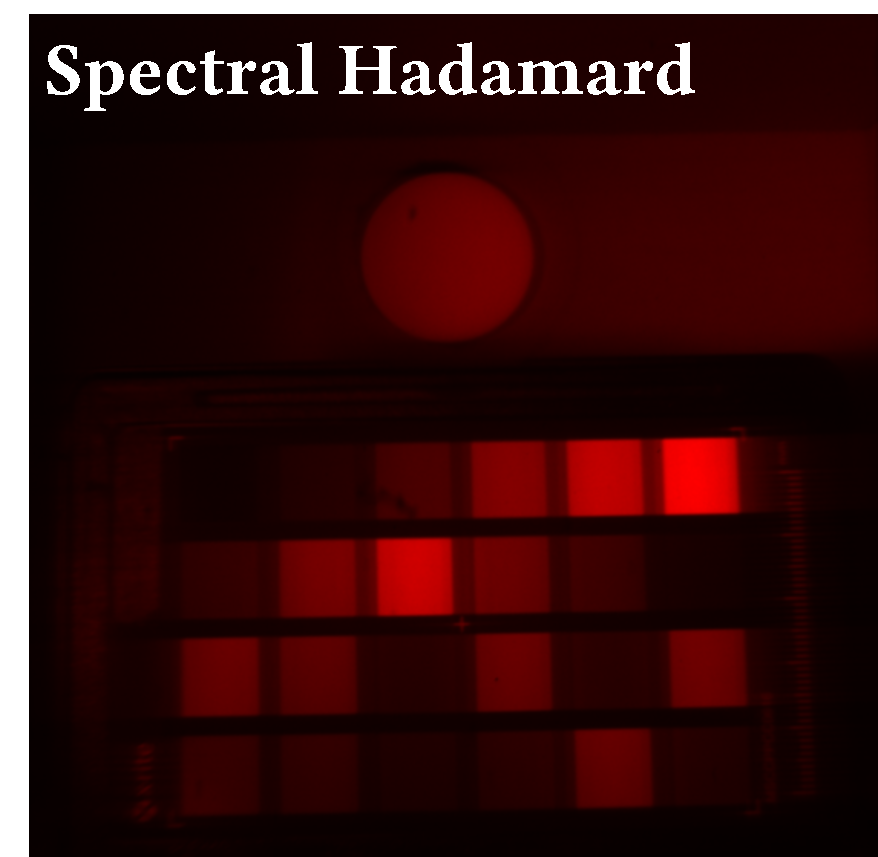}
		\quad
		\includegraphics[width=0.45\textwidth]{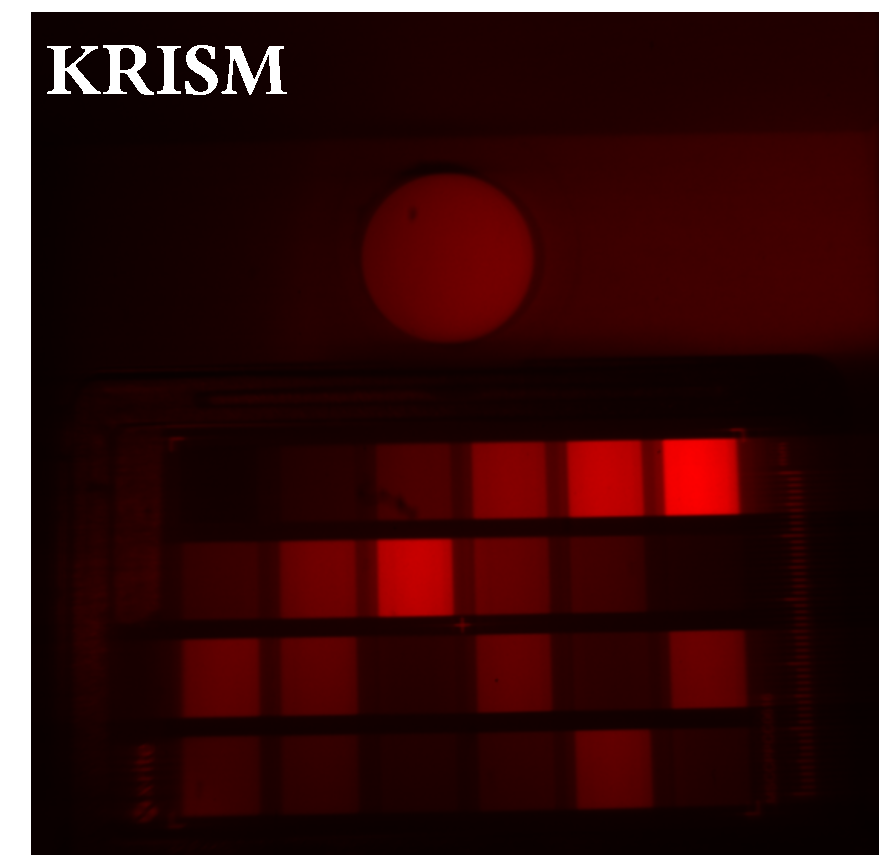}
		\caption{First singular vector}
	\end{subfigure}
	\quad	
	\begin{subfigure}[c]{0.23\textwidth}
		\centering
		\includegraphics[width=\textwidth]{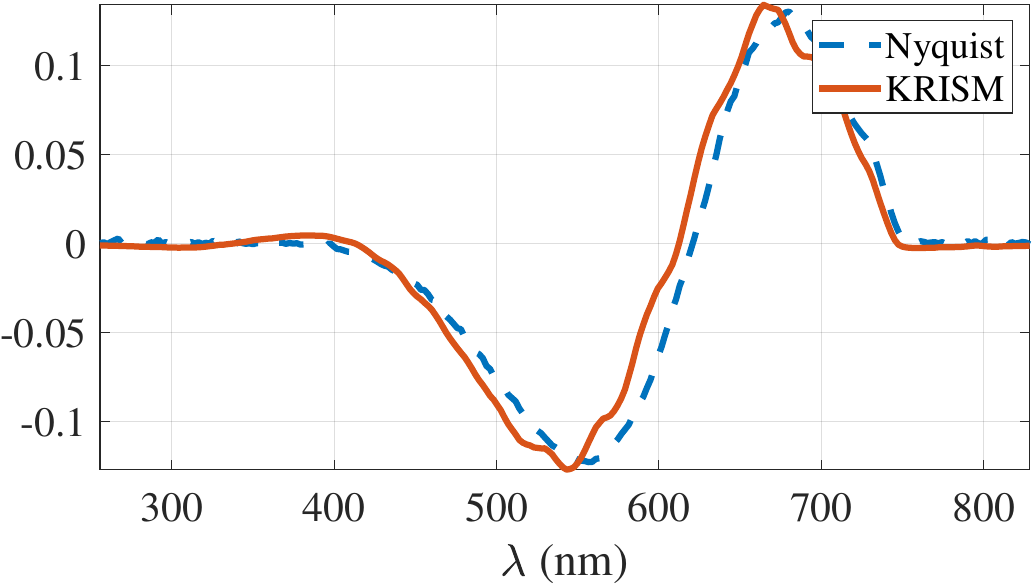}
		\includegraphics[width=0.45\textwidth]{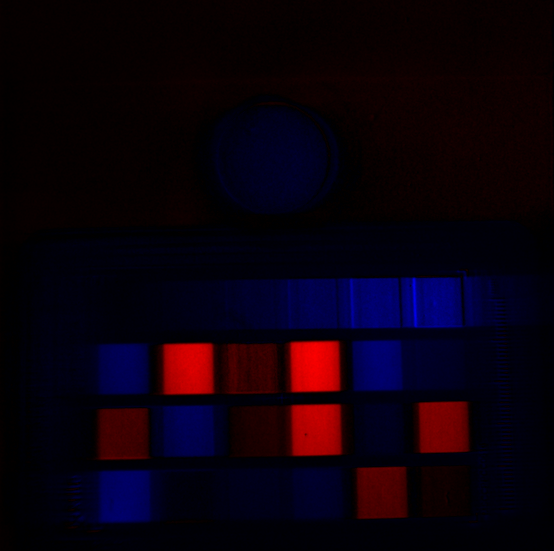}
		\quad
		\includegraphics[width=0.45\textwidth]{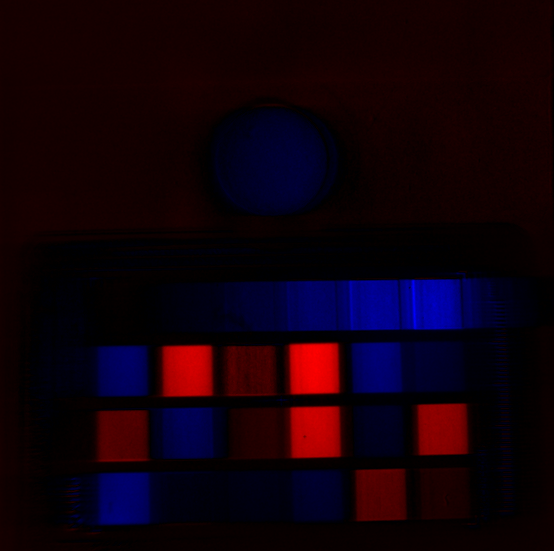}
		\caption{Second singular vector}
	\end{subfigure}
	\quad
	\begin{subfigure}[c]{0.23\textwidth}
		\centering
		\includegraphics[width=\textwidth]{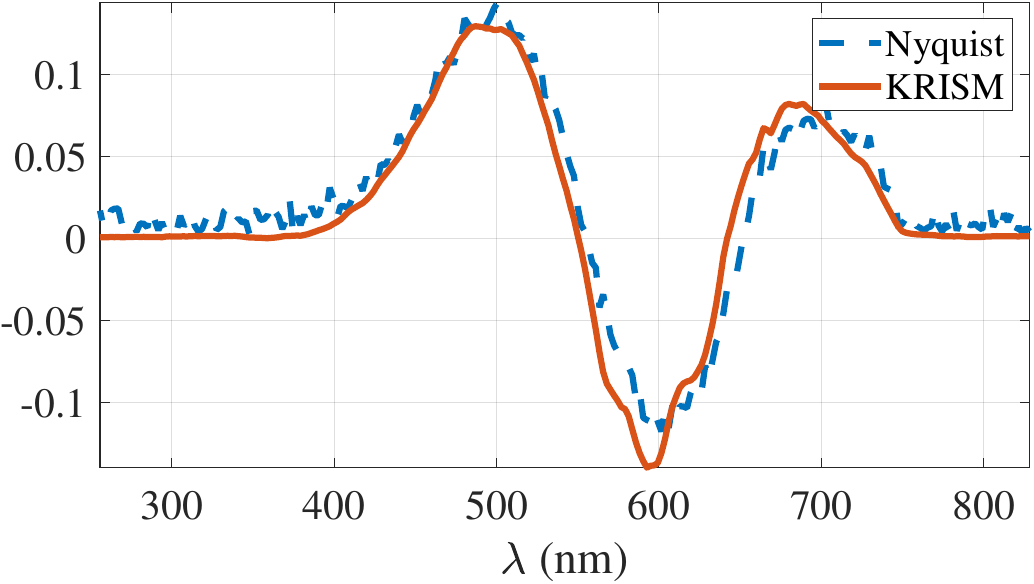}
		\includegraphics[width=0.45\textwidth]{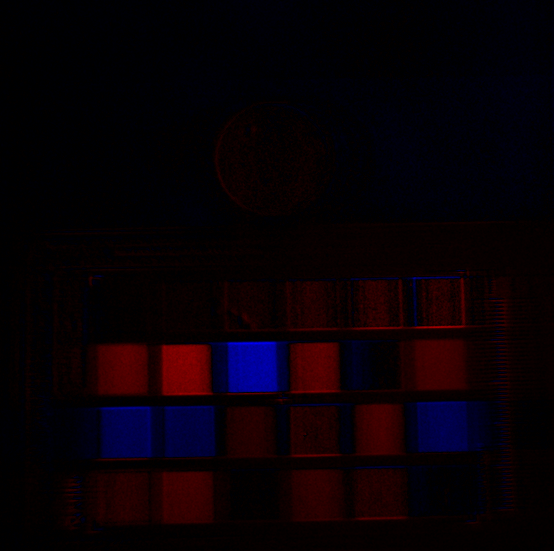}
		\quad
		\includegraphics[width=0.45\textwidth]{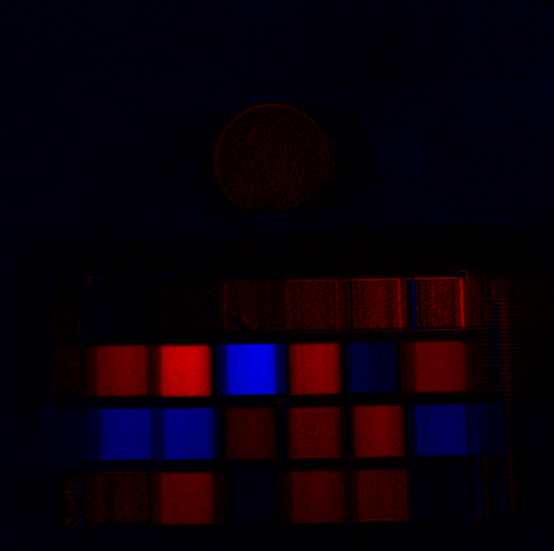}
		\caption{Third singular vector}
	\end{subfigure}
	\vspace{1em}	
	\begin{subfigure}[c]{0.23\textwidth}
		\centering
		\includegraphics[width=\textwidth]{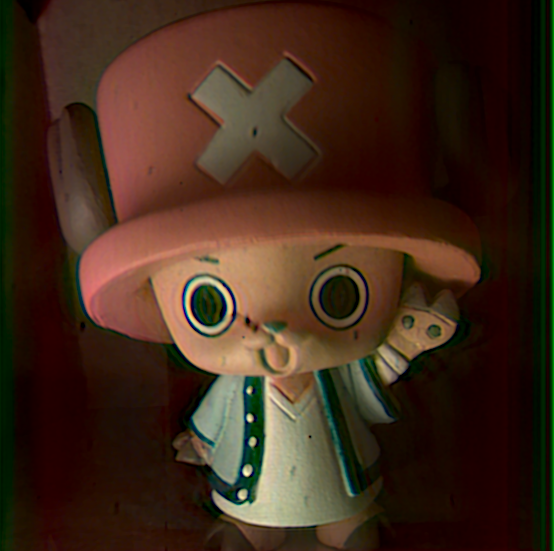}
		\caption{``Chopper" scene}
	\end{subfigure}
	\quad
	\begin{subfigure}[c]{0.23\textwidth}
		\centering
		\includegraphics[width=\textwidth]{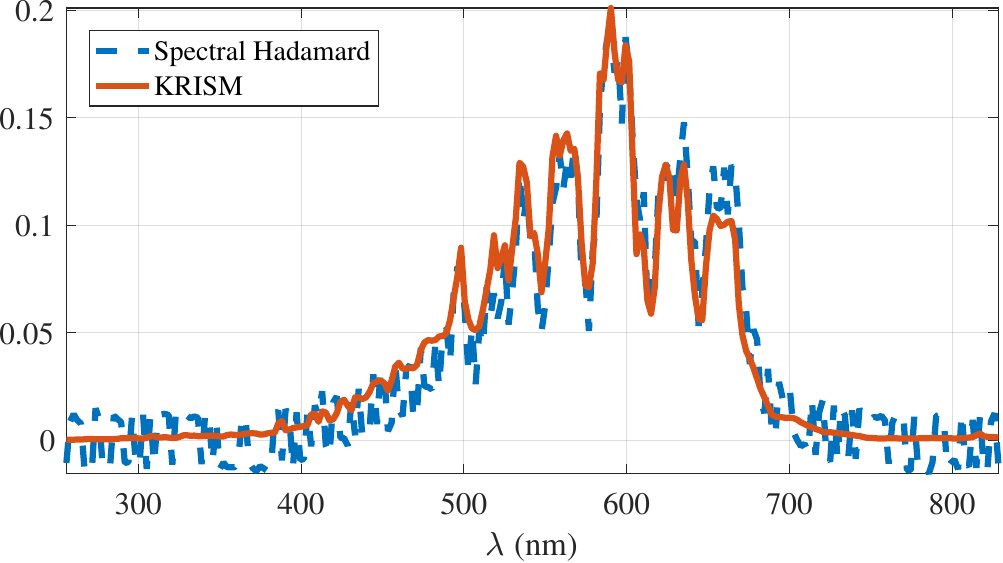}
		\includegraphics[width=0.45\textwidth]{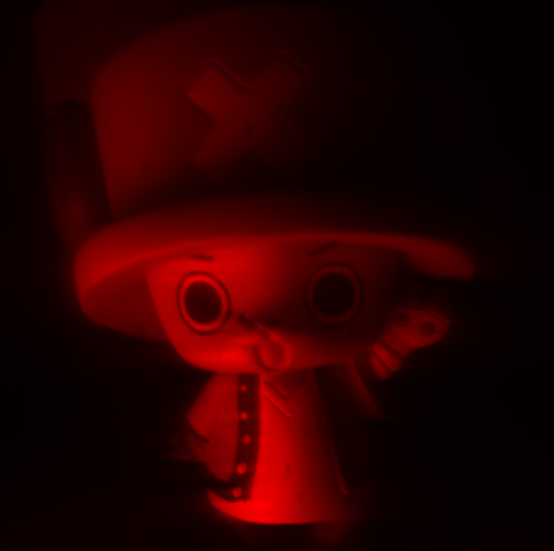}
		\quad
		\includegraphics[width=0.45\textwidth]{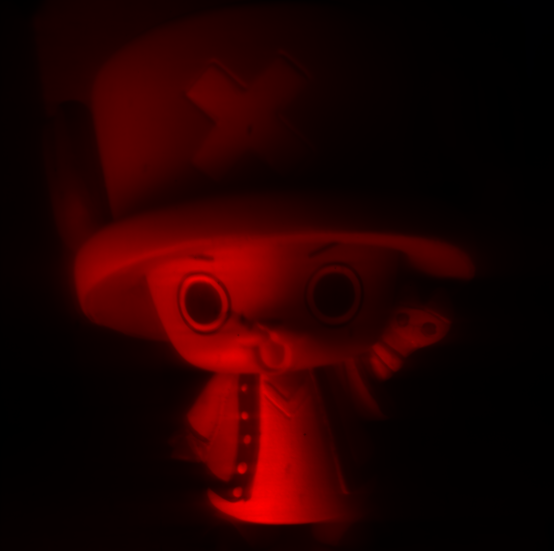}
		\caption{First singular vector}
	\end{subfigure}
	\quad
	\begin{subfigure}[c]{0.23\textwidth}
		\centering
		\includegraphics[width=\textwidth]{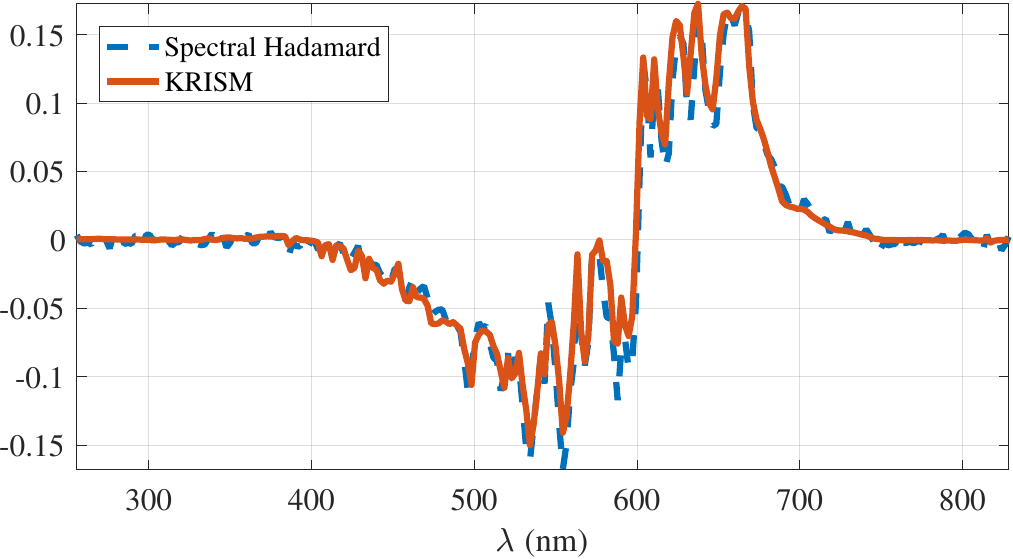}
		\includegraphics[width=0.45\textwidth]{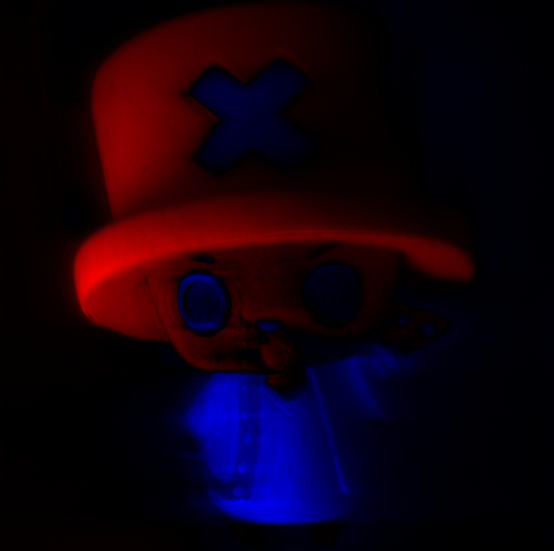}
		\quad
		\includegraphics[width=0.45\textwidth]{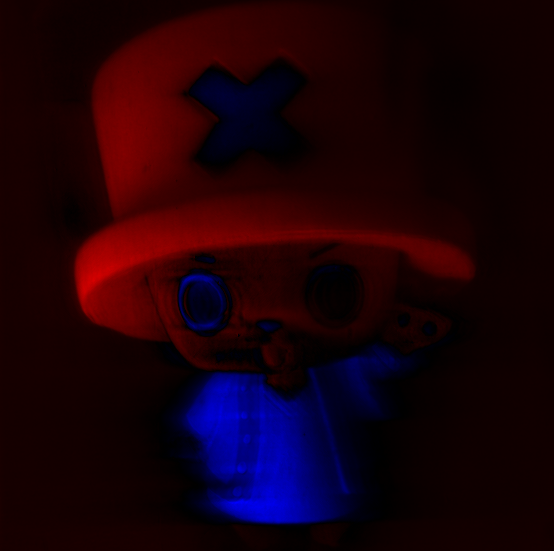}
		\caption{Second singular vector}
	\end{subfigure}
	\quad
	\begin{subfigure}[c]{0.23\textwidth}
		\centering
		\includegraphics[width=\textwidth]{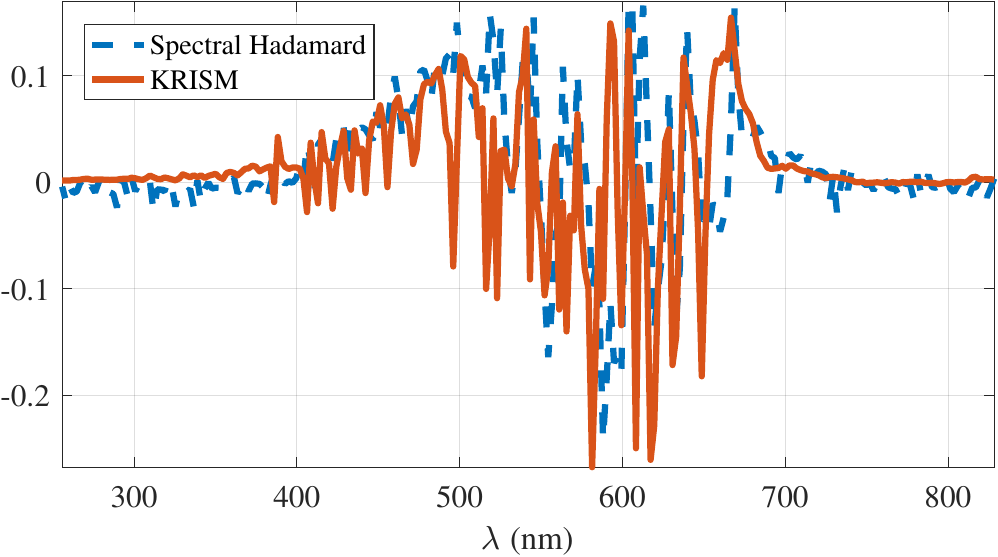}
		\includegraphics[width=0.45\textwidth]{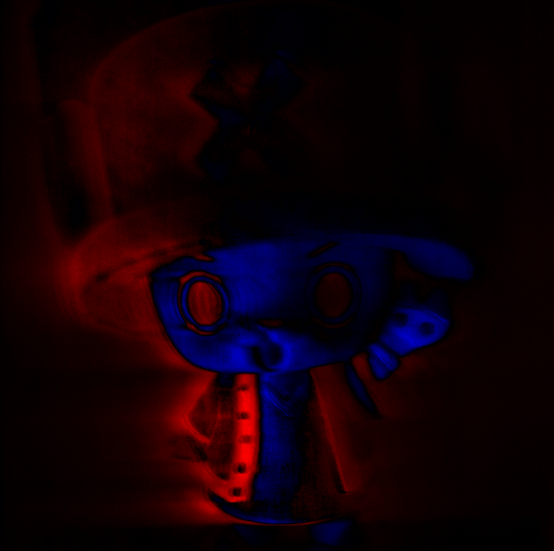}
		\quad
		\includegraphics[width=0.45\textwidth]{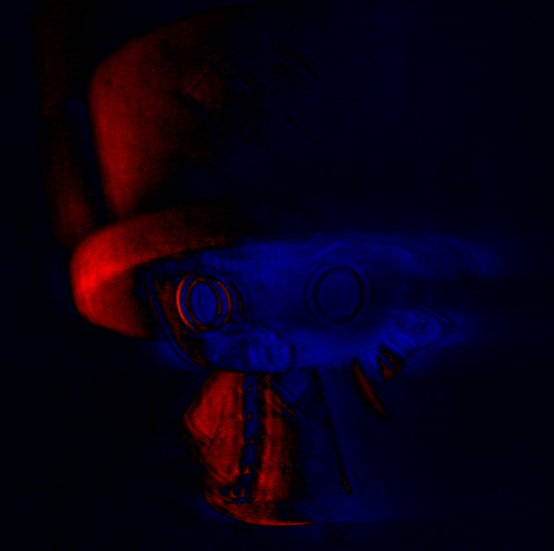}
		\caption{Third singular vector}
	\end{subfigure}
	\vspace{1em}	
	\begin{subfigure}[t]{0.24\textwidth}
		\centering
		\includegraphics[width=\textwidth]{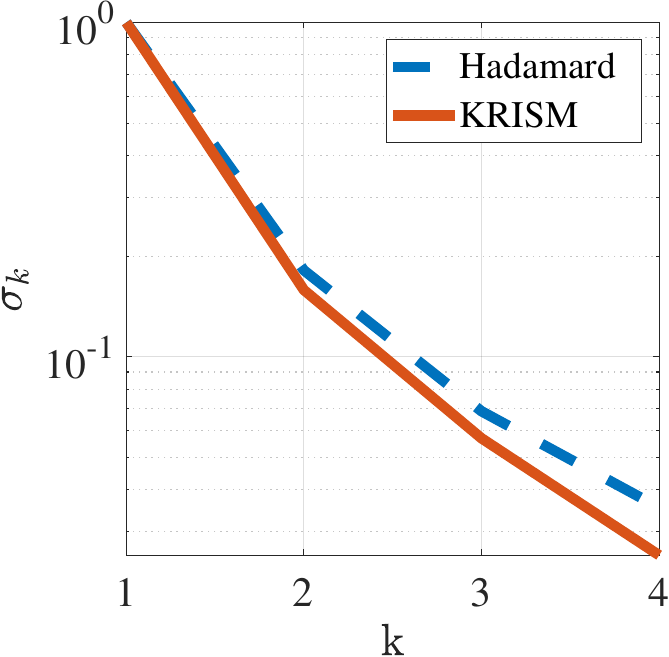}
		\caption{Dice (PSNR = 37.2dB)}
	\end{subfigure}
	\begin{subfigure}[t]{0.24\textwidth}
		\centering
		\includegraphics[width=\textwidth]{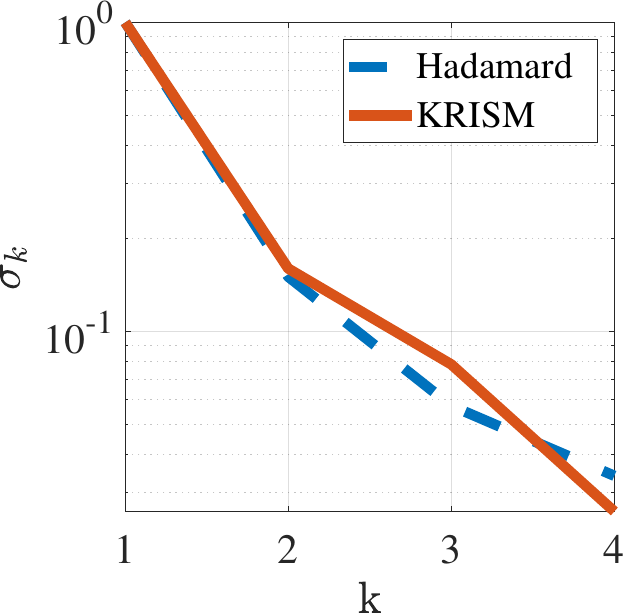}
		\caption{Color checker (PSNR = 45.8dB)}
	\end{subfigure}
	\begin{subfigure}[t]{0.24\textwidth}
		\centering
		\includegraphics[width=\textwidth]{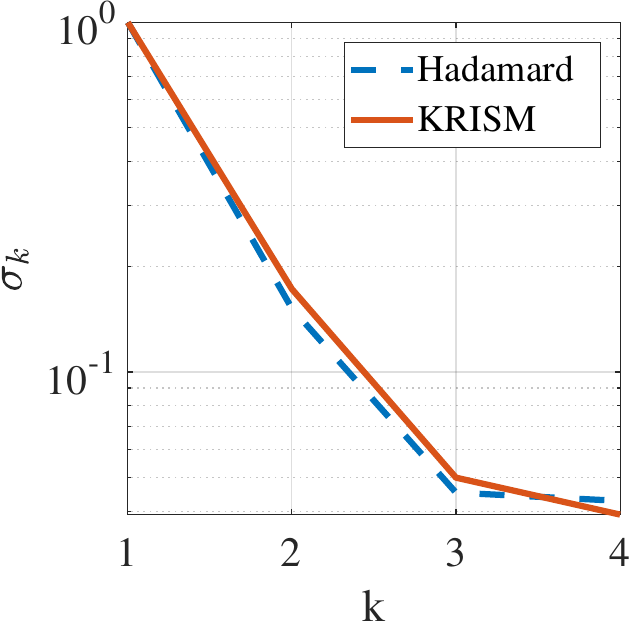}
		\caption{Chopper (PSNR = 39.3dB)}
	\end{subfigure}
	\begin{subfigure}[t]{0.24\textwidth}
		\centering
		\includegraphics[width=\textwidth]{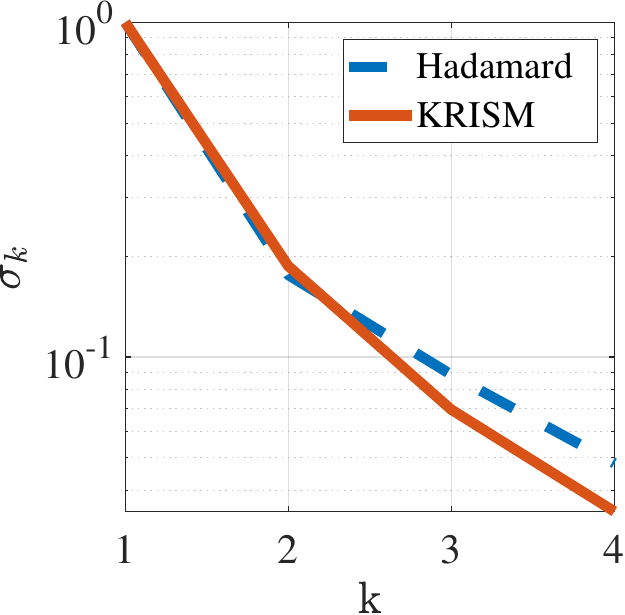}
		\caption{Crayons (PSNR = 37.8dB)}
	\end{subfigure}
	\caption{Comparison of singular values and singular vectors captured via spectrally Hadamard-multiplexed sensing and KRISM. The left image singular vector is from Hadamard multiplexed data and the right one is from KRISM. Blue represents negative values and red represents positive values. KRISM method required capturing a total of 6 spectral and 6 spatial measurements to construct 4 singular vectors. While the Nyquist sampling method took a total of 59 minutes, KRISM took under 5 minutes. Top row shows singular vectors for ``Color checker" scene, the middle row shows singular vectors for ``Chopper" scene, and the last row compares singular values computed by Hadamard multiplexing and KRISM. Overall, our optical setup captures a low-rank approximation of the HSI with high accuracy.}
	\label{fig:eigvec_compare}
\end{figure*}

\subsection{Comparison of singular values and singular vectors}
The ability of KRISM to accurately compute singular vectors has been presented in the main paper.
Here, we present two more experimental measurements to show how KRISM is applicable across various settings.
Comparison is done against spectrally Hadamard multiplexed data, and then computing singular vectors on computer.
We evaluate three metrics, namely, SNR between singular values, SAM between spectral singular vectors and SAM between spatial singular vectors.
``Color checker" experiment (first row in Figure \ref{fig:eigvec_compare}) was captured by placing the Macbeth chart in front of the camera, and illuminating with a tungsten-halogen light source.
The PSNR between singular values was $45.8$dB, average SAM between spectral singular vectors $10^\circ$ and that between spatial singular vectors was $10^\circ$.
``Chopper" experiment (second row in Figure \ref{fig:eigvec_compare}) was captured by placing the Chopper toy in front of the camera, and illuminating it with CFL, a peaky illuminant.
The PSNR between singular values was $39.3$dB, average SAM between spectral singular vectors $10^\circ$ and that between spatial singular vectors was $10^\circ$.
Finally, the last row shows a comparison between singular values from Hadamard multiplexing and singular values from KRISM for some scenes presented in the main paper.
Across the board, KRISM computes the low-rank approximation with very high accuracy, as is evident from the experiments.

	\section{Synthetic experiments} \label{section:sup_synthetic}
	
We showed some simulation results in the main paper. We show several more examples here, with emphasis on diversity of datasets. We tested KRISM via simulations on four different datasets and compared it against competing techniques for hyperspectral imaging. 
\begin{figure*}[!tt]
	\centering
	\begin{subfigure}[c]{0.12\textwidth}
		\includegraphics[width=\textwidth]{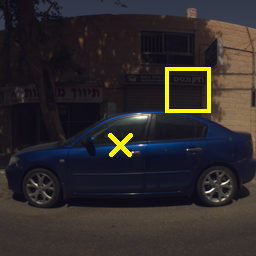}
		\caption{Arad and \\ Ben-Shahar\\($256\times256\times260$)}
	\end{subfigure}
	\begin{subfigure}[c]{0.12\textwidth}
		\includegraphics[width=\textwidth]{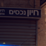}
		\caption{Ground truth\\ \hspace{0.1em}\\ \hspace{0.1em}}
	\end{subfigure}
	\begin{subfigure}[c]{0.12\textwidth}
		\includegraphics[width=\textwidth]{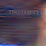}
		\caption{Kittle et al.\\PSNR: 38.6dB, $N/M: 5$}
	\end{subfigure}
	\begin{subfigure}[c]{0.12\textwidth}
		\includegraphics[width=\textwidth]{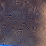}
		\caption{Sun and Kelly\\PSNR: 41.1dB, $N/M: 5$}
	\end{subfigure}
	\begin{subfigure}[c]{0.12\textwidth}
		\includegraphics[width=\textwidth]{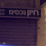}
		\caption{Fazel et al.\\PSNR: 40.3dB, $N/M: 43$}
	\end{subfigure}
	\begin{subfigure}[c]{0.12\textwidth}
		\includegraphics[width=\textwidth]{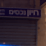}
		\caption{KRISM\\PSNR: 49.1dB, $N/M: 43$}
	\end{subfigure}
	\begin{subfigure}[c]{0.25\textwidth}
		\centering
		\includegraphics[width=\textwidth]{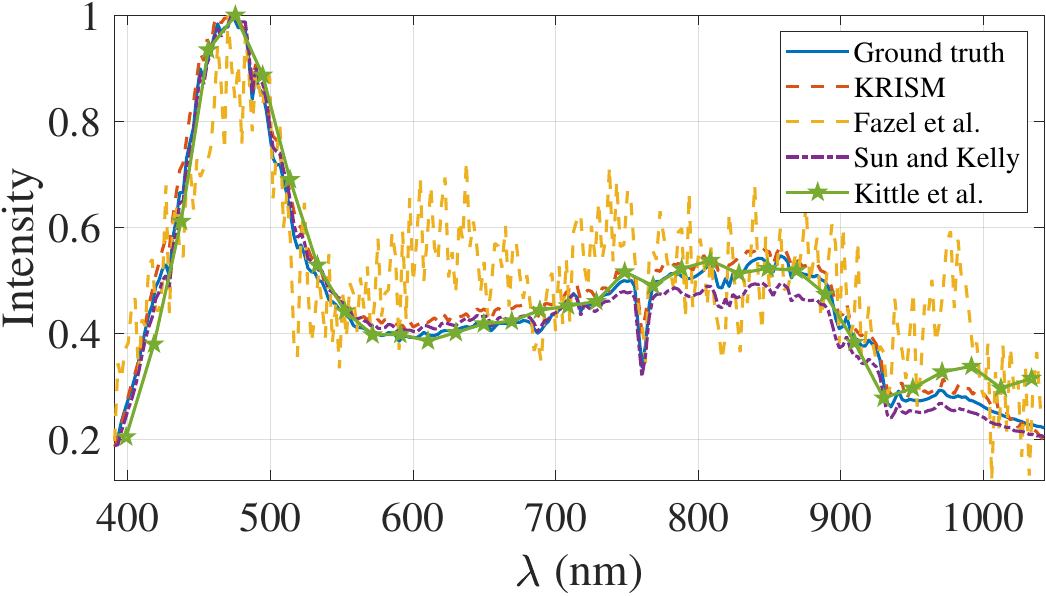}
		\caption{Spectrum at marked point\\\vphantom{1em}}
	\end{subfigure}
	\setcounter{subfigure}{0}
	\\
	\begin{subfigure}[c]{0.12\textwidth}
		\includegraphics[width=\textwidth]{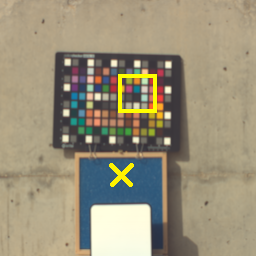}
		\caption{Arad and Ben-Shahar \\ ($256\times256\times260$)}
	\end{subfigure}
	\begin{subfigure}[c]{0.12\textwidth}
		\includegraphics[width=\textwidth]{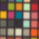}
		\caption{Ground truth\\ \hspace{0.1em}\\ \hspace{0.1em}}
	\end{subfigure}
	\begin{subfigure}[c]{0.12\textwidth}
		\includegraphics[width=\textwidth]{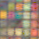}
		\caption{Kittle et al.\\PSNR: 33.3dB, $N/M: 5$}
	\end{subfigure}
	\begin{subfigure}[c]{0.12\textwidth}
		\includegraphics[width=\textwidth]{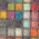}
		\caption{Sun and Kelly\\PSNR: 34.0dB, $N/M: 5$}
	\end{subfigure}
	\begin{subfigure}[c]{0.12\textwidth}
		\includegraphics[width=\textwidth]{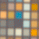}
		\caption{Fazel et al.\\PSNR: 35.4dB, $N/M: 43$}
	\end{subfigure}
	\begin{subfigure}[c]{0.12\textwidth}
		\includegraphics[width=\textwidth]{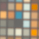}
		\caption{KRISM\\PSNR: 39.9dB, $N/M: 43$}
	\end{subfigure}
	\begin{subfigure}[c]{0.25\textwidth}
		\centering
		\includegraphics[width=\textwidth]{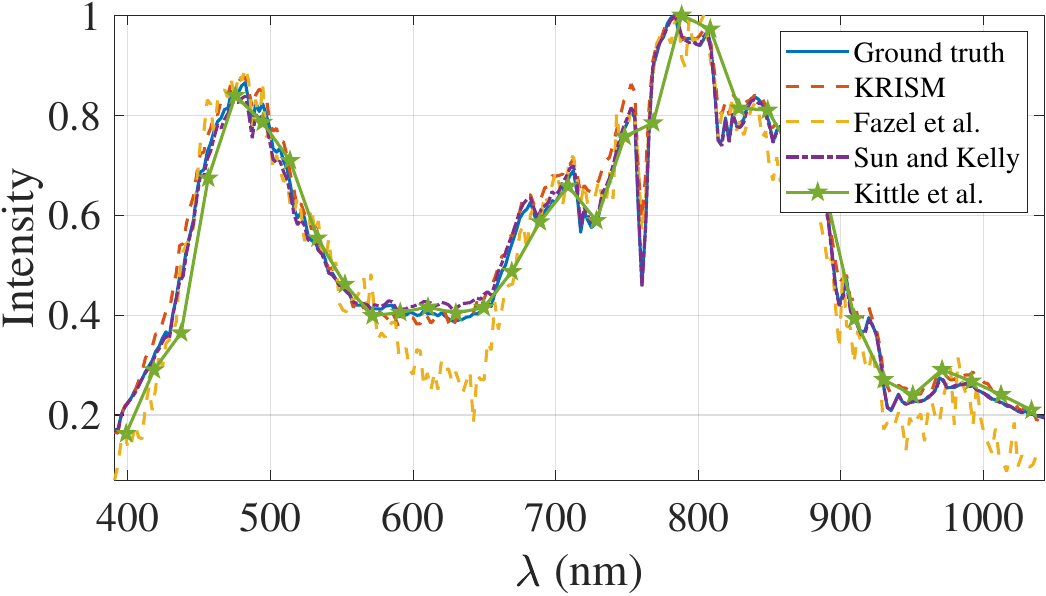}
		\caption{Spectrum at marked point\\\vphantom{1em}}
	\end{subfigure}
	\setcounter{subfigure}{0}
	\\
	\begin{subfigure}[c]{0.12\textwidth}
		\includegraphics[width=\textwidth]{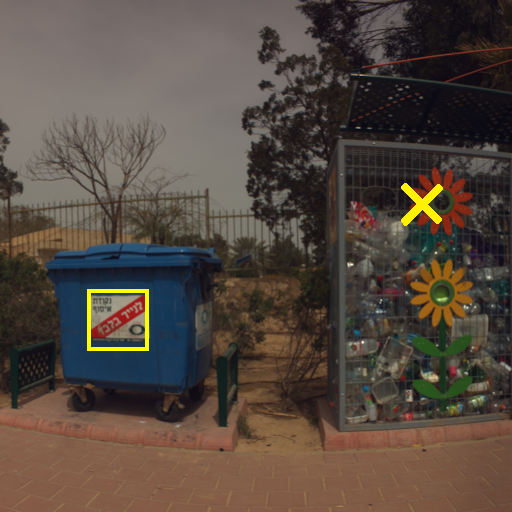}
		\caption{Arad and Ben-Shahar \\ ($512\times512\times260$)}
	\end{subfigure}
	\begin{subfigure}[c]{0.12\textwidth}
		\includegraphics[width=\textwidth]{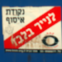}
		\caption{Ground truth\\ \hspace{0.1em}\\ \hspace{0.1em}}
	\end{subfigure}
	\begin{subfigure}[c]{0.12\textwidth}
		\includegraphics[width=\textwidth]{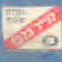}
		\caption{Kittle et al.\\PSNR: 27.1dB, $N/M: 5$}
	\end{subfigure}
	\begin{subfigure}[c]{0.12\textwidth}
		\includegraphics[width=\textwidth]{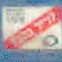}
		\caption{Sun and Kelly\\PSNR: 30.1dB, $N/M: 5$}
	\end{subfigure}
	\begin{subfigure}[c]{0.12\textwidth}
		\includegraphics[width=\textwidth]{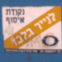}
		\caption{Fazel et al.\\PSNR: 36.8dB, $N/M: 43$}
	\end{subfigure}
	\begin{subfigure}[c]{0.12\textwidth}
		\includegraphics[width=\textwidth]{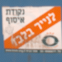}
		\caption{KRISM\\PSNR: 44.1dB, $N/M: 43$}
	\end{subfigure}
	\begin{subfigure}[c]{0.25\textwidth}
		\centering
		\includegraphics[width=\textwidth]{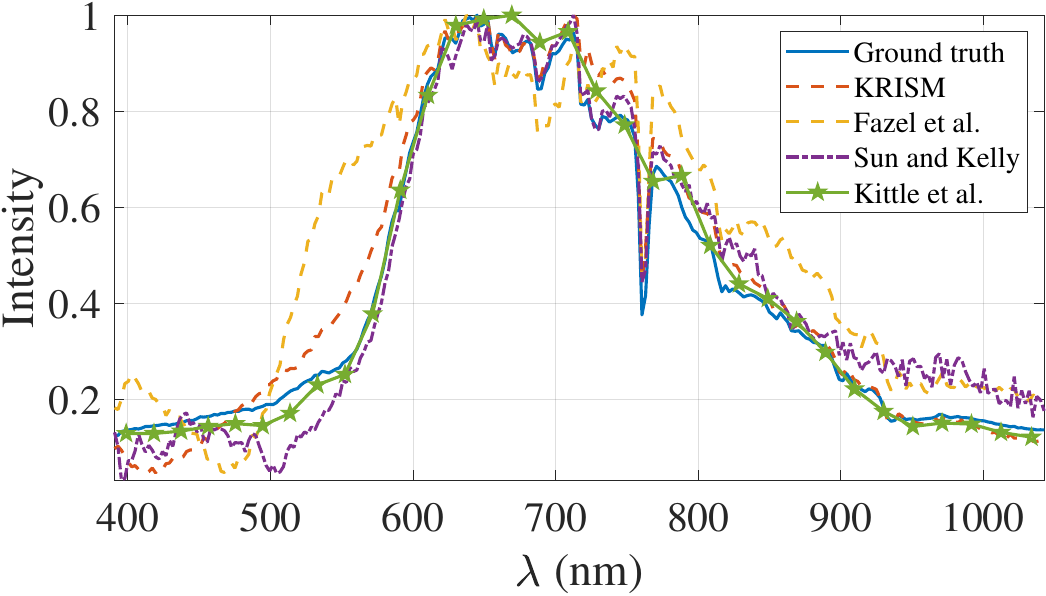}
		\caption{Spectrum at marked point\\\vphantom{1em}}
	\end{subfigure}
	\caption{Visualization of results for the high resolution dataset by \cite{arad_and_ben_shahar_2016_ECCV}. ``CASSI" represents Single Disperser CASSI, recovered using spectral prior \cite{deepcassi2017}. Kittle et al. uses multiple spatio-spectral images \cite{kittle2010multiframe}, and was reconstructed with sparsity in wavelet domain. Sun and Kelly represents spatially-multiplexed measurements \cite{sun2009compressive}, and was reconstructed with sparsity in wavelet domain. Row/Col CS represents random row and column projections \cite{fazel2008compressed}, and KRISM is the proposed method. Simulations were performed with 60dB readout, and photon noise. Row/Col CS and KRISM were simulated with spatio-spectral diffraction blur.  We show zoomed in image patches for each method and spectrum at pixel marked by a cross. Across the board, KRISM outperforms all methods, both qualitatively and quantitatively.}
	\label{fig:synthetic_hr}
\end{figure*}
\paragraph{Datasets.} We used the hyperspectral data set by \citet{arad_and_ben_shahar_2016_ECCV} (ICVL dataset), which consists of several high spatial and spectral resolution hyperspectral images covering 519 bands in visible and near IR wavelengths.
We downsampled the HSI to $256 \times 256 \times 260$ to keep computation with CASSI-type simulations tractable.
We also used datasets from \citet{deepcassi2017,yasuma2010generalized} and \citet{chakrabarti2011statistics} with 31 spectral bands to compare with  learning-based techniques.
Finally, we present one example from NASA AVIRIS to compare KRISM against Row/Column CS proposed in \cite{fazel2008compressed}.
\paragraph{Competing methods.}
We compared KRISM against four competing CS hyperspectral imaging techniques. All methods were simulated with 60dB readout and photon noise and 12-bit quantization. Specifics of each simulation model are given below:
\begin{enumerate}[leftmargin=*]
	\item \textit{KRISM}: We performed a rank-4 approximation of the HSI with 6 spatial and 6 spectral measurements. Diffraction blur due to coded aperture was introduced both in spectral and spatial profiles. Deconvolution was then done using Wiener deconvolution in both spectral and spatial domains.
	\item \textit{Fazel et al.} \citeNN{fazel2008compressed}: As with KRISM, we performed a rank-4 approximation of the HSI by computing random Gaussian projections with 6 spatial and 6 spectral measurements. Diffraction blur due to coded aperture was introduced as well.
	\item \textit{Lin et al.} \citeNN{lin2014spatial}: We recovered HSI from a single snapshot image using technique in \citet{lin2014spatial}. 
	\item \textit{Choi et al.} \citeNN{deepcassi2017}: We recovered HSI from a single snapshot image using technique in \cite{deepcassi2017}. 
	\item \textit{Kittle et al.} \citeNN{kittle2010multiframe}: We used the multi-frame CASSI architecture for obtaining coded images, and recovered the HSI with sparsity prior in wavelet domain. We reduced the number of spectral bands for ICVL dataset to 31 to keep computations tractable.
	\item \textit{Sun and Kelly} \citeNN{sun2009compressive}: We obtained spatially-multiplexed spectral measurements with random permuted Hadamard matrix and recovered the HSI with sparsity in wavelet domain.
\end{enumerate}
We define reconstruction SNR as $\text{rsnr} = 20\log_{10}\left( \frac{\|\bfx\|_F}{\|\bfx - \hat{\bfx}\|_F} \right)$, where $\|\cdot\|_F$ is the Frobenius norm and $\hat{\bfx}$ is the recovered version of $\bfx$.
\begin{figure}[!tt]
	\centering
	\begin{subfigure}[t]{0.22\columnwidth}
		\centering
		\includegraphics[width=\textwidth]{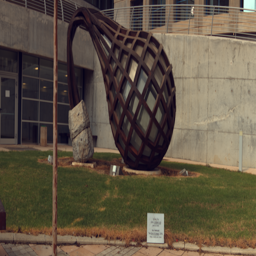}
		\caption{Ground truth}
	\end{subfigure}
	\hspace{0.1em}
	\begin{subfigure}[t]{0.22\columnwidth}
		\centering
		\includegraphics[width=\textwidth]{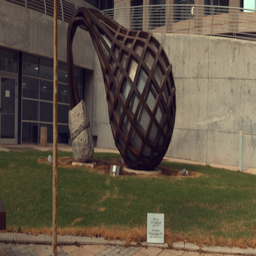}
		\caption{KRISM\\$47.7$dB}
	\end{subfigure}
	\hspace{0.1em}
	\begin{subfigure}[t]{0.22\columnwidth}
		\centering
		\includegraphics[width=\textwidth]{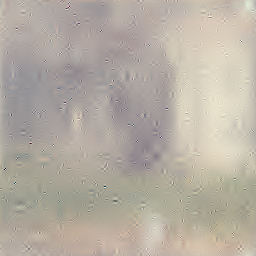}
		\caption{Sun et al.\\$32.5$dB}
	\end{subfigure}
	\hspace{0.1em}
	\begin{subfigure}[t]{0.22\columnwidth}
		\centering
		\includegraphics[width=\textwidth]{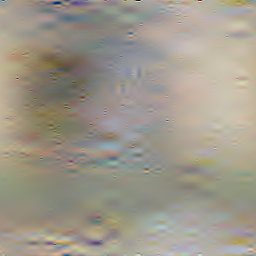}
		\caption{Kittle et al.\\$30.2$dB}
	\end{subfigure}
	\caption{Performance of multi-frame methods at high compression. We show a simulated example of recovery with $N/M = 43$ with (a) KRISM, (b) \citet{sun2009compressive} and (c) \citet{kittle2010multiframe}. Existing multi-frame techniques do not work well under high compression ratio. While KRISM recovers spatial images accurately, \cite{sun2009compressive,kittle2010multiframe} lead to severe loss in resolution.}
	\label{fig:high_comp}
\end{figure}
\begin{figure*}[!tt]
	\centering
	\begin{subfigure}[t]{0.2\textwidth}
		\centering
		\includegraphics[width=\textwidth]{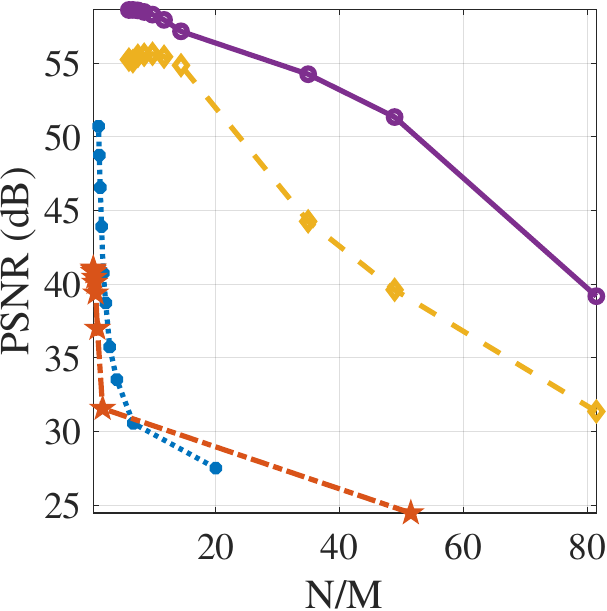}
		\caption{Car scene}
	\end{subfigure}
	\quad
	\begin{subfigure}[t]{0.2\textwidth}
		\centering
		\includegraphics[width=\textwidth]{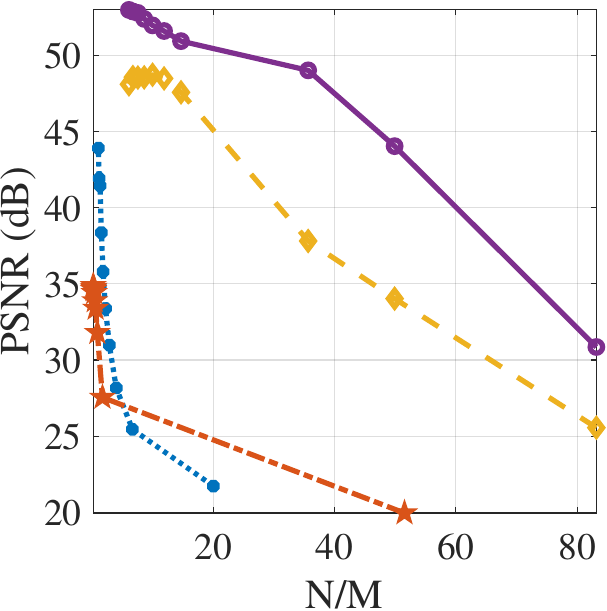}
		\caption{Checker scene}
	\end{subfigure}
	\quad
	\begin{subfigure}[t]{0.2\textwidth}
		\centering
		\includegraphics[width=\textwidth]{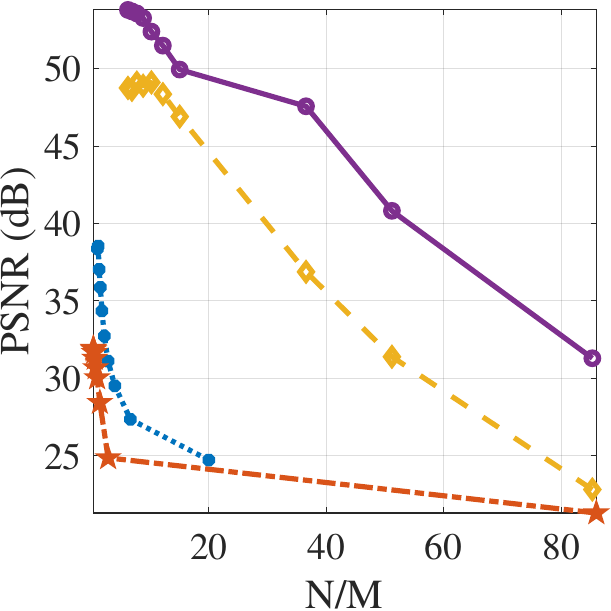}
		\caption{Dustbin scene}
	\end{subfigure}
	\quad
	\begin{subfigure}[t]{0.295\textwidth}
		\centering
		\includegraphics[width=\textwidth]{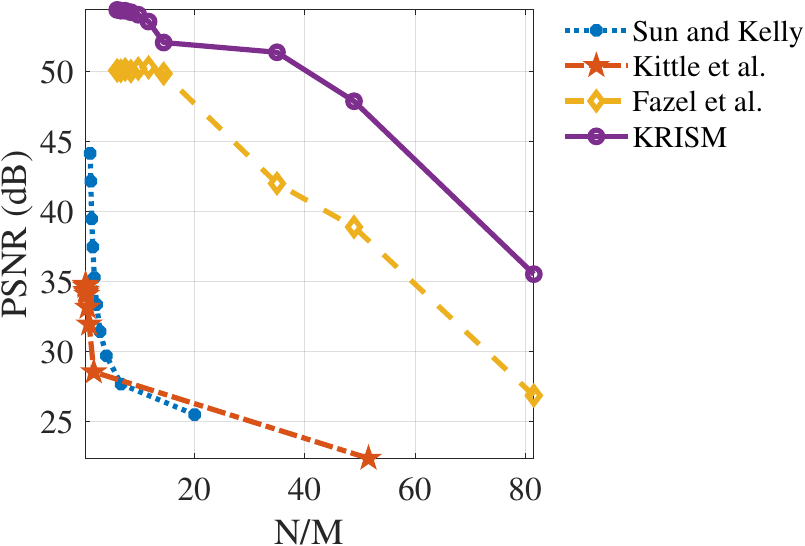}
		\caption{Bulb scene}
	\end{subfigure}
	\caption{Comparison of reconstruction SNR vs compression ratio for various methods on \cite{arad_and_ben_shahar_2016_ECCV} dataset. Simulations were done as described in Figure \ref{fig:synthetic_hr}. KRISM outperforms any other method by a larger margin in both approximation accuracy, as well as compression ratio.}
	\label{fig:compression_hr}
\end{figure*}
\begin{figure*}[!tt]
	\centering
	\begin{subfigure}[c]{0.45\textwidth}
		\begin{subfigure}[t]{0.32\columnwidth}
			\centering
			\includegraphics[width=0.955\textwidth]{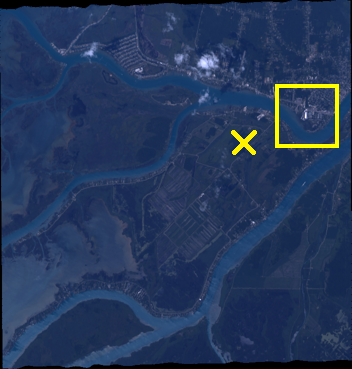}
			\caption{NASA AVIRIS ($352\times369\times224$)}
		\end{subfigure}
		\begin{subfigure}[t]{0.32\columnwidth}
			\centering
			\includegraphics[width=\textwidth]{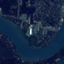}
			\caption{Fazel et al. \\(25.9dB, $N/M = 37$)}
		\end{subfigure}
		\begin{subfigure}[t]{0.32\columnwidth}
			\centering
			\includegraphics[width=\textwidth]{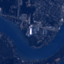}
			\caption{KRISM \\(47.8dB, $N/M = 37$)}
		\end{subfigure}
		\\
		\begin{subfigure}[t]{0.96\columnwidth}
			\centering
			\includegraphics[width=\textwidth]{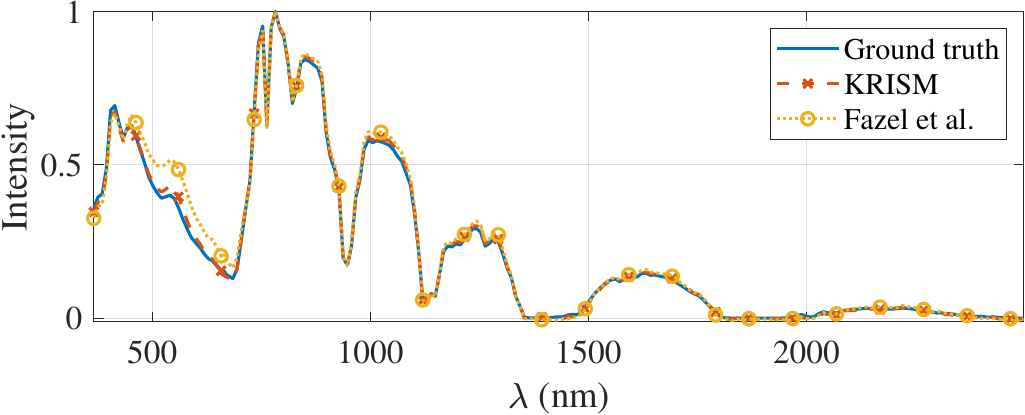}
		\end{subfigure}
	\end{subfigure}
	\quad
	\begin{subfigure}[c]{0.45\textwidth}
		\begin{subfigure}[b]{0.4\columnwidth}
			\centering
			\begin{subfigure}[t]{\columnwidth}
				\centering
				\includegraphics[width=0.955\textwidth]{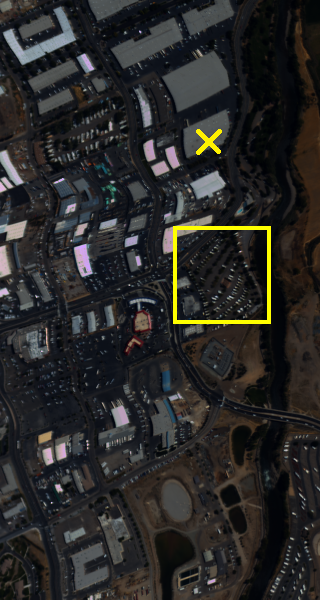}
				\caption{\citet{spectir} ($320\times600\times178$)}
			\end{subfigure}
		\end{subfigure}
		\begin{subfigure}[b]{0.59\columnwidth}
			\begin{subfigure}[t]{0.48\columnwidth}
				\centering
				\includegraphics[width=\textwidth]{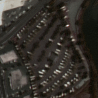}
				\caption{Row/col CS \\(41.3dB, $N/M = 30$)}
			\end{subfigure}
			\begin{subfigure}[t]{0.48\columnwidth}
				\centering
				\includegraphics[width=\textwidth]{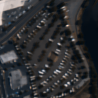}
				\caption{KRISM \\(46.5dB, $N/M = 30$)}
			\end{subfigure}
			\\
			\begin{subfigure}[t]{0.96\columnwidth}
				\centering
				\includegraphics[width=\textwidth]{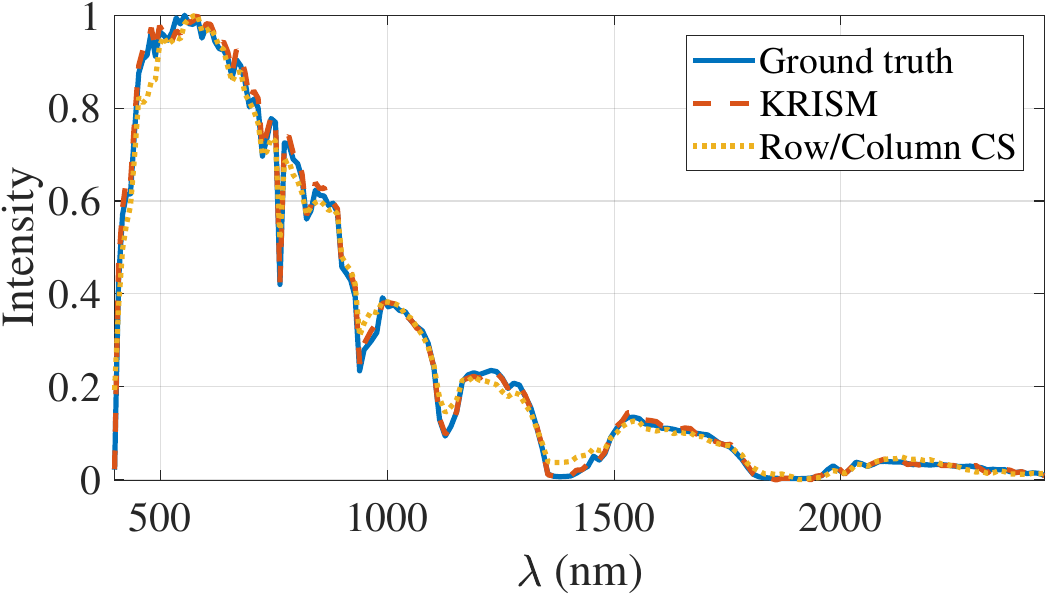}
				\caption{Spectrum at the marked location.}
			\end{subfigure}
		\end{subfigure}
	\end{subfigure}
	\caption{Comparision of Row/Col CS vs KRISM for large number of spectral bands with high spatial and spectral resolution, making a good candidate for KRISM. Simulations were done with 60dB readout noise, photon noise, and diffraction blur on spatial images and spectra. We show zoomed in image patches for each method and spectrum at pixel marked by a cross. For the same compression ratio, KRISM outperforms Row/Col CS by 10dB.}
	\label{fig:spectir}
\end{figure*}
\subsection{Performance with high spectral resolution}
The true potential of KRISM can be exploited when there are a large number of spectral bands, such as the ones in the dataset by  \citet{arad_and_ben_shahar_2016_ECCV}.
We keep compression low for competing methods as the accuracy scaled poorly with higher compressions (see Figure \ref{fig:high_comp}) 
Results on some representative examples have been show in Figure \ref{fig:synthetic_hr}.
Qualitatively, the reconstructed spatial images as well as the spectral signatures are very close to ground truth.
Figure \ref{fig:compression_hr} shows a comparison of reconstruction SNR as a function of compression ratios. 
As is evident, KRISM works significantly better than other methods despite very high compression ratios.
The closes competitor to KRISM is the Row/Column CS approach by \cite{fazel2008compressed}.
We show comparison between KRISM and Row/Column CS on one example from NASA's AVIRIS dataset consisting of of 224 spectral bands between 400-2400nm, and on \citet{spectir} dataset consisting of 178 spectral bands, making it a good example to test our method.
Results are shown in Figure \ref{fig:spectir}. For the same compression ration, KRISM offers a 10dB higher accuracy, and is qualitatively more accurate in both spatial images and spectral profiles. 
\subsection{Performance with low spectral resolution}
\begin{figure*}[!h]
	\begin{subfigure}[c]{0.12\textwidth}
		\includegraphics[width=\textwidth]{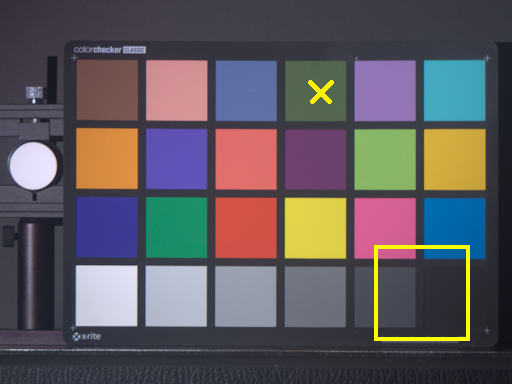}
		\caption{KAIST dataset \\ ($512\times384\times31$)\\\vphantom{1.5em}}
	\end{subfigure}
	\begin{subfigure}[c]{0.12\textwidth}
		\includegraphics[width=\textwidth]{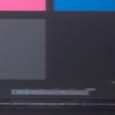}
		\caption{Choi et al.\\PSNR: 36.9dB, $N/M: 29$}
	\end{subfigure}
	\begin{subfigure}[c]{0.12\textwidth}
		\includegraphics[width=\textwidth]{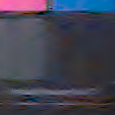}
		\caption{Kittle et al.\\PSNR: 32.9dB, $N/M: 5$}
	\end{subfigure}
	\begin{subfigure}[c]{0.12\textwidth}
		\includegraphics[width=\textwidth]{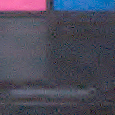}
		\caption{Sun and Kelly\\PSNR: 31.6dB, $N/M: 5$}
	\end{subfigure}
	\begin{subfigure}[c]{0.12\textwidth}
		\includegraphics[width=\textwidth]{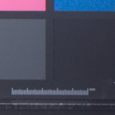}
		\caption{Fazel et al.\\PSNR: 32.5dB, $N/M: 5$}
	\end{subfigure}
	\begin{subfigure}[c]{0.12\textwidth}
		\includegraphics[width=\textwidth]{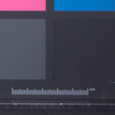}
		\caption{KRISM\\PSNR: 41.8dB, $N/M: 5$}
	\end{subfigure}
	\begin{subfigure}[c]{0.25\textwidth}
		\centering
		\includegraphics[width=\textwidth]{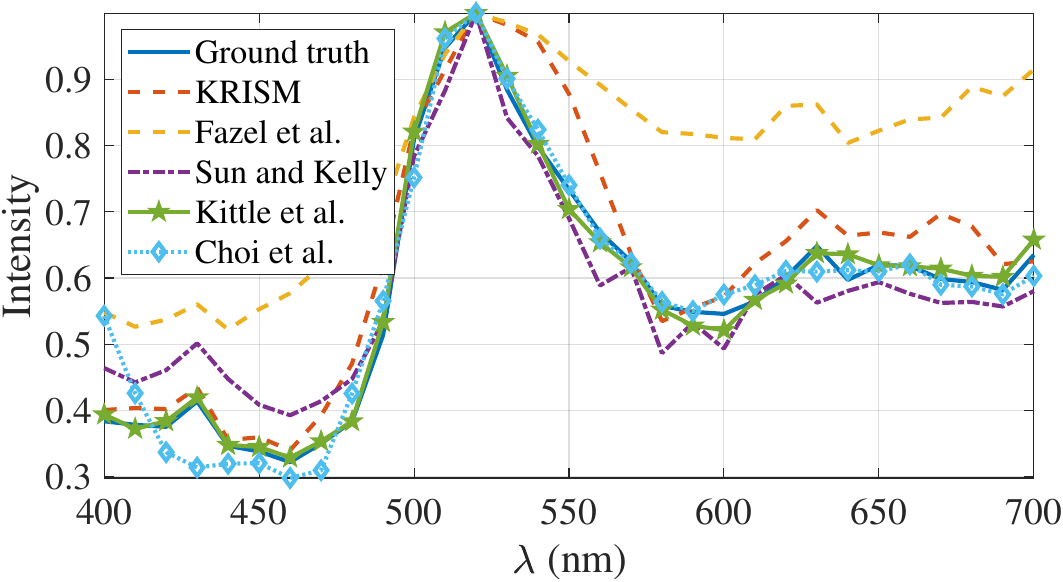}
		\caption{Spectrum at marked point\\\vphantom{1em}}
	\end{subfigure}
	\setcounter{subfigure}{0}
	\\
	\begin{subfigure}[c]{0.12\textwidth}
		\includegraphics[width=\textwidth]{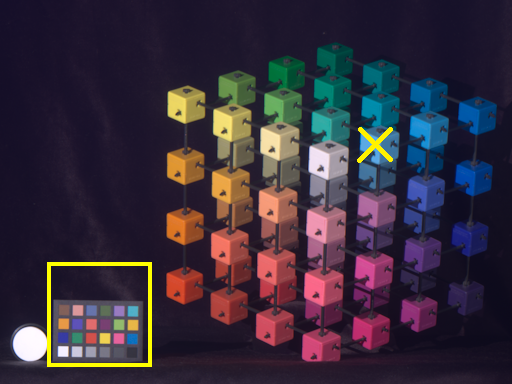}
		\caption{KAIST dataset \\ ($512\times384\times31$)\\\vphantom{1.5em}}
	\end{subfigure}
	\begin{subfigure}[c]{0.12\textwidth}
		\includegraphics[width=\textwidth]{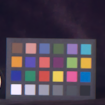}
		\caption{Choi et al.\\PSNR: 35.0dB, $N/M: 29$}
	\end{subfigure}
	\begin{subfigure}[c]{0.12\textwidth}
		\includegraphics[width=\textwidth]{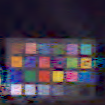}
		\caption{Kittle et al.\\PSNR: 32.4dB, $N/M: 5$}
	\end{subfigure}
	\begin{subfigure}[c]{0.12\textwidth}
		\includegraphics[width=\textwidth]{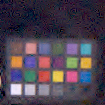}
		\caption{Sun and Kelly\\PSNR: 31.2dB, $N/M: 5$}
	\end{subfigure}
	\begin{subfigure}[c]{0.12\textwidth}
		\includegraphics[width=\textwidth]{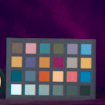}
		\caption{Fazel et al.\\PSNR: 30.6dB, $N/M: 5$}
	\end{subfigure}
	\begin{subfigure}[c]{0.12\textwidth}
		\includegraphics[width=\textwidth]{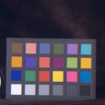}
		\caption{KRISM\\PSNR: 37.5dB, $N/M: 5$}
	\end{subfigure}
	\begin{subfigure}[c]{0.25\textwidth}
		\centering
		\includegraphics[width=\textwidth]{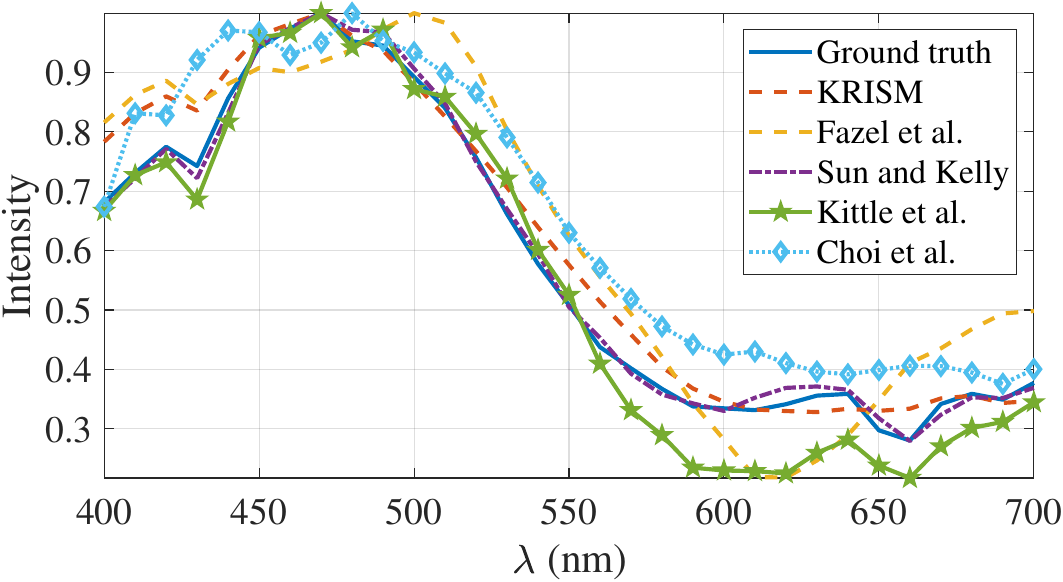}
		\caption{Spectrum at marked point\\\vphantom{1em}}
	\end{subfigure}
	\setcounter{subfigure}{0}
	\\
	\begin{subfigure}[c]{0.12\textwidth}
		\includegraphics[width=\textwidth]{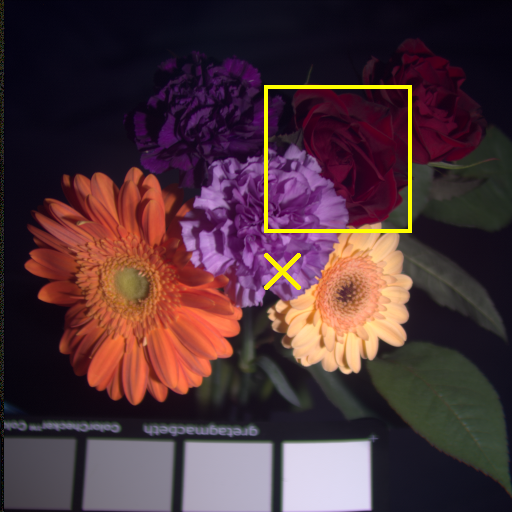}
		\caption{Yasuma et al. \\ ($512\times512\times31$)\\\vphantom{1.5em}}
	\end{subfigure}
	\begin{subfigure}[c]{0.12\textwidth}
		\includegraphics[width=\textwidth]{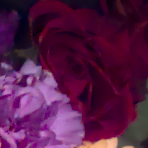}
		\caption{Choi et al.\\PSNR: 37.7dB, $N/M: 29$}
	\end{subfigure}
	\begin{subfigure}[c]{0.12\textwidth}
		\includegraphics[width=\textwidth]{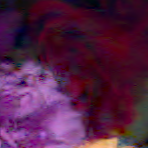}
		\caption{Kittle et al.\\PSNR: 35.2dB, $N/M: 5$}
	\end{subfigure}
	\begin{subfigure}[c]{0.12\textwidth}
		\includegraphics[width=\textwidth]{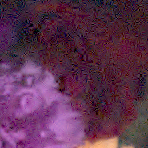}
		\caption{Sun and Kelly\\PSNR: 31.6dB, $N/M: 5$}
	\end{subfigure}
	\begin{subfigure}[c]{0.12\textwidth}
		\includegraphics[width=\textwidth]{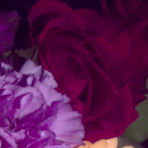}
		\caption{Fazel et al.\\PSNR: 40.6dB, $N/M: 5$}
	\end{subfigure}
	\begin{subfigure}[c]{0.12\textwidth}
		\includegraphics[width=\textwidth]{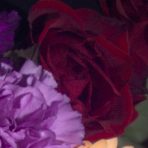}
		\caption{KRISM\\PSNR: 48.2dB, $N/M: 5$}
	\end{subfigure}
	\begin{subfigure}[c]{0.25\textwidth}
		\centering
		\includegraphics[width=\textwidth]{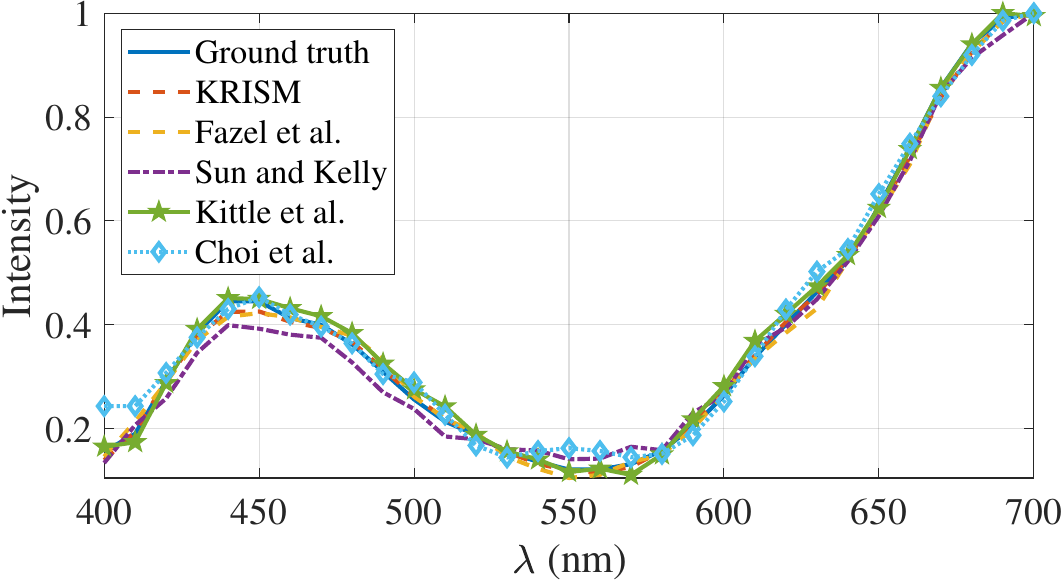}
		\caption{Spectrum at marked point\\\vphantom{1em}}
	\end{subfigure}
	\setcounter{subfigure}{0}
	\\
	\begin{subfigure}[c]{0.12\textwidth}
		\includegraphics[width=\textwidth]{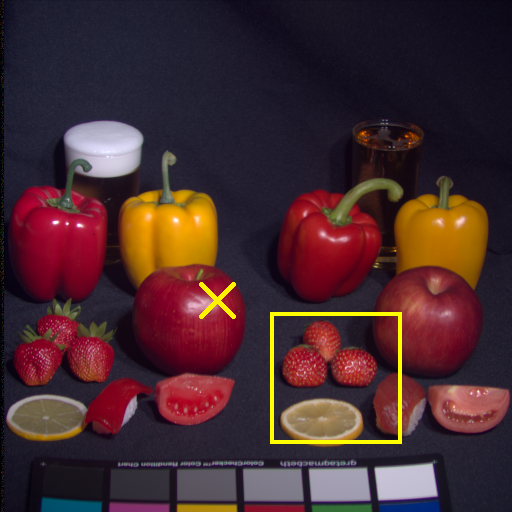}
		\caption{Yasuma et al. \\ ($512\times512\times31$)\\\vphantom{1.5em}}
	\end{subfigure}
	\begin{subfigure}[c]{0.12\textwidth}
		\includegraphics[width=\textwidth]{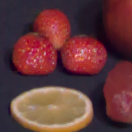}
		\caption{Choi et al.\\PSNR: 37.6dB, $N/M: 29$}
	\end{subfigure}
	\begin{subfigure}[c]{0.12\textwidth}
		\includegraphics[width=\textwidth]{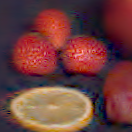}
		\caption{Kittle et al.\\PSNR: 36.4dB, $N/M: 5$}
	\end{subfigure}
	\begin{subfigure}[c]{0.12\textwidth}
		\includegraphics[width=\textwidth]{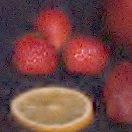}
		\caption{Sun and Kelly\\PSNR: 32.2dB, $N/M: 5$}
	\end{subfigure}
	\begin{subfigure}[c]{0.12\textwidth}
		\includegraphics[width=\textwidth]{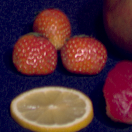}
		\caption{Fazel et al.\\PSNR: 32.8dB, $N/M: 5$}
	\end{subfigure}
	\begin{subfigure}[c]{0.12\textwidth}
		\includegraphics[width=\textwidth]{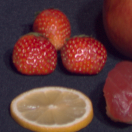}
		\caption{KRISM\\PSNR: 41.2dB, $N/M: 5$}
	\end{subfigure}
	\begin{subfigure}[c]{0.25\textwidth}
		\centering
		\includegraphics[width=\textwidth]{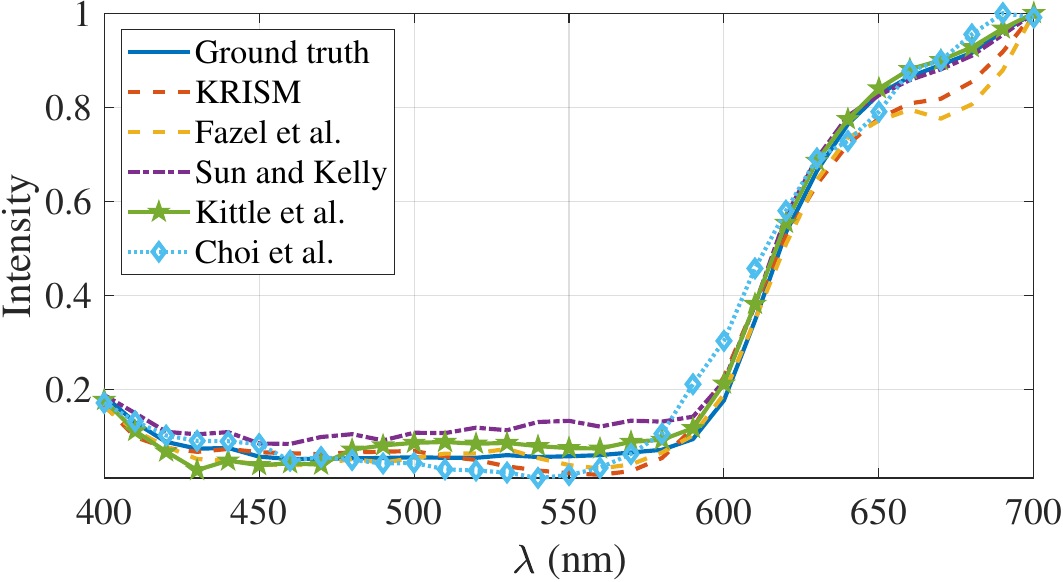}
		\caption{Spectrum at marked point\\\vphantom{1em}}
	\end{subfigure}
	\setcounter{subfigure}{0}
	\\
	\begin{subfigure}[c]{0.12\textwidth}
		\includegraphics[width=\textwidth]{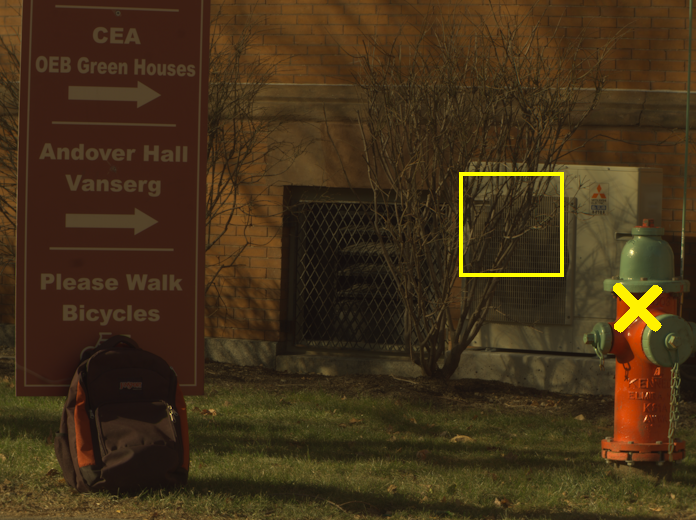}
		\caption{Chakrabarti et al. \\ ($696\times520\times31$)\\\vphantom{1.5em}}
	\end{subfigure}
	\begin{subfigure}[c]{0.12\textwidth}
		\includegraphics[width=\textwidth]{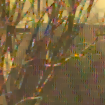}
		\caption{Choi et al.\\PSNR: 42.0dB, $N/M: 29$}
	\end{subfigure}
	\begin{subfigure}[c]{0.12\textwidth}
		\includegraphics[width=\textwidth]{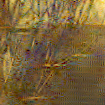}
		\caption{Kittle et al.\\PSNR: 38.9dB, $N/M: 5$}
	\end{subfigure}
	\begin{subfigure}[c]{0.12\textwidth}
		\includegraphics[width=\textwidth]{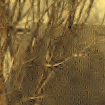}
		\caption{Sun and Kelly\\PSNR: 44.5dB, $N/M: 5$}
	\end{subfigure}
	\begin{subfigure}[c]{0.12\textwidth}
		\includegraphics[width=\textwidth]{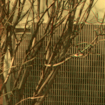}
		\caption{Fazel et al.\\PSNR: 45.6dB, $N/M: 5$}
	\end{subfigure}
	\begin{subfigure}[c]{0.12\textwidth}
		\includegraphics[width=\textwidth]{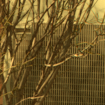}
		\caption{KRISM\\PSNR: 50.7dB, $N/M: 5$}
	\end{subfigure}
	\begin{subfigure}[c]{0.25\textwidth}
		\centering
		\includegraphics[width=\textwidth]{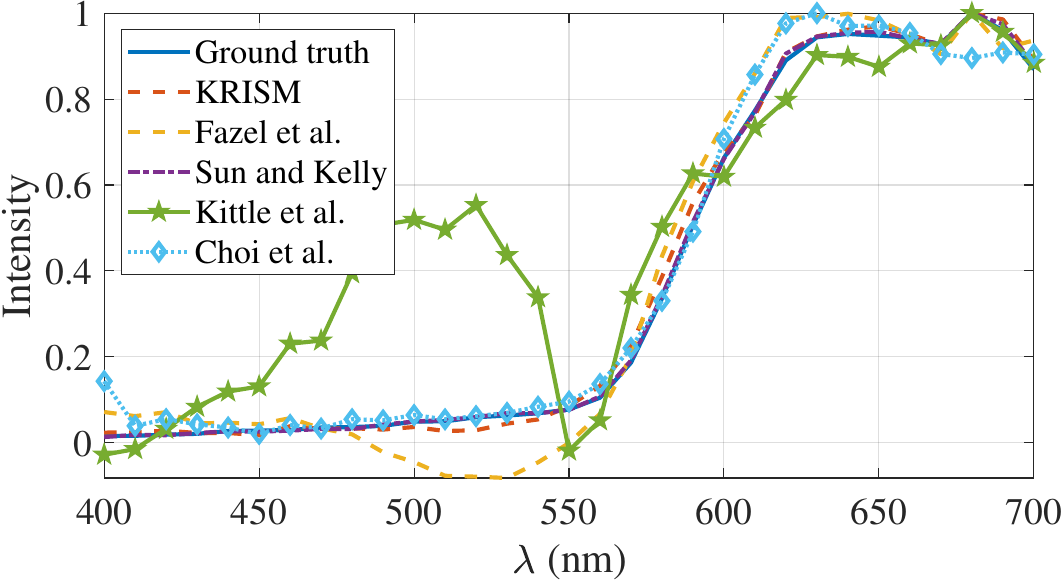}
		\caption{Spectrum at marked point\\\vphantom{1em}}
	\end{subfigure}
	\setcounter{subfigure}{0}
	\\
	\begin{subfigure}[c]{0.12\textwidth}
		\includegraphics[width=\textwidth]{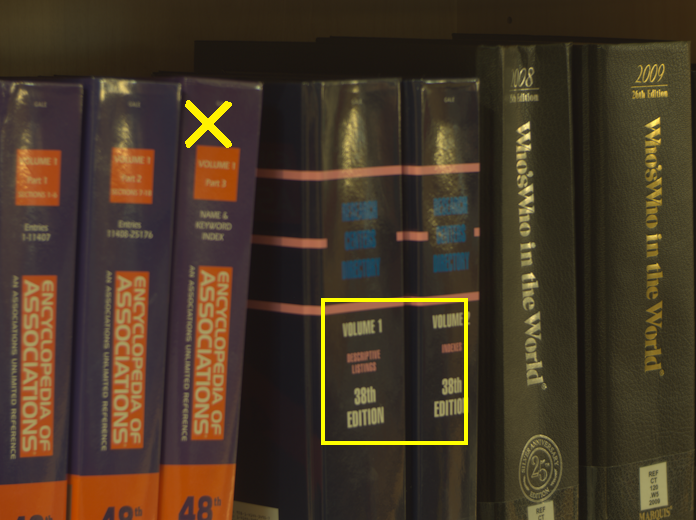}
		\caption{Chakrabarti et al. \\ ($696\times520\times31$)\\\vphantom{1.5em}}
	\end{subfigure}
	\begin{subfigure}[c]{0.12\textwidth}
		\includegraphics[width=\textwidth]{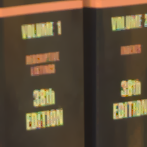}
		\caption{Choi et al.\\PSNR: 41.3dB, $N/M: 29$}
	\end{subfigure}
	\begin{subfigure}[c]{0.12\textwidth}
		\includegraphics[width=\textwidth]{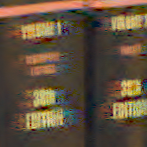}
		\caption{Kittle et al.\\PSNR: 36.0dB, $N/M: 5$}
	\end{subfigure}
	\begin{subfigure}[c]{0.12\textwidth}
		\includegraphics[width=\textwidth]{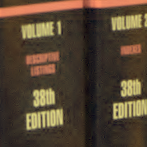}
		\caption{Sun and Kelly\\PSNR: 46.9dB, $N/M: 5$}
	\end{subfigure}
	\begin{subfigure}[c]{0.12\textwidth}
		\includegraphics[width=\textwidth]{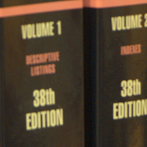}
		\caption{Fazel et al.\\PSNR: 47.9dB, $N/M: 5$}
	\end{subfigure}
	\begin{subfigure}[c]{0.12\textwidth}
		\includegraphics[width=\textwidth]{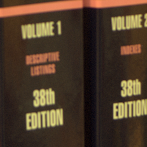}
		\caption{KRISM\\PSNR: 50.2dB, $N/M: 5$}
	\end{subfigure}
	\begin{subfigure}[c]{0.25\textwidth}
		\centering
		\includegraphics[width=\textwidth]{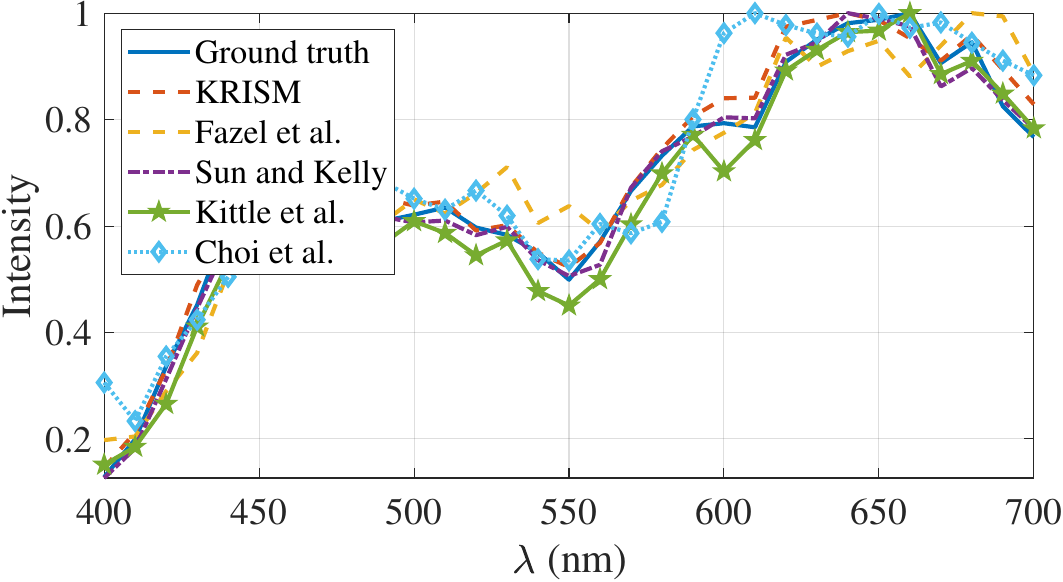}
		\caption{Spectrum at marked point\\\vphantom{1em}}
	\end{subfigure}
	\caption{Comparison of reconstructed images low spectral resolution. All experiments were performed with 60dB readout noise and poisson noise.  We show zoomed in image patches for each method and spectrum at pixel marked by a cross. For lower spectral resolution, KRISM offers limited benefits in compression ratios but is superior in terms of spatial and spectral reconstructions and overall accuracy.}
	\label{fig:synthetic_lr}
\end{figure*}
\begin{figure}[!tt]
	\centering
	\begin{subfigure}[t]{0.23\columnwidth}
		\centering
		\includegraphics[width=\columnwidth]{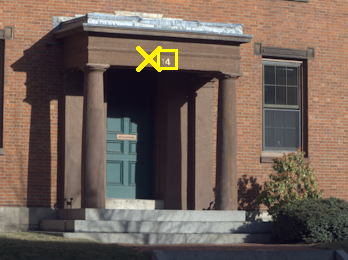}
		\caption{Lin et al. \\ ($512\times384\times31$)\\}
	\end{subfigure}
	\hspace{0.1em}
	\begin{subfigure}[t]{0.23\columnwidth}
		\centering
		\includegraphics[width=\columnwidth]{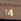}
		\caption{Ground truth}
	\end{subfigure}
	\hspace{0.1em}
	\begin{subfigure}[t]{0.23\columnwidth}
		\centering
		\includegraphics[width=\columnwidth]{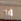}
		\caption{Lin et. al\\$N/M: 29$,\\ PSNR: $26.6$dB}
	\end{subfigure}
	\hspace{0.1em}
	\begin{subfigure}[t]{0.23\columnwidth}
		\centering
		\includegraphics[width=\columnwidth]{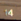}		\caption{Choi et al.\\$N/M: 29$,\\ PSNR: $33.3$dB}
	\end{subfigure}
	\\
	\begin{subfigure}[t]{0.23\columnwidth}
		\centering
		\includegraphics[width=\columnwidth]{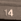}
		\caption{KRISM $K: 1$\\$N/M: 10$,\\ PSNR: $28.3$dB}
	\end{subfigure}
	\hspace{0.1em}
	\begin{subfigure}[t]{0.23\columnwidth}
		\centering
		\includegraphics[width=\columnwidth]{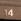}
		\caption{KRISM $K: 2$\\$N/M: 8$,\\ PSNR: $37.7$dB}
	\end{subfigure}
	\hspace{0.1em}
	\begin{subfigure}[t]{0.23\columnwidth}
		\centering
		\includegraphics[width=\columnwidth]{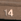}
		\caption{KRISM $K: 3$\\$N/M: 6$,\\ PSNR: $41.1$dB}
	\end{subfigure}
	\hspace{0.1em}
	\begin{subfigure}[t]{0.23\columnwidth}
		\centering
		\includegraphics[width=\columnwidth]{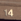}
		\caption{KRISM $K: 4$\\$N/M: 5$,\\ PSNR: $44.2$dB}
	\end{subfigure}
	\hspace{0.1em}
	\\
	\begin{subfigure}[t]{\columnwidth}
		\centering
		\includegraphics[width=\columnwidth]{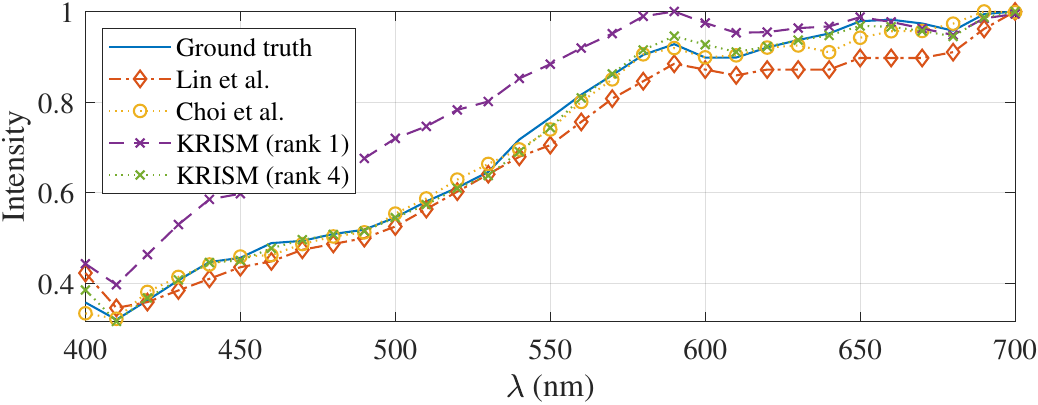}
	\end{subfigure}
	\caption{Evaluation on Door dataset from \cite{chakrabarti2011statistics}. We compared KRISM against methods from \cite{lin2014spatial} and \cite{deepcassi2017} on low-resolution spectra. We show zoomed in image patches for each method and spectrum at pixel marked by a cross. At settings close to snapshot sensing ($K = 1$), data-driven techniques perform better; with more iterations, KRISM achieves higher quality in spatial and spectral resolution.}
	\label{fig:door_sim}
\end{figure}
\begin{figure}[!tt]
	\centering
	\begin{subfigure}[t]{0.37\columnwidth}
		\centering
		\includegraphics[width=\textwidth]{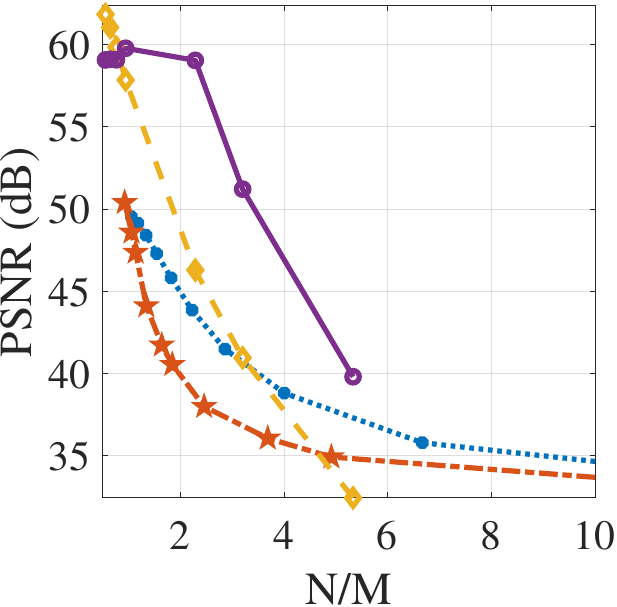}
		\caption{Outdoor scene}
	\end{subfigure}
	\quad
	\begin{subfigure}[t]{0.57\columnwidth}
		\centering
		\includegraphics[width=\textwidth]{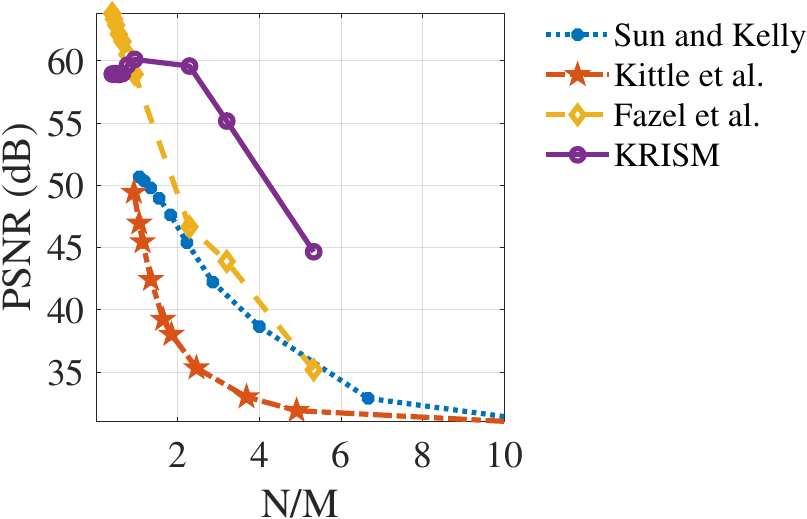}
		\caption{Books scene}
	\end{subfigure}
	\caption{Comparison of reconstruction SNR vs compression ratio for various methods on \cite{chakrabarti2011statistics} dataset. Simulations were done as described in Figure \ref{fig:synthetic_lr}. Despite lower compression ratios, KRISM promises greater overall performance.}
	\label{fig:compression_lr}
\end{figure}

Most of the visible HSI datasets contains 31-33 spectral bands between 400 - 700nm. In this regime, learning-based snapshot techniques such as \citet{deepcassi2017} and \cite{lin2014spatial} have better performance.
We used the dataset by \citet{chakrabarti2011statistics}, \citet{deepcassi2017} and \citet{yasuma2010generalized} for simulations with 31 spectral bands. Spatial resolution has been specified for individual images in Figure \ref{fig:synthetic_lr}.
We compare KRISM with varying number of measurements against snapshot techniques \cite{deepcassi2017,lin2014spatial} in Figure \ref{fig:door_sim}.
\FloatBarrier
We observe that in the setting closest to snapshot mode, \citet{deepcassi2017} and \citet{lin2014spatial} do outperform KRISM; this is to be expected since after a single iteration, KRISM provides only a rank-1 approximation.
As the number of KRISM iterations are increased (which allows approximations of higher ranks), KRISM performance improves.
We note that simulations for \cite{lin2014spatial} were performed with downsampled dataset and only on a select set of scenes as the recovery required several days for each scene even with parallelization.
Figure \ref{fig:compression_lr} shows recovery SNR as a function of compression for multi-frame techniques.
As was discussed in the main paper, KRISM is particularly effective for high resolution imaging. 
However, even with small number of bands, performance is superior in terms of spatial and spectral resolutions.
It is worth noting that learning-based snapshot technique by \cite{deepcassi2017} outperforms \cite{kittle2010multiframe} with fewer measurements.
This is expected, as it exploits the smooth nature of underlying spectra.
\end{appendices}

\end{document}